\documentstyle[12pt,bezier]{article}

\newcommand{\newsection}{    
\setcounter{equation}{0}
\section}
\def\appendix#1{
  \addtocounter{section}{1}
  \setcounter{equation}{0}
  \renewcommand{\thesection}{\Alph{section}}
  \section*{Appendix \thesection\protect\indent \parbox[t]{11.705cm} {#1} }
  \addcontentsline{toc}{section}{Appendix \thesection\ \ \ #1}
  }
\newcommand{\tr}[1]{\,{\rm tr}\,#1}
\newcommand{\ntr}[1]{\,\frac {\rm tr}{N}\,#1}
\def\e{{\,\rm e}\,}

\def\eop{\vspace*{\fill}\pagebreak}
\def\be{\begin{equation}}
\def\ee{\end{equation}}
\def\bea{\begin{eqnarray}}
\def\eea{\end{eqnarray}}
\def\LA{\left\langle}
\def\RA{\right\rangle}
\newcommand{\Di}{\,\hbox{Disc}_\nu\,}
\newcommand{\Co}{\,\hbox{Cont}_\nu\,}

\newcommand{\Cow}{\,\hbox{Cont}_\omega\,}
\newcommand{\rf}[1]{(\ref{#1})}
\newcommand{\eq}[1]{Eq.~(\ref{#1})}

\def\a{\alpha}
\def\b{\beta}
\def\d{\partial}
\def\L{\Lambda}
\def\l{\lambda}
\def\om{\omega}
\def\C{\Gamma}
\def\ka{\kappa}
\newcommand{\ie}{{\it i.e.}\ }
\newcommand{\p}{{\prime}}
\newcommand{\ra}{\rightarrow}
\newcommand{\fr}[2]{{\textstyle {#1 \over #2}}}

\newcommand{\ci}{\int_{C_1}\frac{d\omega}{2\pi i}}

\newcommand{\eps}{\varepsilon}
\newcommand{\non}{\nonumber \\*}
\newcommand{\tV}{\tilde{V}}
\newcommand{\VVp}{{\cal V}^\prime}
\newcommand{\hV}{\hat{V}}
\newcommand{\htV}{\hat{\tilde{V}}}
\newcommand{\tx}{\hat{x}_-}
\newcommand{\ty}{\hat{x}_+}
\newcommand{\tl}{\hat{\lambda}}
\newcommand{\tphi}{\hat{\phi}}
\newcommand{\tG}{\tilde{G}}
\newcommand{\hC}{\hat{C}}
\newcommand{\re}{\,\hbox{Re}\,}
\newcommand{\im}{\,\hbox{Im}\,}

\def\fun#1#2{\lower3.6pt\vbox{\baselineskip0pt\lineskip.9pt
\ialign{$\mathsurround=0pt#1\hfil##\hfil$\crcr#2\crcr\sim\crcr}}}

\begin{document}

\begin{titlepage}
\begin{flushright}
YM-4-94 \\
July, 1994 \\
{\small hep-th/9408029}
\end{flushright}
\vspace{0.6cm}

\begin{center}
 {\LARGE  Critical Scaling and Continuum Limits}\\
\vspace{0.6cm}
{\LARGE in the $D>1$ Kazakov--Migdal Model}
\end{center} \vspace{0.5cm}
\begin{center}
{\large Yu.\ Makeenko}\footnote{E--mail:
\ makeenko@vxitep.itep.msk.su \ / \
 makeenko@nbivax.nbi.dk \ }
\\ \mbox{} \\
{\it Institute of Theoretical and Experimental Physics,}
\\ {\it B. Cheremushkinskaya 25, 117259 Moscow, Russian Federation}
\\ \vskip .2 cm
and  \\  \vskip .2 cm
{\it The Niels Bohr Institute,} \\
{\it Blegdamsvej 17, 2100 Copenhagen, Denmark}
\end{center}

\vskip 1.0 cm
\begin{abstract}
I investigate the Kazakov--Migdal (KM) model --- the Hermitean
gauge-invariant matrix model on a $D$-dimensional lattice.
I utilize an exact large-$N$ solution of the KM model with a logarithmic
potential to examine its critical behavior. I find
critical lines associated with $\gamma_{str}=-1/2$ and $\gamma_{str}=0$
as well as
a tri-critical point associated with  a Kosterlitz--Thouless phase transition.
The continuum theories are constructed expanding around the critical points.
The one associated with \mbox{$\gamma_{str}=0$}
coincides with the standard $d=1$
string while the Kosterlitz--Thouless phase
transition separates it from that with $\gamma_{str}=-1/2$ which is
indistinguishable from pure $2D$ gravity for local
observables but has a continuum limit for correlators of extended
Wilson loops at large distances due to a singular behavior of
the Itzykson--Zuber correlator of the gauge fields.
I reexamine the KM model with an arbitrary potential
in the large-$D$ limit and show that it reduces at large $N$
to a one-matrix model whose potential is determined self-consistently.
A relation with discretized random surfaces is established via the gauged Potts
model which is equivalent to the KM model at large $N$ providing the
coordination numbers coincide.  \end{abstract}


\eop
\end{titlepage}
\setcounter{page}{2}

\newsection{Introduction}

Matrix models are usually associated with discretized random
surfaces or strings in a $d$$\leq$$1$-dimensional embedding space and, in
particular, with two-dimensional quantum gravity.
The simplest Hermitean one-matrix model corresponds
to pure gravity while a chain of Hermitean matrices
describes two-dimensional gravity interacting with $d$$\leq$$1$ matter.
The long-standing problem with this approach is that the stringy phase does not
exist above the $d$$=$$1$ barrier (see Ref.~\cite{Amb93} for a review).

A natural multi-dimensional extension of this construction is
the Kazakov--Migdal (KM) model~\cite{KM92} which
is defined by the partition function
\be
Z_{KM}=\int \prod_{\{x,y\}} dU_{xy} \prod_x
 d\phi_x \e^{  N \tr{}\left(- \sum_x V(\phi_x)+ c
\sum_{\{x,y\}} \phi_x U_{xy} \phi_{y} U_{xy}^\dagger \right)}\,.
\label{spartition}
\ee
Here the integration over the gauge field $U_{xy}$ is over the Haar measure
on $SU(N)$ at each link of a $D$-dimensional
hypercubic lattice with $x$ labeling
its sites and $\{xy\}$ labeling the link from the site $x$ to the neighbor site
$y$. The model~\rf{spartition} obviously recovers the standard
open matrix chain if
the lattice is just a one-dimensional sequence of points for which the gauge
field can be absorbed by a unitary transformation of $\phi_x$.

The KM model was originally introduced in the context of
induced lattice gauge
theories~\cite{KM92}. Its remarkable property is an existence of
self-consistent scaling solutions with nontrivial critical
indices~\cite{Mig92a} for a quartic potential $V$ in~\rf{spartition}
at large $N$ and any $D$. Some other investigations of a critical behavior of
the KM model have been performed recently~\cite{Bou93,MM94}.  There is however
a number of problems with the scaling solutions.  In particular, it is not
clear what physical system are they associated with.

{}From this point of view it is instructive to look at exact large-$N$
solutions of the KM model and to pass to the continuum by approaching
critical points. The only exact solutions of
the KM model are known for quadratic~\cite{Gro92} and
logarithmic~\cite{Mak93} potentials. The spectral density describing
the distribution of eigenvalues of a $x$-independent saddle-point matrix,
which dominates the path integral~\rf{spartition} at large $N$, is
quite similar in both cases to that for a one-matrix model
and was obtained by standard methods (see Ref.~\cite{Mak93a} for a review).
The solution of the KM model with the logarithmic potential
reduced~\cite{Mak93} to algebraic equations for the end points of the
eigenvalue support --- the boundary equations.

In the present paper I investigate the boundary equations of the KM model
with the logarithmic potential which explicitly determine the solution%
\footnote{A similar analysis is performed in Ref.~\cite{PW94}.}
and calculate specific heat (or string susceptibility) at large $N$.
The model has a rich critical behavior: a critical line associated
with $\gamma_{str}=-1/2$, a critical line associated with $\gamma_{str}=0$
and a tri-critical point where a Kosterlitz--Thouless phase transition
between these two phases occurs.
In order to construct the continuum theories I perform an expansion
around the critical points which is quite similar to the one for
two-dimensional gravity. The continuum limit associated with $\gamma_{str}=0$
coincides with the standard $d=1$ string
(or $2D$ gravity plus critical matter).
A Kosterlitz--Thouless phase transition separates this phase from the one
with $\gamma_{str}=-1/2$ which is indistinguishable from pure $2D$
gravity (or a $d=0$ string) for local observables which live at the same
site of the lattice. There is, however, another type of observables ---
extended Wilson loops --- for which the continuum limit sets up
at distances $L\sim 1/\sqrt{\eps}$ (where $\eps$ characterizes deviation from
the critical point) due to a singular behavior of the Itzykson--Zuber
correlator of the gauge fields.
This phenomenon occurs in the vicinity of the tri-critical point
where the Itzykson--Zuber correlator changes its behavior.

In order to compare with previous results I reexamine the KM model
in the large-$D$ limit and show that at large $N$ it reduces
for arbitrary $V$ in~\rf{spartition} and $c\sim 1/D$
to a one-matrix model whose potential is determined self-consistently.
While this one-matrix model has generically $\gamma_{str}=-1/2$,
one can obtain $\gamma_{str}=1/(k+1)$ ($k\geq1$) by tuning the value of $c$
quite similarly to Refs.~\cite{DDSW90,Kor92,ABC93}.
For the involved logarithmic potential one obtains in this way
$\gamma_{str}=0$ in agreement with exact solution.

In order to relate the KM model with discretized
random surfaces and strings, I propose a matrix model
---  the gauged Potts model --- which is equivalent to the KM model at large
$N$ providing the coordination numbers coincide.
The gauged Potts model has a natural connection with
discretized random surfaces and is convenient for interpreting the results
obtained in this paper.
The proof of equivalence of the gauged Potts and KM models
is given via loop equations which reduce at
large $N$ to a one-link equation that is similar to the one for
the Hermitean two-matrix model~\cite{2mamo,Alf93,Sta93}.
This reduction and, therefore, the equivalence hold in the strong coupling
phase where the vacuum expectation values of the closed Wilson loops
of the gauge field vanish except for those of vanishing minimal area.

This paper is organized as follows.
Sect.~2 is devoted to the description of the KM model with the logarithmic
potential, its one-cut solution at large $N$ and the calculation of
string susceptibility. Critical points and the phase structure of
the model are obtained. In Sect.~3 an expansion around
the critical points which specifies the continuum limits of the model
is performed. The behavior of the Itzykson--Zuber
correlator of gauge fields in the continuum is studied.
In Sect.~4 the large-$D$ limit of the KM model with an arbitrary potential is
discussed. In Sect.~5 the gauged Potts model is introduced and its equivalence
to the KM model at large $N$ is proven. This section also contains a
description of the large-$N$ solution of these models.
In Sect.~6 I discuss the results and some related problems for future
investigations.
Appendix~A is devoted to a non-standard behavior of the eigenvalue support
which leads to $\gamma_{str}=0$.
Appendix~B contains a proof of the convolution formula for the continuum
Itzykson--Zuber correlators which is used in Sect.~3.
Appendix~C contains the derivation of the loop equations which are considered
in Sect.~5.

\newsection{Explicit solution for logarithmic potential}

The distribution of eigenvalues of $\phi_x$ for the KM model
with the logarithmic potential coincides at large $N$
with that for a one-matrix model
which interpolates between cubic and Penner potentials
and is solved in this section by the standard technique.
The string susceptibility of the KM model with the logarithmic
potential is calculated at large $N$. It reveals
a rich critical behavior: a critical line associated
with $\gamma_{str}=-1/2$, a critical line associated with $\gamma_{str}=0$
and a tri-critical point where a Kosterlitz--Thouless phase transition
occurs.

\subsection{The KM model with logarithmic potential \label{logpot}}

Besides the Gaussian case~\cite{Gro92} the only exact solution of the
KM model~\rf{spartition} is known for the logarithmic potential~\cite{Mak93}
\footnote{The parameter $\a$ is related to the original parameters of
Ref.~\cite{Mak93} as follows: $\a=ab+c$. }
\be
V(\phi_x)= -\a \log{(b-\phi_x)}- (2D-1) (\a+1) \log{(a+\phi_x)}
+[(2D-1)b-a] \phi_x \,.
\label{V}
\ee
We put for simplicity the coefficient in front of the kinetic term
in~\rf{spartition} $c=1$.
The eigenvalue distribution of the ($x$-independent) saddle-point
configuration coincides for the model~\rf{spartition}
with the potential~\rf{V} and for the Hermitean
one-matrix model with the potential
\be
\tV(\phi)= -\a \log{(b-\phi)}+(\a+1) \log{(a+\phi)}
-(b+a) \phi \,,
\label{tV}
\ee
which is recovered by the potential~\rf{V} at $D=0$.

The potentials~\rf{V} and \rf{tV} can be simplified shifting $\phi_x$ by a
constant value:
\be
\phi_x = \tphi_x +\frac{b-a}{2}\,\hbox{I}  \,.
\label{tphi}
\ee
The corresponding shift of the kinetic term in the action is
\be
\ntr \phi_x U_{xy} \phi_y U^\dagger_{xy} =
\ntr \tphi_x U_{xy} \tphi_y U^\dagger_{xy} + \frac{b-a}{2}
\left( \ntr \tphi_x + \ntr \tphi_y \right) \,.
\ee
Absorbing the last term on the r.h.s.\ into the new potential, $\hV(\tphi)$,
and introducing
\be
\b = \frac{a+b}{2} \,,
\label{beta}
\ee
one gets
\be
\hV(\tphi_x)= -\a \log{(\b-\tphi_x)}- (2D-1) (\a+1) \log{(\b+\tphi_x)}
+2(D-1)\b \tphi_x
\label{hV}
\ee
and
\be
\htV(\tphi)= -\a \log{(\b-\tphi)}+(\a+1) \log{(\b+\tphi)}
- 2\b \tphi \,.
\label{htV}
\ee
Note, that both Eqs.~\rf{hV} and \rf{htV} can be straightforwardly obtained
from Eqs.~\rf{V} and \rf{tV} substituting $a=b=\b$.

The potential~\rf{hV} depends on two parameters $\a$ and $\b$. However, one
more parameter $\eta$ appears in the perturbative expansion when \be \tphi_x
= \eta \, \hbox{I}\, + \phi_x \label{eta} \ee and the expansion goes in
$\phi_x$. Comparing  with \eq{tphi}, one identifies $\eta$ with $(a-b)/2$.
Therefore, the expression~\rf{V} is convenient for the perturbative expansion.

The perturbative expansion starts from the Gaussian model which is
characterized by the quadratic potential
\be
V(\phi_x) = \frac{m_0^2}{2} \phi^2_x
\label{qpot}
\ee
where
\be
m^2_0 = \frac ab + (2D-1) \frac ba
\label{m0}
\ee
as it follows from~\rf{V} in the limit~\cite{Mak93}
\be
\a=ab\,,~~~~ a \sim b\ra\infty~~~~\hbox{and} ~~\phi_x\sim1
{}~~~~~~\hbox{(quadratic potential)}\,.
\label{gausslimit}
\ee
The analogous parameter for the one-matrix model~\rf{tV} is
\be
\mu = \frac ab - \frac ba \,.
\label{mu}
\ee
Solving the quadratic equation~\rf{m0} for $a/b$ versus $m_0^2$ and
substituting into \eq{mu}, one finds
\be
\mu=\frac{(D-1)m_0^2+D\sqrt{m^{4}_0-4 (2D-1)}}{(2D-1)}
\label{mum0}
\ee
which coincides with the solution of Ref.~\cite{Gro92}.

A more narrow region of the parameters than~\rf{gausslimit}:
\be
\a=ab~,~~~a,b\ra \infty~,~~~
(a-b) \sim \phi \sim b^{\frac 13} \,,
\label{cubiclimit}
\ee
is of special interest because the potential~\rf{tV}
of the associated one-matrix model then reduces to a cubic one.

It is worth mentioning that the normalization of the parameters of the
potential~\rf{V} is chosen to make \rf{tV} to be $D$ independent. It were be
more conventional to have instead a $D$-independent potential $V$ moving the
$D$-dependence to $\tV$. The Gaussian formulas~\rf{m0} and \rf{mum0} show how
this can be done.

For purposes of the perturbative expansion it is useful to restore the hopping
parameter $c$ in front of the kinetic term, like it enters the exponent in
\eq{spartition}. This can be achieved by the rescaling
\be
\{\phi_x,b\}\ra \sqrt{c} \{\phi_x,b\}~,~~~a \ra \frac{1}{\sqrt{c}} a
\label{resc}
\ee
at fixed $\a$. This results in the potentials
\be
V(\phi_x)= -\a \log{(b-\phi_x)}- (2D-1)
(\a+1) \log{(a+c\phi_x)} +[(2D-1)cb-a] \phi_x
\label{Vc}
\ee
and \be \tV(\phi)= -\a
\log{(b-\phi)}+(\a+1) \log{(a+c\phi)} -(a+cb) \phi \,.
\label{tVc}
\ee
It is easy to see that the potential~\rf{tVc} of the associated one-matrix
model becomes the Penner one~\cite{Pen87}
\be
\tV_{Penner}(\phi) = -\a \log{(b-\phi)} -a \phi
\label{Pennerpot}
\ee
as $c\ra0$.

\subsection{One-cut solution}

The one-cut solution for
\be
E_\l \equiv \LA
\ntr{}\Big( \frac{1}{\l-\phi_x} \Big) \RA ,
\label{2.19}
\ee
where the average is defined with the same measure as in~\rf{spartition},
is given by the general formula~\cite{Mig83}
\be
E_\l = \int_{C_1} \frac{d\om}{4\pi i} \frac{\tilde{V}^\p(\om)}{(\l-\om)}
\frac{\sqrt{(\l-x_-)(\l-x_+)}}{\sqrt{(\om-x_-)(\om-x_+)}}
\label{EvstV}
\ee
where the ends of the cut, $x_\pm$, are determined by the asymptotic
conditions
\be
\ci \frac{\tilde{V}^\p(\om)}{\sqrt{(\om-x_-)(\om-x_+)}} =0\,,~~~~
\ci \frac{\om \tilde{V}^\p(\om)}{\sqrt{(\om-x_-)(\om-x_+)}} =2 \,.
\label{xandy}
\ee
The contour $C_1$ encircles counterclockwise the cut
leaving outside singularities of $\tilde{V}^\p(\om)$
and the pole at $\l=\om$ so that the
integration over $\omega$ on the l.h.s.\ of \eq{EvstV} plays the role of a
projector picking up negative powers of $\l$.
It implies that the branch cut of $E_\l$ does not pass through
singularities (or branch cuts) of $\tV'(\l)$. This will be the case for our
meromorphic $\tV'(\l)$, which has poles at $\l=-a$ and $\l=b$, providing they
lie outside of the cut of $E_\l$.

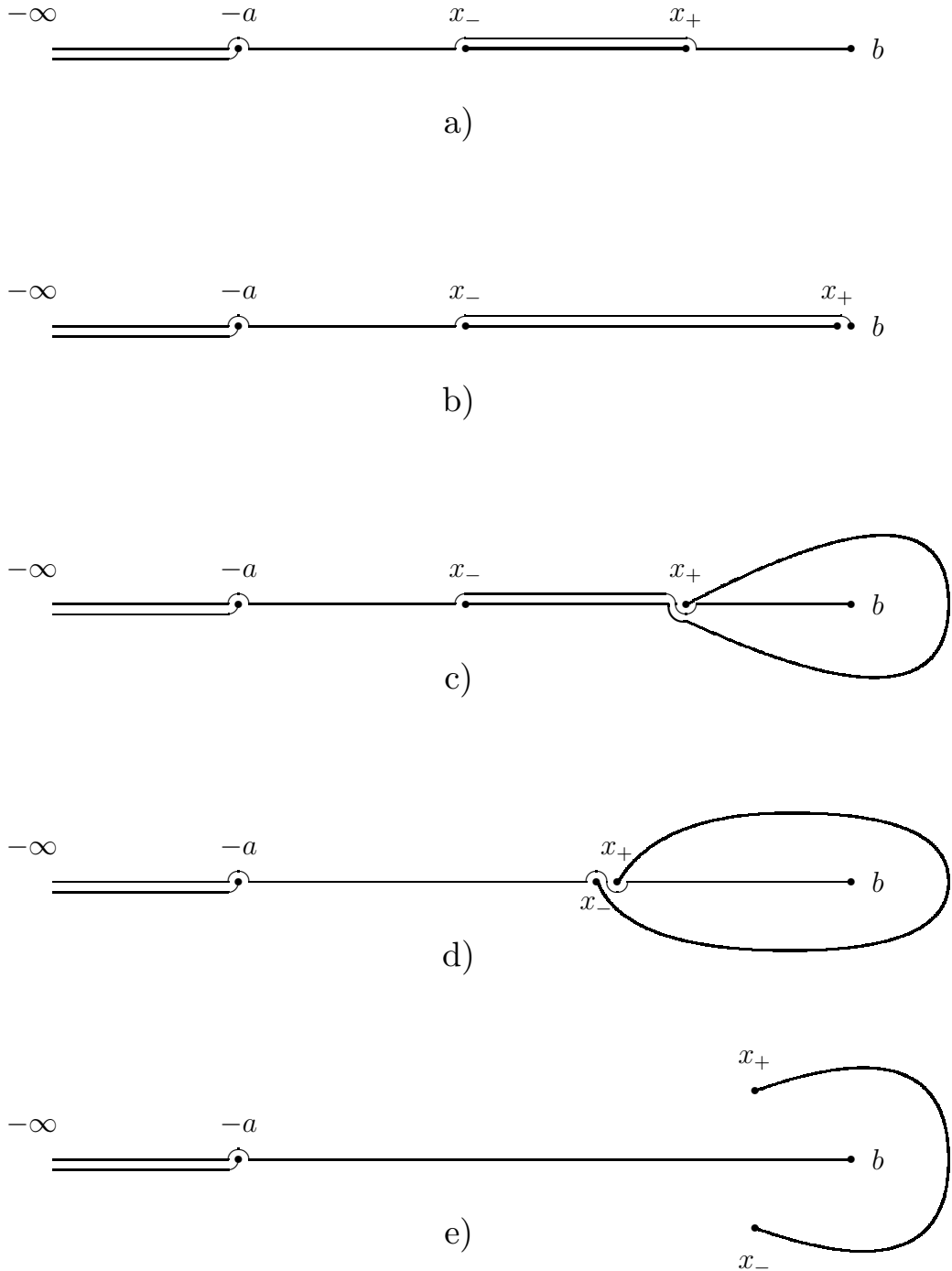
\begin{figure}[p]
\unitlength=1.00mm
\linethickness{0.6pt}
\begin{picture}(160.1,40.00)(-30,80)
\put(104.00,100.00){\circle*{1.00}}
\put(104.00,100.00){\line(-1,0){22.50}}
\put(80.00,100.00){\oval(3.00,3.00)[rt]}
\put(80.00,101.50){\line(-1,0){32.00}}
\put(48.00,100.00){\oval(3.00,3.00)[lt]}
\put(46.50,100.00){\line(-1,0){30.00}}
\put(15.00,100.00){\oval(3.00,3.00)[t]}
\put(15.00,100.00){\circle*{1.00}}
\put(13.50,100.00){\line(-1,0){25.50}}
\put(13.50,100.00){\oval(3.00,3.00)[rb]}
\put(13.50,98.50){\line(-1,0){25.50}}
\thicklines
\put(48.00,100.00){\circle*{1.00}}
\put(48.00,100.00){\line(1,0){32.00}}
\put(80.00,100.00){\circle*{1.00}}
\put(48.00,104.00){\makebox(0,0)[cb]{$x_-$}}
\put(80.00,104.00){\makebox(0,0)[cb]{$x_+$}}
\put(107.00,100.00){\makebox(0,0)[lc]{$b$}}
\put(15.00,104.00){\makebox(0,0)[cb]{$-a$}}
\put(-15.00,104.00){\makebox(0,0)[cb]{$-\infty$}}
\put(47.00,89.00){\makebox(0,0)[cc]{{\large a)}}}
\end{picture}
\begin{picture}(160.1,40.00)(-30,80)
\put(104.00,100.00){\circle*{1.00}}
\put(102.50,100.00){\oval(3.00,3.00)[rt]}
\put(102.50,101.50){\line(-1,0){54.50}}
\put(48.00,100.00){\oval(3.00,3.00)[lt]}
\put(46.50,100.00){\line(-1,0){30.00}}
\put(15.00,100.00){\oval(3.00,3.00)[t]}
\put(15.00,100.00){\circle*{1.00}}
\put(13.50,100.00){\line(-1,0){25.50}}
\put(13.50,100.00){\oval(3.00,3.00)[rb]}
\put(13.50,98.50){\line(-1,0){25.50}}
\thicklines
\put(48.00,100.00){\circle*{1.00}}
\put(48.00,100.00){\line(1,0){54.00}}
\put(102.00,100.00){\circle*{1.00}}
\put(48.00,104.00){\makebox(0,0)[cb]{$x_-$}}
\put(102.00,104.00){\makebox(0,0)[cb]{$x_+$}}
\put(107.00,100.00){\makebox(0,0)[lc]{$b$}}
\put(15.00,104.00){\makebox(0,0)[cb]{$-a$}}
\put(-15.00,104.00){\makebox(0,0)[cb]{$-\infty$}}
\put(47.00,89.00){\makebox(0,0)[cc]{{\large b)}}}
\end{picture}
\begin{picture}(160.1,40.00)(-30,80)
\put(104.00,100.00){\circle*{1.00}}
\put(104.00,100.00){\line(-1,0){22.50}}
\put(80.00,100.00){\oval(3.00,3.00)[b]}
\put(77.00,100.00){\oval(3.00,3.00)[rt]}
\put(77.00,101.50){\line(-1,0){29.00}}
\put(48.00,100.00){\oval(3.00,3.00)[lt]}
\put(46.50,100.00){\line(-1,0){30.00}}
\put(15.00,100.00){\oval(3.00,3.00)[t]}
\put(15.00,100.00){\circle*{1.00}}
\put(13.50,100.00){\line(-1,0){25.50}}
\put(13.50,100.00){\oval(3.00,3.00)[rb]}
\put(13.50,98.50){\line(-1,0){25.50}}
\thicklines
\put(48.00,100.00){\circle*{1.00}}
\put(48.00,100.00){\line(1,0){29.50}}
\put(80.00,100.00){\oval(5.00,5.00)[lb]}
\bezier{256}(80.00,97.50)(118.00,80.00)(118.00,100.00)
\bezier{244}(80.00,100.00)(118.00,120.00)(118.00,100.00)
\put(80.00,100.00){\circle*{1.00}}
\put(48.00,104.00){\makebox(0,0)[cb]{$x_-$}}
\put(80.00,104.00){\makebox(0,0)[cb]{$x_+$}}
\put(107.00,100.00){\makebox(0,0)[lc]{$b$}}
\put(15.00,104.00){\makebox(0,0)[cb]{$-a$}}
\put(-15.00,104.00){\makebox(0,0)[cb]{$-\infty$}}
\put(47.00,89.00){\makebox(0,0)[cc]{{\large c)}}}
\end{picture}
\begin{picture}(160.1,40.00)(-30,80)
\put(104.00,100.00){\circle*{1.00}}
\put(104.00,100.00){\line(-1,0){32.50}}
\put(70.00,100.00){\oval(3.00,3.00)[b]}
\put(67.00,100.00){\oval(3.00,3.00)[t]}
\put(65.50,100.00){\line(-1,0){49.00}}
\put(15.00,100.00){\oval(3.00,3.00)[t]}
\put(15.00,100.00){\circle*{1.00}}
\put(13.50,100.00){\line(-1,0){25.50}}
\put(13.50,100.00){\oval(3.00,3.00)[rb]}
\put(13.50,98.50){\line(-1,0){25.50}}
\thicklines
\put(67.00,100.00){\circle*{1.00}}
\bezier{160}(67.00,100.00)(71.00,90.00)(95.00,90.00)
\bezier{160}(95.00,90.00)(118.00,90.00)(118.00,100.00)
\bezier{155}(95.00,110.00)(118.00,110.00)(118.00,100.00)
\bezier{155}(70.00,100.00)(76.00,110.00)(95.00,110.00)
\put(70.00,100.00){\circle*{1.00}}
\put(67.00,96.00){\makebox(0,0)[cb]{$x_-$}}
\put(70.00,104.00){\makebox(0,0)[cb]{$x_+$}}
\put(107.00,100.00){\makebox(0,0)[lc]{$b$}}
\put(15.00,104.00){\makebox(0,0)[cb]{$-a$}}
\put(-15.00,104.00){\makebox(0,0)[cb]{$-\infty$}}
\put(47.00,89.00){\makebox(0,0)[cc]{{\large d)}}}
\end{picture}
\begin{picture}(160.1,40.00)(-30,80)
\put(104.00,100.00){\circle*{1.00}}
\put(104.00,100.00){\line(-1,0){87.50}}
\put(15.00,100.00){\oval(3.00,3.00)[t]}
\put(15.00,100.00){\circle*{1.00}}
\put(13.50,100.00){\line(-1,0){25.50}}
\put(13.50,100.00){\oval(3.00,3.00)[rb]}
\put(13.50,98.50){\line(-1,0){25.50}}
\thicklines
\put(90.00,90.00){\circle*{1.00}}
\bezier{216}(90.00,90.00)(118.00,80.00)(118.00,100.00)
\bezier{200}(90.00,110.00)(118.00,120.00)(118.00,100.00)
\put(90.00,110.00){\circle*{1.00}}
\put(90.00,86.00){\makebox(0,0)[ct]{$x_-$}}
\put(90.00,114.00){\makebox(0,0)[cb]{$x_+$}}
\put(107.00,100.00){\makebox(0,0)[lc]{$b$}}
\put(15.00,104.00){\makebox(0,0)[cb]{$-a$}}
\put(-15.00,104.00){\makebox(0,0)[cb]{$-\infty$}}
\put(47.00,89.00){\makebox(0,0)[cc]{{\large e)}}}
\end{picture}
\caption[x]   {\hspace{0.2cm}\parbox[t]{13cm}
{\small
   The eigenvalue support of the spectral density (the bold
   line) and the branch cuts of the logarithms (the thin lines):
   a) for $\a>0$, b) for $\a\ra+0$, c) for $-1<\a<0$, d)
   for $\a\ra-1$ and e) for $\a<-1$.  }}
\label{Fig.1}
\end{figure}

For $\tilde{V}$ given by~\rf{tV} the contour integral can easily
be calculated taking the residues at $\om=\l,b$ and $-a$ while the residue at
infinity vanishes since $E_\om$ falls down as $1/\om$.  One gets
\bea
E_\l = \frac 12 \left( \frac{\a}{b-\l}-a + \frac{\a+1}{a+\l}-b \right) -
  \frac 12 \sqrt{(\l-x_-)(\l-x_+)} \non \cdot
\left( \frac{|\a|}{b-\l} \frac{1}{\sqrt{(b-x_-)(b-x_+)}}
- \frac{\a+1}{a+\l}\frac{1}{\sqrt{(a+x_-)(a+x_+)}}\right)
\label{one-cut}
\eea
where we use positive numerical values for $\sqrt{(b-x_-)(b-x_+)}$
and $\sqrt{(a+x_-)(a+x_+)}$ which are obtained from
the analytic function $\sqrt{(\l-x_-)(\l-x_+)}$ whose cut is depicted for
positive $\a$ in Figs.~\ref{Fig.1}a), \ref{Fig.1}b)
and for negative $-1<\a<0$ in Fig.~\ref{Fig.1}c).
When $\a\ra-1$ from above, the point $x_-$ approach $x_+$ along the real axis
as is depicted in Fig.~\ref{Fig.1}d).
For $\a<-1$ the points $x_-$ and $x_+$  become complex conjugate as is depicted
in Fig.~\ref{Fig.1}e).

Some comments concerning \eq{one-cut} and Fig.~\ref{Fig.1} are in
order. In the Gaussian limit~\rf{gausslimit} \eq{one-cut} recovers the
semicircle distribution of eigenvalues which has the support of the
type depicted in Fig.~\ref{Fig.1}a).
The analytic function $\sqrt{(\l-x_-)(\l-x_+)}$ with this branch cut
takes at $\l=b$ on the value
\be
\sqrt{(\l-x_-)(\l-x_+)} \Big|_{\l=b} = \sqrt{(b-x_-)(b-x_+)} \,.
\label{plussign}
\ee
This solution realizes for all $\a>0$.

When $\a\ra+0$, the end of the cut,
$x_+$, approaches $b$ as is depicted in Fig.~\ref{Fig.1}b).
For $\a<0$ the boundary equations~\rf{xandy} were {\it not}\/ have a solution
for~\rf{plussign}. For this reason the branch cut
of the square root in Eqs.~\rf{EvstV} and
\rf{xandy} should encircle the point $b$ as is depicted in
Fig.~\ref{Fig.1}c) which
provides
\be
\sqrt{(\l-x_-)(\l-x_+)} \Big|_{\l=b} = - \sqrt{(b-x_-)(b-x_+)} \,.
\label{minussign}
\ee
Then the boundary equations~\rf{xandy} can be explicitly written as
\bea
\frac{|\a|}{\sqrt{(b-x_-)(b-x_+)}}=
a +\frac{1}{a+b} + \frac{x_-+x_+}{2}~, \non
\frac{\a+1}{\sqrt{(a+x_-)(a+x_+)}} =b-\frac{1}{a+b} - \frac{x_-+x_+}{2}\,.
\label{be1}
\eea
They possess a solution both for positive and negative $\a$.

\eq{minussign} which is associated with the cut structure of
Fig.~\ref{Fig.1}c) is analogous to that for
the Penner model~\cite{CDL91} and
for the generalized Penner model~\cite{AKM94}. As is shown
in the next subsection, the
boundary equations~\rf{be1} reduce to those for the Penner model
under the rescaling~\rf{resc} with $c\ra0$
 at fixed $\a$ and the potential~\rf{tVc} smoothly interpolates
between the Gaussian and Penner ones when $\a$ is decreased.

\eq{minussign} always holds
when the branch cut of the square root in Eqs.~\rf{EvstV} and
\rf{xandy} encircles the point $b$ and does not impose further restrictions
on the location of the cut. It is convenient to take it along
the support of the spectral density
whose position in the complex plane is unambiguously determined by
the criterion of Ref.~\cite{Dav91}.
The proper contour for $\a=-1$ is constructed in Appendix~A.

\subsection{Solution of the boundary equations
\label{appA}}

The boundary equations~\rf{be1} can be conveniently
rewritten introducing the variable
\be
z = \frac{a-b}{2}+ \frac{1}{a+b} + \frac{x_-+x_+}{2}
\label{defz}
\ee
and shifting
\be
\{\l,x_-,x_+\}  =
\{\tl+\frac{b-a}{2},\tx+\frac{b-a}{2},\ty+\frac{b-a}{2}\}
\label{tildexyl}
\ee
which is analogous to~\rf{tphi}.
After this all quantities depend on $\a$ and $\b$ which is defined by \eq{beta}
while the dependence on $a-b$ formally disappears from the equations.
This was already discussed in Subsect.~\ref{logpot}.
In particular, \eq{defz} takes the form
\be
z = \frac{1}{2\b} + \frac{\tx+\ty}{2}
\label{defhz}
\ee
while Eqs.~\rf{be1} read
\be
\frac{|\a|}{\sqrt{(\b-\tx)(\b-\ty)}}=\b +z~,~~~~~
\frac{\a+1}{\sqrt{(\b+\tx)(\b+\ty)}} =\b-z \,.
\label{absnewbc}
\ee

The product $\tx\ty$ can now be expressed via $z$ by
\be
\tx \ty = \frac 12 \left[ \frac{\a^2}{(z+\b)^2}
+ \frac{(\a+1)^2}{(z-\b)^2} \right] - \b^2
\label{xpm}
\ee
while $z$ is determined via $\a$ and $\b$ by the $5^{th}$ order
algebraic equation
\be
4\b z -2 = \frac{(\a+1)^2}{(z-\b)^2}
-\frac{\a^2}{(z+\b)^2}
\label{bcp}
\ee
which can be rewritten as
\be
(\b^2-z^2)^2(4\b z -2) -(1+2\a)(\b+z)^2-4\a^2 \b z =0\,.
\label{bc}
\ee
Finally, $E_\l$ given by \eq{one-cut} takes the simple form
\be
E_\l = \frac 12 \left( \frac{\a}{\b-\tl} + \frac{\a+1}{\b+\tl} \right) -
\b - \sqrt{(\tl-\tx)(\tl-\ty)} \frac{\b(z+\tl)}{\b^2-\tl^2} \,.
\label{zbone-cut}
\ee

\eq{bc} is quadratic w.r.t.\ $\a$. The two solutions read
\be
\a_+=(z+\b)(\b-z-\frac{1}{2z})=\b^2-z^2-\frac{\b}{2z}-\frac 12
\label{a+}
\ee
and
\be
\a_-=(z+\b)(z-\b-\frac{1}{2\b})=z^2-\b^2-\frac{z}{2\b}-\frac 12 \,.
\label{a-}
\ee
It is easy to see from Eqs.~\rf{a+} and \rf{a-} that
\bea
\a_+ +1=(\b-z)(\b+z-\frac{1}{2z})=\b^2-z^2-\frac{\b}{2z}+\frac 12 \,,
\non
\a_- +1=(z-\b)(z+\b-\frac{1}{2\b})=z^2-\b^2-\frac{z}{2\b}+\frac 12
\label{a+1}
\eea
so that \eq{bcp} is obviously satisfied.

Then $\hat{x}_\pm$ are determined from Eqs.~\rf{defhz} and \rf{xpm} to be
\be
 \hat{x}_\pm \Big|_{\a_+}= z - \frac{1}{2\b} \pm
\frac{ \sqrt{\left(\b^2-z^2\right)(4\b z -1)} }{2\b z}
\label{roots}
\ee
and
\be
 \hat{x}_\pm \Big|_{\a_-}= z - \frac{1}{2\b}
\label{false}
\ee
for the solutions~\rf{a+} and \rf{a-}, respectively.

The solution~\rf{false} looks like a meromorphic one
and not of the one-cut type. The spectral density has,
nevertheless, a nontrivial support for this solution at some values of the
parameters quite similarly for the solution~\rf{roots} at $\a=-1$.%
\footnote{For the solution~\rf{false} one gets
$$
E_\l = \frac{1}{2}\left[\frac{\a}{\b-\tl} +\frac{\a+1}{\b+\tl}\right]
- \b \pm \frac 12\left[ \frac{\a}{\b-\tl} +\frac{\a+1}{\b+\tl} + 2\b
\right] = \left\{ \begin{array}{l} \frac{\a}{\b-\tl} +\frac{\a+1}{\b+\tl}  \\
-2\b \end{array} \right.
$$
and the support of the spectral density is along the closed contour $C$
in the complex $\tl$-plane which is determined by the equation
$$
 -2 \b \l + \a \log{(\b-\l)} -(\a+1) \log{(\b+\l)}
\Big|_{z-\fr{1}{2\b}}^{\tl} = 0
$$
with $z$ given by the solution of the quadratic equation~\rf{a-}.
The plus sign should be substituted outside of $C$ and the minus
sign should be substituted inside $C$.}
The solution~\rf{false} can {\it not}\/ be obtained, however, from the
Gaussian case varying the parameters
and the susceptibility~\rf{susc} is strictly divergent
at any values of the parameters $\a$ and $\b$.  We shall consider below
for this reason only
the solution~\rf{roots} denoting $\a_+$ just as $\a$.

The equation~\rf{a+} which determines $z$ versus $\a$ and $\b$ is cubic
and can be rewritten in the standard form
\be
z^3 - z\left(\b^2 -\a -\fr 12 \right) +\frac{\b}{2} =0 \,.
\label{ce}
\ee
The discriminant of this equation is non-positive for
\be
\a \leq \b^2 -3 \left(\frac{\b}{4}\right)^{\frac 23} -\fr 12 \,.
\label{discriminant}
\ee
When the inequality \rf{discriminant} is satisfied, \eq{ce}
possesses three real solutions.
If the inequality \rf{discriminant} is not satisfied,
then \eq{ce} has one real solution.

As we shall see in a moment the one-cut
solution to our one-matrix model
exists only in the domain of the parameters~\rf{discriminant} where \eq{ce} has
three real solutions. We choose the following one of them which is
represented in a parametric form as
\be
z =  \frac{2}{\sqrt{3}} \sqrt{\b^2-\a-\fr 12} \;
\cos{\left(\frac \pi6 + \frac{\theta}{3} \right)}~,~~~~
{\theta} =
\arcsin{\left\{\frac \b4
\left( \frac13 \left(\b^2-\a-\fr 12\right)\right)^{-\frac 32}
\right\}}
\label{1of3}
\ee
and $0\leq\theta\leq\frac \pi2$.
The solution is well-defined in the domain of the parameters
$\a$ and $\b$ restricted by the inequality~\rf{discriminant}
which is always satisfied for this solution while
the equality sign corresponds to $\sin{\theta}=1$.
This  solution
is connected to the perturbative expansion around the Gaussian
model when $\theta\ra0$.

In the Gaussian limit ($\a\sim\b^2\sim(\b^2-\a)\ra \infty$) one gets from
Eqs.~\rf{1of3}, \rf{roots}:
\be
z =\sqrt{\b^2-\a}~,~~~~~~
\frac{\ty -\tx}{2}=\frac{\sqrt{\a}}{\sqrt{\b}\sqrt[4]{\b^2-\a}} \,,
\ee
while \rf{zbone-cut} recovers the one for the semi-circle distribution
with the parameter
\be
\mu = \frac{2\b}{\a}\sqrt{\b^2-\a} \,.
\ee
$\mu$ exists for $\a<\b^2$ and vanishes at $\a=\b^2$ which  exactly
coincides with the criterion based on \rf{discriminant}.

The limit of the cubic potential is described by the solution~\rf{1of3}
when $\a\ra\infty$ in the vicinity of $\b^2$ as $\b\ra\infty$
so that
\be
\b^2-\a\sim\b^{\frac 23}\,.
\ee
No simplifications occurs in this
limit in \rf{1of3} and \rf{discriminant} which recover the known
results for the cubic potential~\cite{BIPZ78}.

In the Penner limit ($\b\ra\infty$, $\a\sim 1$) Eqs.~\rf{1of3} and
\rf{roots} yields
\be
z =\b-\frac{\a+1}{2\b}~,~~~~~~
\frac{\tx-\ty}{2}=\frac{\sqrt{\a+1}}{\b}
\ee
and \eq{zbone-cut} recovers the solution for the Penner model.
The inequality~\rf{discriminant} is always satisfied.

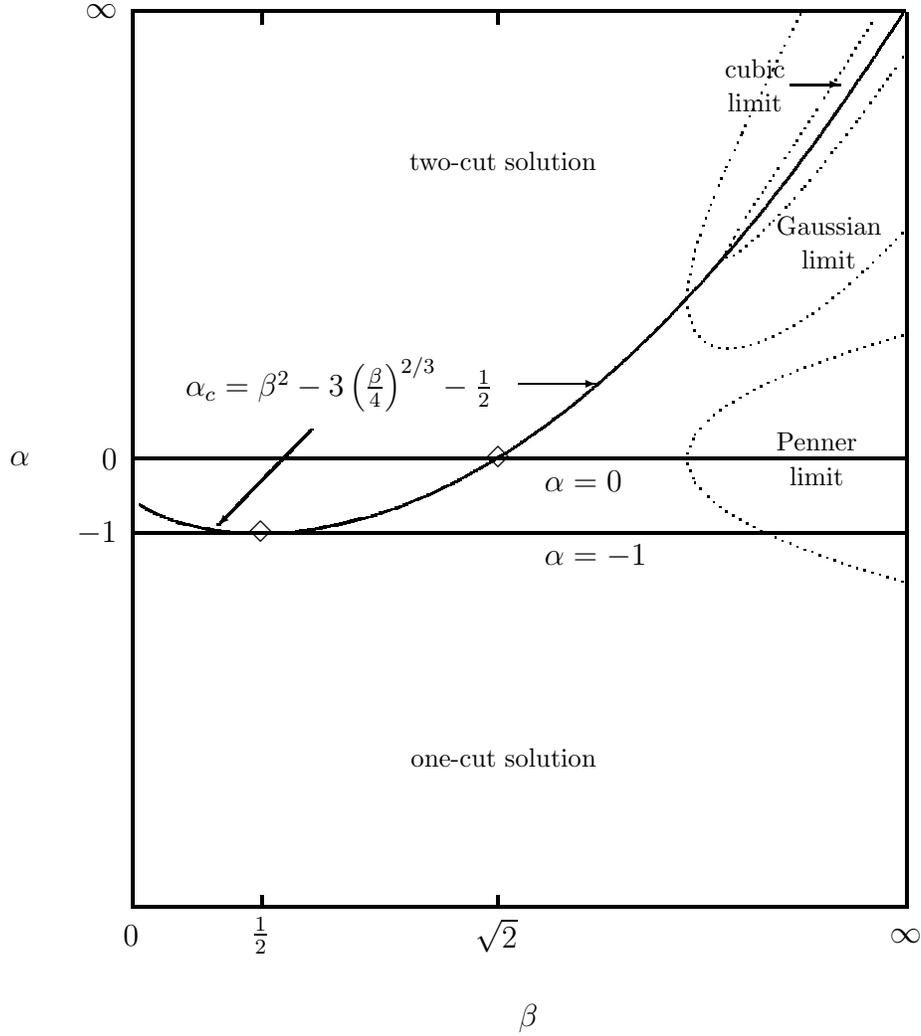
\begin{figure}[p]
\centering
\setlength{\unitlength}{0.240900pt}
\ifx\plotpoint\undefined\newsavebox{\plotpoint}\fi
\sbox{\plotpoint}{\rule[-0.500pt]{1.000pt}{1.000pt}}%
\begin{picture}(1500,1500)(0,-200)
\font\gnuplot=cmr10 at 10pt
\gnuplot
\sbox{\plotpoint}{\rule[-0.500pt]{1.000pt}{1.000pt}}%
\put(220.0,772.0){\rule[-0.500pt]{292.934pt}{1.000pt}}
\put(220.0,68.0){\rule[-0.500pt]{1.000pt}{339.187pt}}
\put(220.0,655.0){\rule[-0.500pt]{4.818pt}{1.000pt}}
\put(198,655){\makebox(0,0)[r]{$-1$}}
\put(1416.0,655.0){\rule[-0.500pt]{4.818pt}{1.000pt}}
\put(220.0,1476.0){\rule[-0.500pt]{4.818pt}{1.000pt}}
\put(198,1476){\makebox(0,0)[r]{$\infty$}}
\put(1416.0,1476.0){\rule[-0.500pt]{4.818pt}{1.000pt}}
\put(220.0,772.0){\rule[-0.500pt]{4.818pt}{1.000pt}}
\put(198,772){\makebox(0,0)[r]{$0$}}
\put(1416.0,772.0){\rule[-0.500pt]{4.818pt}{1.000pt}}
\put(793.0,68.0){\rule[-0.500pt]{1.000pt}{4.818pt}}
\put(793,23){\makebox(0,0){$\sqrt{2}$}}
\put(793.0,1456.0){\rule[-0.500pt]{1.000pt}{4.818pt}}
\put(423.0,68.0){\rule[-0.500pt]{1.000pt}{4.818pt}}
\put(423,23){\makebox(0,0){$\frac 12$}}
\put(423.0,1456.0){\rule[-0.500pt]{1.000pt}{4.818pt}}
\put(220.0,68.0){\rule[-0.500pt]{1.000pt}{4.818pt}}
\put(220,23){\makebox(0,0){$0$}}
\put(220.0,1456.0){\rule[-0.500pt]{1.000pt}{4.818pt}}
\put(1436.0,68.0){\rule[-0.500pt]{1.000pt}{4.818pt}}
\put(1436,23){\makebox(0,0){$\infty$}}
\put(1436.0,1456.0){\rule[-0.500pt]{1.000pt}{4.818pt}}
\put(220.0,68.0){\rule[-0.500pt]{292.934pt}{1.000pt}}
\put(1436.0,68.0){\rule[-0.500pt]{1.000pt}{339.187pt}}
\put(220.0,1476.0){\rule[-0.500pt]{292.934pt}{1.000pt}}
\put(45,772){\makebox(0,0){$\alpha$}}
\put(828,-107){\makebox(0,0)[l]{$\beta$}}
\put(1233,772){\makebox(0,0)[l]{\shortstack{Penner \\ \mbox{} \\ limit}}}
\put(1233,1112){\makebox(0,0)[l]{\shortstack{Gaussian \\ \mbox{} \\limit}}}
\put(1152,1359){\makebox(0,0)[l]{\shortstack{cubic \\ \mbox{} \\limit}}}
\put(787,889){\makebox(0,0)[r]{$\alpha_c=\beta^2-3\left(\frac{\beta}{4}\right)^{2/3} -\frac 12$}}
\put(869,737){\makebox(0,0)[l]{$\alpha=0$}}
\put(869,619){\makebox(0,0)[l]{$\alpha=-1$}}
\put(950,303){\makebox(0,0)[r]{one-cut solution}}
\put(950,1241){\makebox(0,0)[r]{two-cut solution}}
\put(220.0,68.0){\rule[-0.500pt]{1.000pt}{339.187pt}}
\put(1254,1359){\vector(1,0){79}}
\put(828,889){\vector(1,0){122}}
\multiput(501.67,813.70)(-0.499,-0.511){286}{\rule{0.120pt}{1.277pt}}
\multiput(501.92,816.35)(-147.000,-148.349){2}{\rule{1.000pt}{0.639pt}}
\put(357,668){\vector(-1,-1){0}}
\sbox{\plotpoint}{\rule[-0.300pt]{0.600pt}{0.600pt}}%
\put(232,700){\usebox{\plotpoint}}
\multiput(232.00,698.50)(0.833,-0.501){11}{\rule{1.125pt}{0.121pt}}
\multiput(232.00,698.75)(10.665,-8.000){2}{\rule{0.563pt}{0.600pt}}
\multiput(245.00,690.50)(1.066,-0.501){7}{\rule{1.350pt}{0.121pt}}
\multiput(245.00,690.75)(9.198,-6.000){2}{\rule{0.675pt}{0.600pt}}
\multiput(257.00,684.50)(1.350,-0.502){5}{\rule{1.590pt}{0.121pt}}
\multiput(257.00,684.75)(8.700,-5.000){2}{\rule{0.795pt}{0.600pt}}
\multiput(269.00,679.50)(1.350,-0.502){5}{\rule{1.590pt}{0.121pt}}
\multiput(269.00,679.75)(8.700,-5.000){2}{\rule{0.795pt}{0.600pt}}
\multiput(281.00,674.50)(2.141,-0.503){3}{\rule{2.100pt}{0.121pt}}
\multiput(281.00,674.75)(8.641,-4.000){2}{\rule{1.050pt}{0.600pt}}
\put(294,669.25){\rule{2.550pt}{0.600pt}}
\multiput(294.00,670.75)(6.707,-3.000){2}{\rule{1.275pt}{0.600pt}}
\put(306,666.25){\rule{2.550pt}{0.600pt}}
\multiput(306.00,667.75)(6.707,-3.000){2}{\rule{1.275pt}{0.600pt}}
\put(318,663.25){\rule{2.750pt}{0.600pt}}
\multiput(318.00,664.75)(7.292,-3.000){2}{\rule{1.375pt}{0.600pt}}
\put(331,660.75){\rule{2.891pt}{0.600pt}}
\multiput(331.00,661.75)(6.000,-2.000){2}{\rule{1.445pt}{0.600pt}}
\put(343,658.75){\rule{2.891pt}{0.600pt}}
\multiput(343.00,659.75)(6.000,-2.000){2}{\rule{1.445pt}{0.600pt}}
\put(355,657.25){\rule{2.891pt}{0.600pt}}
\multiput(355.00,657.75)(6.000,-1.000){2}{\rule{1.445pt}{0.600pt}}
\put(367,655.75){\rule{3.132pt}{0.600pt}}
\multiput(367.00,656.75)(6.500,-2.000){2}{\rule{1.566pt}{0.600pt}}
\put(392,654.25){\rule{2.891pt}{0.600pt}}
\multiput(392.00,654.75)(6.000,-1.000){2}{\rule{1.445pt}{0.600pt}}
\put(380.0,656.0){\rule[-0.300pt]{2.891pt}{0.600pt}}
\put(441,654.25){\rule{2.891pt}{0.600pt}}
\multiput(441.00,653.75)(6.000,1.000){2}{\rule{1.445pt}{0.600pt}}
\put(404.0,655.0){\rule[-0.300pt]{8.913pt}{0.600pt}}
\put(466,655.25){\rule{2.891pt}{0.600pt}}
\multiput(466.00,654.75)(6.000,1.000){2}{\rule{1.445pt}{0.600pt}}
\put(478,656.75){\rule{2.891pt}{0.600pt}}
\multiput(478.00,655.75)(6.000,2.000){2}{\rule{1.445pt}{0.600pt}}
\put(490,658.75){\rule{3.132pt}{0.600pt}}
\multiput(490.00,657.75)(6.500,2.000){2}{\rule{1.566pt}{0.600pt}}
\put(503,660.25){\rule{2.891pt}{0.600pt}}
\multiput(503.00,659.75)(6.000,1.000){2}{\rule{1.445pt}{0.600pt}}
\put(515,662.25){\rule{2.550pt}{0.600pt}}
\multiput(515.00,660.75)(6.707,3.000){2}{\rule{1.275pt}{0.600pt}}
\put(527,664.75){\rule{2.891pt}{0.600pt}}
\multiput(527.00,663.75)(6.000,2.000){2}{\rule{1.445pt}{0.600pt}}
\put(539,667.25){\rule{2.750pt}{0.600pt}}
\multiput(539.00,665.75)(7.292,3.000){2}{\rule{1.375pt}{0.600pt}}
\put(552,670.25){\rule{2.550pt}{0.600pt}}
\multiput(552.00,668.75)(6.707,3.000){2}{\rule{1.275pt}{0.600pt}}
\put(564,673.25){\rule{2.550pt}{0.600pt}}
\multiput(564.00,671.75)(6.707,3.000){2}{\rule{1.275pt}{0.600pt}}
\put(576,676.25){\rule{2.550pt}{0.600pt}}
\multiput(576.00,674.75)(6.707,3.000){2}{\rule{1.275pt}{0.600pt}}
\multiput(588.00,679.99)(2.141,0.503){3}{\rule{2.100pt}{0.121pt}}
\multiput(588.00,677.75)(8.641,4.000){2}{\rule{1.050pt}{0.600pt}}
\multiput(601.00,683.99)(1.953,0.503){3}{\rule{1.950pt}{0.121pt}}
\multiput(601.00,681.75)(7.953,4.000){2}{\rule{0.975pt}{0.600pt}}
\multiput(613.00,687.99)(1.953,0.503){3}{\rule{1.950pt}{0.121pt}}
\multiput(613.00,685.75)(7.953,4.000){2}{\rule{0.975pt}{0.600pt}}
\multiput(625.00,691.99)(2.141,0.503){3}{\rule{2.100pt}{0.121pt}}
\multiput(625.00,689.75)(8.641,4.000){2}{\rule{1.050pt}{0.600pt}}
\multiput(638.00,695.99)(1.350,0.502){5}{\rule{1.590pt}{0.121pt}}
\multiput(638.00,693.75)(8.700,5.000){2}{\rule{0.795pt}{0.600pt}}
\multiput(650.00,700.99)(1.350,0.502){5}{\rule{1.590pt}{0.121pt}}
\multiput(650.00,698.75)(8.700,5.000){2}{\rule{0.795pt}{0.600pt}}
\multiput(662.00,705.99)(1.350,0.502){5}{\rule{1.590pt}{0.121pt}}
\multiput(662.00,703.75)(8.700,5.000){2}{\rule{0.795pt}{0.600pt}}
\multiput(674.00,710.99)(1.475,0.502){5}{\rule{1.710pt}{0.121pt}}
\multiput(674.00,708.75)(9.451,5.000){2}{\rule{0.855pt}{0.600pt}}
\multiput(687.00,715.99)(1.066,0.501){7}{\rule{1.350pt}{0.121pt}}
\multiput(687.00,713.75)(9.198,6.000){2}{\rule{0.675pt}{0.600pt}}
\multiput(699.00,721.99)(1.066,0.501){7}{\rule{1.350pt}{0.121pt}}
\multiput(699.00,719.75)(9.198,6.000){2}{\rule{0.675pt}{0.600pt}}
\multiput(711.00,727.99)(1.163,0.501){7}{\rule{1.450pt}{0.121pt}}
\multiput(711.00,725.75)(9.990,6.000){2}{\rule{0.725pt}{0.600pt}}
\multiput(724.00,733.99)(1.066,0.501){7}{\rule{1.350pt}{0.121pt}}
\multiput(724.00,731.75)(9.198,6.000){2}{\rule{0.675pt}{0.600pt}}
\multiput(736.00,739.99)(0.888,0.501){9}{\rule{1.179pt}{0.121pt}}
\multiput(736.00,737.75)(9.554,7.000){2}{\rule{0.589pt}{0.600pt}}
\multiput(748.00,746.99)(0.888,0.501){9}{\rule{1.179pt}{0.121pt}}
\multiput(748.00,744.75)(9.554,7.000){2}{\rule{0.589pt}{0.600pt}}
\multiput(760.00,753.99)(0.969,0.501){9}{\rule{1.264pt}{0.121pt}}
\multiput(760.00,751.75)(10.376,7.000){2}{\rule{0.632pt}{0.600pt}}
\multiput(773.00,760.99)(0.888,0.501){9}{\rule{1.179pt}{0.121pt}}
\multiput(773.00,758.75)(9.554,7.000){2}{\rule{0.589pt}{0.600pt}}
\multiput(785.00,767.99)(0.764,0.501){11}{\rule{1.050pt}{0.121pt}}
\multiput(785.00,765.75)(9.821,8.000){2}{\rule{0.525pt}{0.600pt}}
\multiput(797.00,775.99)(0.969,0.501){9}{\rule{1.264pt}{0.121pt}}
\multiput(797.00,773.75)(10.376,7.000){2}{\rule{0.632pt}{0.600pt}}
\multiput(810.00,782.99)(0.764,0.501){11}{\rule{1.050pt}{0.121pt}}
\multiput(810.00,780.75)(9.821,8.000){2}{\rule{0.525pt}{0.600pt}}
\multiput(822.00,790.99)(0.764,0.501){11}{\rule{1.050pt}{0.121pt}}
\multiput(822.00,788.75)(9.821,8.000){2}{\rule{0.525pt}{0.600pt}}
\multiput(834.00,798.99)(0.671,0.501){13}{\rule{0.950pt}{0.121pt}}
\multiput(834.00,796.75)(10.028,9.000){2}{\rule{0.475pt}{0.600pt}}
\multiput(846.00,807.99)(0.732,0.501){13}{\rule{1.017pt}{0.121pt}}
\multiput(846.00,805.75)(10.890,9.000){2}{\rule{0.508pt}{0.600pt}}
\multiput(859.00,816.99)(0.764,0.501){11}{\rule{1.050pt}{0.121pt}}
\multiput(859.00,814.75)(9.821,8.000){2}{\rule{0.525pt}{0.600pt}}
\multiput(871.00,825.00)(0.599,0.501){15}{\rule{0.870pt}{0.121pt}}
\multiput(871.00,822.75)(10.194,10.000){2}{\rule{0.435pt}{0.600pt}}
\multiput(883.00,834.99)(0.732,0.501){13}{\rule{1.017pt}{0.121pt}}
\multiput(883.00,832.75)(10.890,9.000){2}{\rule{0.508pt}{0.600pt}}
\multiput(896.00,843.99)(0.671,0.501){13}{\rule{0.950pt}{0.121pt}}
\multiput(896.00,841.75)(10.028,9.000){2}{\rule{0.475pt}{0.600pt}}
\multiput(908.00,853.00)(0.599,0.501){15}{\rule{0.870pt}{0.121pt}}
\multiput(908.00,850.75)(10.194,10.000){2}{\rule{0.435pt}{0.600pt}}
\multiput(920.00,863.00)(0.599,0.501){15}{\rule{0.870pt}{0.121pt}}
\multiput(920.00,860.75)(10.194,10.000){2}{\rule{0.435pt}{0.600pt}}
\multiput(932.00,873.00)(0.590,0.501){17}{\rule{0.859pt}{0.121pt}}
\multiput(932.00,870.75)(11.217,11.000){2}{\rule{0.430pt}{0.600pt}}
\multiput(945.00,884.00)(0.599,0.501){15}{\rule{0.870pt}{0.121pt}}
\multiput(945.00,881.75)(10.194,10.000){2}{\rule{0.435pt}{0.600pt}}
\multiput(957.00,894.00)(0.541,0.501){17}{\rule{0.805pt}{0.121pt}}
\multiput(957.00,891.75)(10.330,11.000){2}{\rule{0.402pt}{0.600pt}}
\multiput(969.00,905.00)(0.590,0.501){17}{\rule{0.859pt}{0.121pt}}
\multiput(969.00,902.75)(11.217,11.000){2}{\rule{0.430pt}{0.600pt}}
\multiput(982.00,916.00)(0.541,0.501){17}{\rule{0.805pt}{0.121pt}}
\multiput(982.00,913.75)(10.330,11.000){2}{\rule{0.402pt}{0.600pt}}
\multiput(994.00,927.00)(0.541,0.501){17}{\rule{0.805pt}{0.121pt}}
\multiput(994.00,924.75)(10.330,11.000){2}{\rule{0.402pt}{0.600pt}}
\multiput(1006.00,938.00)(0.494,0.500){19}{\rule{0.750pt}{0.121pt}}
\multiput(1006.00,935.75)(10.443,12.000){2}{\rule{0.375pt}{0.600pt}}
\multiput(1018.00,950.00)(0.538,0.500){19}{\rule{0.800pt}{0.121pt}}
\multiput(1018.00,947.75)(11.340,12.000){2}{\rule{0.400pt}{0.600pt}}
\multiput(1031.00,962.00)(0.494,0.500){19}{\rule{0.750pt}{0.121pt}}
\multiput(1031.00,959.75)(10.443,12.000){2}{\rule{0.375pt}{0.600pt}}
\multiput(1043.00,974.00)(0.494,0.500){19}{\rule{0.750pt}{0.121pt}}
\multiput(1043.00,971.75)(10.443,12.000){2}{\rule{0.375pt}{0.600pt}}
\multiput(1055.00,986.00)(0.494,0.500){21}{\rule{0.750pt}{0.121pt}}
\multiput(1055.00,983.75)(11.443,13.000){2}{\rule{0.375pt}{0.600pt}}
\multiput(1069.00,998.00)(0.500,0.538){19}{\rule{0.121pt}{0.800pt}}
\multiput(1066.75,998.00)(12.000,11.340){2}{\rule{0.600pt}{0.400pt}}
\multiput(1081.00,1011.00)(0.500,0.538){19}{\rule{0.121pt}{0.800pt}}
\multiput(1078.75,1011.00)(12.000,11.340){2}{\rule{0.600pt}{0.400pt}}
\multiput(1093.00,1024.00)(0.500,0.538){19}{\rule{0.121pt}{0.800pt}}
\multiput(1090.75,1024.00)(12.000,11.340){2}{\rule{0.600pt}{0.400pt}}
\multiput(1104.00,1038.00)(0.494,0.500){21}{\rule{0.750pt}{0.121pt}}
\multiput(1104.00,1035.75)(11.443,13.000){2}{\rule{0.375pt}{0.600pt}}
\multiput(1118.00,1050.00)(0.500,0.582){19}{\rule{0.121pt}{0.850pt}}
\multiput(1115.75,1050.00)(12.000,12.236){2}{\rule{0.600pt}{0.425pt}}
\multiput(1130.00,1064.00)(0.500,0.582){19}{\rule{0.121pt}{0.850pt}}
\multiput(1127.75,1064.00)(12.000,12.236){2}{\rule{0.600pt}{0.425pt}}
\multiput(1142.00,1078.00)(0.500,0.582){19}{\rule{0.121pt}{0.850pt}}
\multiput(1139.75,1078.00)(12.000,12.236){2}{\rule{0.600pt}{0.425pt}}
\multiput(1154.00,1092.00)(0.500,0.535){21}{\rule{0.121pt}{0.796pt}}
\multiput(1151.75,1092.00)(13.000,12.348){2}{\rule{0.600pt}{0.398pt}}
\multiput(1167.00,1106.00)(0.500,0.626){19}{\rule{0.121pt}{0.900pt}}
\multiput(1164.75,1106.00)(12.000,13.132){2}{\rule{0.600pt}{0.450pt}}
\multiput(1179.00,1121.00)(0.500,0.626){19}{\rule{0.121pt}{0.900pt}}
\multiput(1176.75,1121.00)(12.000,13.132){2}{\rule{0.600pt}{0.450pt}}
\multiput(1191.00,1136.00)(0.500,0.575){21}{\rule{0.121pt}{0.842pt}}
\multiput(1188.75,1136.00)(13.000,13.252){2}{\rule{0.600pt}{0.421pt}}
\multiput(1204.00,1151.00)(0.500,0.626){19}{\rule{0.121pt}{0.900pt}}
\multiput(1201.75,1151.00)(12.000,13.132){2}{\rule{0.600pt}{0.450pt}}
\multiput(1216.00,1166.00)(0.500,0.670){19}{\rule{0.121pt}{0.950pt}}
\multiput(1213.75,1166.00)(12.000,14.028){2}{\rule{0.600pt}{0.475pt}}
\multiput(1228.00,1182.00)(0.500,0.626){19}{\rule{0.121pt}{0.900pt}}
\multiput(1225.75,1182.00)(12.000,13.132){2}{\rule{0.600pt}{0.450pt}}
\multiput(1240.00,1197.00)(0.500,0.616){21}{\rule{0.121pt}{0.888pt}}
\multiput(1237.75,1197.00)(13.000,14.156){2}{\rule{0.600pt}{0.444pt}}
\multiput(1253.00,1213.00)(0.500,0.670){19}{\rule{0.121pt}{0.950pt}}
\multiput(1250.75,1213.00)(12.000,14.028){2}{\rule{0.600pt}{0.475pt}}
\multiput(1265.00,1229.00)(0.500,0.714){19}{\rule{0.121pt}{1.000pt}}
\multiput(1262.75,1229.00)(12.000,14.924){2}{\rule{0.600pt}{0.500pt}}
\multiput(1277.00,1246.00)(0.500,0.616){21}{\rule{0.121pt}{0.888pt}}
\multiput(1274.75,1246.00)(13.000,14.156){2}{\rule{0.600pt}{0.444pt}}
\multiput(1290.00,1262.00)(0.500,0.714){19}{\rule{0.121pt}{1.000pt}}
\multiput(1287.75,1262.00)(12.000,14.924){2}{\rule{0.600pt}{0.500pt}}
\multiput(1302.00,1279.00)(0.500,0.714){19}{\rule{0.121pt}{1.000pt}}
\multiput(1299.75,1279.00)(12.000,14.924){2}{\rule{0.600pt}{0.500pt}}
\multiput(1314.00,1296.00)(0.500,0.714){19}{\rule{0.121pt}{1.000pt}}
\multiput(1311.75,1296.00)(12.000,14.924){2}{\rule{0.600pt}{0.500pt}}
\multiput(1326.00,1313.00)(0.500,0.696){21}{\rule{0.121pt}{0.981pt}}
\multiput(1323.75,1313.00)(13.000,15.964){2}{\rule{0.600pt}{0.490pt}}
\multiput(1339.00,1331.00)(0.500,0.758){19}{\rule{0.121pt}{1.050pt}}
\multiput(1336.75,1331.00)(12.000,15.821){2}{\rule{0.600pt}{0.525pt}}
\multiput(1351.00,1349.00)(0.500,0.758){19}{\rule{0.121pt}{1.050pt}}
\multiput(1348.75,1349.00)(12.000,15.821){2}{\rule{0.600pt}{0.525pt}}
\multiput(1363.00,1367.00)(0.500,0.696){21}{\rule{0.121pt}{0.981pt}}
\multiput(1360.75,1367.00)(13.000,15.964){2}{\rule{0.600pt}{0.490pt}}
\multiput(1376.00,1385.00)(0.500,0.758){19}{\rule{0.121pt}{1.050pt}}
\multiput(1373.75,1385.00)(12.000,15.821){2}{\rule{0.600pt}{0.525pt}}
\multiput(1388.00,1403.00)(0.500,0.802){19}{\rule{0.121pt}{1.100pt}}
\multiput(1385.75,1403.00)(12.000,16.717){2}{\rule{0.600pt}{0.550pt}}
\multiput(1400.00,1422.00)(0.500,0.758){19}{\rule{0.121pt}{1.050pt}}
\multiput(1397.75,1422.00)(12.000,15.821){2}{\rule{0.600pt}{0.525pt}}
\multiput(1412.00,1440.00)(0.500,0.737){21}{\rule{0.121pt}{1.027pt}}
\multiput(1409.75,1440.00)(13.000,16.869){2}{\rule{0.600pt}{0.513pt}}
\multiput(1425.00,1459.00)(0.501,0.867){15}{\rule{0.121pt}{1.170pt}}
\multiput(1422.75,1459.00)(10.000,14.572){2}{\rule{0.600pt}{0.585pt}}
\put(453.0,656.0){\rule[-0.300pt]{3.132pt}{0.600pt}}
\sbox{\plotpoint}{\rule[-0.250pt]{0.500pt}{0.500pt}}%
\put(1092,780){\usebox{\plotpoint}}
\multiput(1092,780)(4.625,11.563){3}{\usebox{\plotpoint}}
\multiput(1104,810)(8.156,9.411){2}{\usebox{\plotpoint}}
\put(1124.45,831.83){\usebox{\plotpoint}}
\put(1133.82,840.02){\usebox{\plotpoint}}
\put(1143.49,847.87){\usebox{\plotpoint}}
\multiput(1153,855)(10.965,5.904){2}{\usebox{\plotpoint}}
\put(1174.91,867.94){\usebox{\plotpoint}}
\put(1185.82,873.91){\usebox{\plotpoint}}
\put(1196.85,879.69){\usebox{\plotpoint}}
\put(1207.89,885.44){\usebox{\plotpoint}}
\put(1219.16,890.73){\usebox{\plotpoint}}
\put(1230.54,895.77){\usebox{\plotpoint}}
\put(1241.79,901.07){\usebox{\plotpoint}}
\put(1253.40,905.58){\usebox{\plotpoint}}
\put(1264.90,910.37){\usebox{\plotpoint}}
\multiput(1276,915)(11.903,3.662){2}{\usebox{\plotpoint}}
\put(1299.83,923.51){\usebox{\plotpoint}}
\put(1311.61,927.54){\usebox{\plotpoint}}
\put(1323.42,931.47){\usebox{\plotpoint}}
\put(1335.07,935.87){\usebox{\plotpoint}}
\put(1346.84,939.95){\usebox{\plotpoint}}
\put(1358.65,943.88){\usebox{\plotpoint}}
\put(1370.53,947.62){\usebox{\plotpoint}}
\put(1382.54,950.89){\usebox{\plotpoint}}
\put(1394.46,954.49){\usebox{\plotpoint}}
\put(1406.27,958.42){\usebox{\plotpoint}}
\put(1418.28,961.68){\usebox{\plotpoint}}
\put(1430.24,965.08){\usebox{\plotpoint}}
\put(1436,967){\usebox{\plotpoint}}
\put(1092,764){\usebox{\plotpoint}}
\multiput(1092,764)(4.625,-11.563){3}{\usebox{\plotpoint}}
\multiput(1104,734)(8.156,-9.411){2}{\usebox{\plotpoint}}
\put(1124.45,712.17){\usebox{\plotpoint}}
\put(1133.82,703.98){\usebox{\plotpoint}}
\put(1143.49,696.13){\usebox{\plotpoint}}
\multiput(1153,689)(10.965,-5.904){2}{\usebox{\plotpoint}}
\put(1174.91,676.06){\usebox{\plotpoint}}
\put(1185.82,670.09){\usebox{\plotpoint}}
\put(1196.85,664.31){\usebox{\plotpoint}}
\put(1207.89,658.56){\usebox{\plotpoint}}
\put(1219.16,653.27){\usebox{\plotpoint}}
\put(1230.54,648.23){\usebox{\plotpoint}}
\put(1241.79,642.93){\usebox{\plotpoint}}
\put(1253.40,638.42){\usebox{\plotpoint}}
\put(1264.90,633.63){\usebox{\plotpoint}}
\multiput(1276,629)(11.903,-3.662){2}{\usebox{\plotpoint}}
\put(1299.83,620.49){\usebox{\plotpoint}}
\put(1311.61,616.46){\usebox{\plotpoint}}
\put(1323.42,612.53){\usebox{\plotpoint}}
\put(1335.07,608.13){\usebox{\plotpoint}}
\put(1346.84,604.05){\usebox{\plotpoint}}
\put(1358.65,600.12){\usebox{\plotpoint}}
\put(1370.53,596.38){\usebox{\plotpoint}}
\put(1382.54,593.11){\usebox{\plotpoint}}
\put(1394.46,589.51){\usebox{\plotpoint}}
\put(1406.27,585.58){\usebox{\plotpoint}}
\put(1418.28,582.32){\usebox{\plotpoint}}
\put(1430.24,578.92){\usebox{\plotpoint}}
\put(1436,577){\usebox{\plotpoint}}
\put(1092,1037){\usebox{\plotpoint}}
\multiput(1092,1037)(2.330,12.233){6}{\usebox{\plotpoint}}
\multiput(1104,1100)(4.031,11.783){3}{\usebox{\plotpoint}}
\multiput(1117,1138)(4.256,11.704){3}{\usebox{\plotpoint}}
\multiput(1129,1171)(4.625,11.563){2}{\usebox{\plotpoint}}
\multiput(1141,1201)(4.762,11.507){3}{\usebox{\plotpoint}}
\multiput(1153,1230)(5.402,11.220){2}{\usebox{\plotpoint}}
\multiput(1166,1257)(5.058,11.380){3}{\usebox{\plotpoint}}
\multiput(1178,1284)(5.219,11.307){2}{\usebox{\plotpoint}}
\multiput(1190,1310)(5.745,11.049){2}{\usebox{\plotpoint}}
\multiput(1203,1335)(5.389,11.227){2}{\usebox{\plotpoint}}
\multiput(1215,1360)(5.389,11.227){3}{\usebox{\plotpoint}}
\multiput(1227,1385)(5.389,11.227){2}{\usebox{\plotpoint}}
\multiput(1239,1410)(5.745,11.049){2}{\usebox{\plotpoint}}
\multiput(1252,1435)(5.569,11.139){2}{\usebox{\plotpoint}}
\multiput(1264,1459)(5.827,11.006){2}{\usebox{\plotpoint}}
\put(1273,1476){\usebox{\plotpoint}}
\put(1092,1009){\usebox{\plotpoint}}
\multiput(1092,1009)(3.662,-11.903){4}{\usebox{\plotpoint}}
\put(1110.37,963.63){\usebox{\plotpoint}}
\put(1119.66,955.45){\usebox{\plotpoint}}
\put(1130.55,949.48){\usebox{\plotpoint}}
\put(1142.44,945.88){\usebox{\plotpoint}}
\put(1154.85,945.14){\usebox{\plotpoint}}
\put(1167.26,946.11){\usebox{\plotpoint}}
\put(1179.63,947.41){\usebox{\plotpoint}}
\put(1191.69,950.52){\usebox{\plotpoint}}
\put(1203.57,954.24){\usebox{\plotpoint}}
\multiput(1215,959)(11.495,4.790){2}{\usebox{\plotpoint}}
\put(1237.71,969.36){\usebox{\plotpoint}}
\put(1248.70,975.22){\usebox{\plotpoint}}
\put(1259.52,981.38){\usebox{\plotpoint}}
\put(1270.04,988.03){\usebox{\plotpoint}}
\put(1280.51,994.77){\usebox{\plotpoint}}
\multiput(1289,1000)(9.963,7.472){2}{\usebox{\plotpoint}}
\put(1310.91,1016.43){\usebox{\plotpoint}}
\put(1320.56,1024.30){\usebox{\plotpoint}}
\put(1330.29,1032.07){\usebox{\plotpoint}}
\multiput(1338,1038)(9.180,8.415){2}{\usebox{\plotpoint}}
\put(1358.37,1056.67){\usebox{\plotpoint}}
\put(1367.75,1064.86){\usebox{\plotpoint}}
\multiput(1375,1071)(8.806,8.806){2}{\usebox{\plotpoint}}
\put(1394.70,1090.70){\usebox{\plotpoint}}
\put(1403.51,1099.51){\usebox{\plotpoint}}
\multiput(1411,1107)(8.806,8.806){2}{\usebox{\plotpoint}}
\put(1429.68,1126.16){\usebox{\plotpoint}}
\put(1436,1133){\usebox{\plotpoint}}
\put(1153,1097){\usebox{\plotpoint}}
\multiput(1153,1097)(5.931,10.950){3}{\usebox{\plotpoint}}
\multiput(1166,1121)(6.179,10.812){2}{\usebox{\plotpoint}}
\put(1183.76,1151.12){\usebox{\plotpoint}}
\multiput(1190,1161)(7.032,10.278){2}{\usebox{\plotpoint}}
\multiput(1203,1180)(6.908,10.362){2}{\usebox{\plotpoint}}
\multiput(1215,1198)(6.650,10.529){2}{\usebox{\plotpoint}}
\multiput(1227,1217)(6.908,10.362){2}{\usebox{\plotpoint}}
\put(1245.57,1244.60){\usebox{\plotpoint}}
\multiput(1252,1254)(6.908,10.362){2}{\usebox{\plotpoint}}
\multiput(1264,1272)(6.650,10.529){2}{\usebox{\plotpoint}}
\multiput(1276,1291)(7.032,10.278){2}{\usebox{\plotpoint}}
\multiput(1289,1310)(6.650,10.529){2}{\usebox{\plotpoint}}
\put(1306.92,1338.38){\usebox{\plotpoint}}
\multiput(1313,1348)(6.650,10.529){2}{\usebox{\plotpoint}}
\multiput(1325,1367)(7.032,10.278){2}{\usebox{\plotpoint}}
\multiput(1338,1386)(6.407,10.679){2}{\usebox{\plotpoint}}
\multiput(1350,1406)(6.407,10.679){2}{\usebox{\plotpoint}}
\multiput(1362,1426)(7.032,10.278){2}{\usebox{\plotpoint}}
\multiput(1375,1445)(6.179,10.812){2}{\usebox{\plotpoint}}
\put(1392.74,1475.57){\usebox{\plotpoint}}
\put(1393,1476){\usebox{\plotpoint}}
\put(1153,1088){\usebox{\plotpoint}}
\multiput(1153,1088)(12.134,2.800){2}{\usebox{\plotpoint}}
\put(1175.25,1097.94){\usebox{\plotpoint}}
\put(1184.65,1106.09){\usebox{\plotpoint}}
\put(1193.96,1114.35){\usebox{\plotpoint}}
\multiput(1203,1122)(8.806,8.806){2}{\usebox{\plotpoint}}
\put(1221.05,1140.05){\usebox{\plotpoint}}
\multiput(1227,1146)(8.447,9.151){2}{\usebox{\plotpoint}}
\put(1246.95,1166.95){\usebox{\plotpoint}}
\multiput(1252,1172)(8.105,9.455){2}{\usebox{\plotpoint}}
\put(1271.67,1194.95){\usebox{\plotpoint}}
\multiput(1276,1200)(8.156,9.411){2}{\usebox{\plotpoint}}
\put(1296.07,1223.24){\usebox{\plotpoint}}
\multiput(1301,1229)(7.780,9.724){2}{\usebox{\plotpoint}}
\put(1319.34,1252.46){\usebox{\plotpoint}}
\multiput(1325,1260)(8.156,9.411){2}{\usebox{\plotpoint}}
\put(1342.85,1281.46){\usebox{\plotpoint}}
\multiput(1350,1291)(7.472,9.963){2}{\usebox{\plotpoint}}
\multiput(1362,1307)(7.565,9.892){2}{\usebox{\plotpoint}}
\put(1380.16,1331.31){\usebox{\plotpoint}}
\multiput(1387,1341)(7.182,10.174){2}{\usebox{\plotpoint}}
\multiput(1399,1358)(7.182,10.174){2}{\usebox{\plotpoint}}
\multiput(1411,1375)(7.565,9.892){2}{\usebox{\plotpoint}}
\put(1430.82,1402.23){\usebox{\plotpoint}}
\put(1436,1410){\usebox{\plotpoint}}
\sbox{\plotpoint}{\rule[-0.500pt]{1.000pt}{1.000pt}}%
\put(220,655){\usebox{\plotpoint}}
\put(220.0,655.0){\rule[-0.500pt]{292.934pt}{1.000pt}}
\sbox{\plotpoint}{\rule[-0.175pt]{0.350pt}{0.350pt}}%
\put(423,655){\raisebox{-.8pt}{\makebox(0,0){$\Diamond$}}}
\put(796,772){\raisebox{-.8pt}{\makebox(0,0){$\Diamond$}}}
\end{picture}
\caption[x]   {\hspace{0.2cm}\parbox[t]{13cm}
{\small
   The phase diagram of the Kazakov--Migdal model with the
   potential~\rf{V}. The one-cut solution realizes for $\a<\a_c$.
   The critical lines $\a=\a_c$ and $\a=-1$
   correspond to $\gamma_{str}=-1/2$ and $\gamma_{str}=0$, respectively,
   while the tri-critical point $\b=1/2$, $\a=-1$ is associated with
   the Kosterlitz--Thouless phase transition.
   }}
\label{phases}
\end{figure}

The proper regions of the parameters are depicted in Fig.~\ref{phases}.
The bold line which starts at $\a=\b=0$ corresponds to the equality sign
in~\rf{discriminant}. The solution~\rf{1of3} is well-defined below this
line.

\subsection{The critical behavior}

The critical behavior of our model can emerge when:
\begin{itemize} \vspace{-8pt}
\addtolength{\itemsep}{-8pt}
\item[i)]
The spectral density ceases to be positive at the interval $[\tx,\ty]$
as for the one-matrix models with a polynomial potential.
\item[ii)]
Either $\tx$ approaches $-\b$ or $\ty$ approaches $\b$.
\item[iii)]
$\tx$ approaches $\ty$ like for the Penner model~\cite{Tan91}.
\vspace{-8pt}
\end{itemize}

Since we start from the Gaussian solution,
we shall verify that
a phase transition of the type $i)$, which restricts the one-cut solution,
does not occur before those of the type $ii)$ or $iii)$.
The proper criterion can be obtained from \eq{zbone-cut}
where the factor $z+\tl$ should vanish for some $\tl \in [\tx,\ty]$
for this phase transition to occur.
The equation for the critical points associated with $\tx = -z$
can be easily obtained from \eq{roots}. It has the following solution
for the critical value of $\a$ at given $\b$
\be
\tx = -z~~~~~\hbox{ for }~~~~
\a_c = \b^2 -3 \left(\frac{\b}{4}\right)^{\frac 23} -\fr 12
\label{critical}
\ee
which exactly coincides with equality sign in~\rf{discriminant}
when the discriminant of the cubic equation~\rf{ce} vanishes.
For this reason the inequality $\tx> -z$ which means the existence of
the one-cut solution~\rf{zbone-cut} is equivalent to~\rf{discriminant}.
This is why the expression~\rf{1of3} fully describes our one-cut solution
of the one-matrix model.

It is easy to understand why the critical value~\rf{critical} is
associated with the vanishing of the discriminant. We can identically
rewrite~\eq{ce} as
\be
\a_c-\a = \left( z - \left(\frac{\b}{4} \right)^{\frac 13} \right)^2
\left( 1+ \frac{2}{z} \left(\frac{\b}{4} \right)^{\frac 13}\right)
\label{stringequation}
\ee
which looks like the genus zero string equation in polynomial one-matrix
models~\cite{Kaz89}. For the critical behavior to occur, two roots of the
cubic equation should coincide which happens when the discriminant
vanishes. The critical value of $z$ extracted from \eq{stringequation} is
\be
z_c =  \left(\frac{\b}{4} \right)^{\frac 13}
\ee
and near the critical point
\be
\a_c-\a = (z-z_c)^2 \left(1+\frac{2z_c}{z} \right) \approx 3 (z-z_c)^2
\ee
in the full analogy to one-matrix models with polynomial potentials.

It follows also from the above formulas that $z$ given by \eq{1of3} is a
monotone function of both $\a$ and $\b$ in the region where the
inequality~\rf{discriminant} is satisfied since
\be
\frac{\d \a}{ \d z} = \frac {2}{z^2}(z_c^3-z^3)
\label{stringequationp}
\ee
and
\be
\frac{\d \b}{ \d z} =
 \frac{z^3-z_c^3}{z^2 \sqrt{z^2+\a+\frac 12 +\frac{1}{16z^2}}}
\ee
while $z \geq z_c$ in the domain~\rf{discriminant}.
The expression under the square root in the latter formula is
always non-negative for $\a \geq -1$.

Another set of singular points is when $\ty = \b$. From \eq{roots} we
find that this happens for $z$ given by \eq{1of3} only when $\a=0$:
\be
\ty = \b~~~~~\hbox{ for }~~~~\a=0 \,,
\label{alpha=0}
\ee
while $z$ is expressed via $\b$ along the line $\a=0$ by
\be
z = \frac \b2 + \sqrt{\frac{\b^2}{4}-\frac 12}
{}~~~~~\hbox{ for }~~~~\a=0 \,.
\label{alongalpha=0}
\ee
\eq{alongalpha=0} is applicable for $\b\geq\b_c=\sqrt{2}$. The value
$\b_c=\sqrt{2}$ satisfies \eq{critical} at $\a=0$. Therefore, the
critical line~\rf{alpha=0}
which is depicted in Fig.~\ref{phases}
by the solid line terminates at
the intersection with the critical line~\rf{critical}.
The tri-critical point is given by
\be
\a_c = 0~,~~~~~\b_c=\sqrt{2} \,.
\label{tri-critical0}
\ee

Analogously, $\tx$ and $\ty$ coincide for the solution~\rf{1of3}
when $\a=-1$:
\be
\tx = \ty~~~~~\hbox{ for }~~~~\a=-1 \,,
\label{alpha=-1}
\ee
while $z$ is expressed via $\b$ along the line $\a=-1$ by
\be
z = \b
{}~~~~~\hbox{ for }~~~~\a=-1 \,.
\label{alongalpha=-1}
\ee
\eq{alongalpha=-1} is applicable for $\b\geq\b_c=1/2$. The value
$\b_c=1/2$ satisfies \eq{critical} at $\a=-1$. Therefore, the
critical line~\rf{alpha=-1}
which is depicted in Fig.~\ref{phases} by the solid
line terminates at the intersection with the critical line~\rf{critical}.
The corresponding tri-critical point reads
\be
\a_c = -1~,~~~~~\b_c=\frac{1}{2} \,.
\label{tri-critical}
\ee
Note that the critical lines~\rf{alpha=0} and \rf{alpha=-1} lie in the allowed
region below the line~\rf{critical}.

\subsection{Calculation of susceptibility}

The critical behavior of the model~\rf{spartition} with
the potential~\rf{V} is characterized by the susceptibility
\be
\chi \equiv - \frac{1}{N^2\; \hbox{Vol.}} \frac{d^2}{d\a^2} \log{Z}
\ee
where\ \/Vol.\ \/stands for the volume of the system
\be
\hbox{Vol.} = \sum_x 1\,.
\ee
By differentiating~\rf{spartition} one gets explicitly
\bea
\chi = -\frac{d}{d\a}\LA \ntr \left[\log{(b-\phi_x)} + (2D-1)\log{(a+\phi_x)}
\right]\RA = \non -
\oint_{C_1} \frac{d\om}{2\pi i} [\log{(b-\om)} +(2D-1) \log{(a+\om)}
\; \dot{E}_\om ]
\label{chi}
\eea
where
\be
\dot{E}_\om\equiv \frac{dE_\om}{d\a}\,.
\ee

Compressing the contour $C_1$ in \eq{chi}
to the cuts of the logarithms, one gets
\be
\chi = - \left\{ \int_{-\infty}^b dt\dot{E}_t
+(2D-1) \int_{-\infty}^{-a}  dt \dot{E}_t   \right\}  \,.
\label{integral}
\ee
This formula can alternatively be derived from \rf{chi}
using the identities
\be
\frac{d}{d\a}\LA \ntr \log{(b- \phi_x)} \RA =
\frac{d}{d\a} \int_{-\infty}^b dt\; \LA \ntr \frac{1}{(t- \phi_x)} \RA =
\int_{-\infty}^b dt \; \dot{E}_t
\label{altchib}
\ee
and
\be
\frac{d}{d\a}\LA \ntr \log{(a+ \phi_x)} \RA =
\frac{d}{d\a} \int_{-\infty}^{-a} dt\; \LA \ntr \frac{1}{(t- \phi_x)} \RA =
\int_{-\infty}^{-a} dt \; \dot{E}_t \,.
\label{altchia}
\ee

To calculate $\chi$ in genus zero, we use the formula
\be
\dot{E}_\l = \int_{C_1} \frac{d\om}{4\pi i}
\frac{\dot{\tilde{V}}{}'(\om)}{(\l-\om)}
\frac{\sqrt{(\om-x_-)(\om-x_+)}}{\sqrt{(\l-x_-)(\l-x_+)}}
\label{dotEvstV}
\ee
which can be proven by a direct differentiation of Eqs.~\rf{EvstV}
and \rf{xandy} and holds for the derivative w.r.t.\ {\it
any}\/ parameter of the potential $\tV$, in particular w.r.t.\ $\a$.
For our logarithmic potential~\rf{tV} one gets explicitly
\bea
\dot{E_\l} = \frac 12 \left\{
\frac{1}{b-\l}+\frac{1}{a+\l} + \frac{1}{\sqrt{(\l-x_-)(\l-x_+)}}\right.\non
\times \left.\left[ \pm \frac{\sqrt{(b-x_-)(b-x_+)}}{\l-b}
+ \frac{\sqrt{(a+x_-)(a+x_+)}}{\l+a}\right] \right\}.
\label{dotone-cut}
\eea
Here and below the plus sign in front of the first term in the square brackets
is associated with $\a>0$ while the minus sign corresponds to $\a<0$.
The result of Ref.~\cite{AKM94} for the generalized Penner model is recovered
by this formula as $a\ra\infty$.

The integral in \eq{integral} with $\dot{E}_t$ given by~\rf{dotone-cut}
is easily calculable using the following formula for an
indefinite integral
\be
\int dt \dot{E}_t = \frac 12 \log{\left\{
\frac{\Big(\sqrt{(t-x_-)(a+x_+)}+ \sqrt{(t-x_+)(a+x_-)}\Big)^2}
{\Big(\sqrt{(t-x_-)(b-x_+)} \pm \sqrt{(t-x_+)(b-x_-)}\Big)^2}
\right\}}
\ee
which yields for the susceptibility in genus zero
\bea \chi_0 = (D-1)
\log{\left\{ \frac 14 \left( \sqrt[4]{\frac{(a+x_-)(b-x_+)}{(a+x_+)(b-x_-)}}
\pm \sqrt[4]{\frac{(a+x_+)(b-x_-)}{(a+x_-)(b-x_+)}} \right)^2\right\}} + \non
D \log{\left\{  \left( \sqrt[4]{\frac{(a+x_-)}{(a+x_+)}} +
\sqrt[4]{\frac{(a+x_+)}{(a+x_-)}}
\right)^2\right\}} -
D \log{\left\{ \left(
\sqrt[4]{\frac{(b-x_+)}{(b-x_-)}} \pm
\sqrt[4]{\frac{(b-x_-)}{(b-x_+)}}
\right)^2\right\}}.
\label{susc}
\eea
As before the positive sign in $\pm$ corresponds to $\a>0$ while the
minus sign should be substituted for $\a<0$. As is already mentioned in the
previous subsection, the $\a<0$ case can be obtained from the $\a>0$ case
by changing the sign of $\sqrt{(b-x_-)(b-x_+)}$.

Let us briefly discuss some properties of
the expression~\rf{susc} which will be used in the next section for studying
the continuum limits of our model.  If $\a>0$ the contour of integrations over
$t$, which coincide with the branch cuts of the logarithms depicted in
Fig.~\ref{Fig.1}a),
can be moved along the complex plane. In particular, one can
integrate in the first term on the r.h.s.\ of \eq{integral} from $b$ to
$+\infty$ along the positive real axis. The result is the same and is not
singular at $x_-=x_+$ if one substitutes the plus sign in \eq{susc}.

On the contrary,
the points $x_-$ and $x_+$ can pinch the integration contour for $\a<0$
as is depicted in Fig.~\ref{Fig.1}d) for $\a=-1$.
This results in a logarithmic
singularity of~\rf{susc} at $x_-=x_+$ when one substitutes the minus sign in
\eq{susc}:
\be
\chi_0 \approx - \log{(x_--x_+)^2}\,.
\label{sing1}
\ee
An
analogous logarithmic singularity emerges when $x_+\approx b$ as is depicted
in Fig.~\ref{Fig.1}b) at $\a\approx0$:
\be
\chi_0 \approx \log{(b-x_+)} \,.
\label{sing2}
\ee
Note, that
\be
\chi_0 = - \log{(1+\a)} + \log \a
\label{penchi}
\ee
in the Penner limit~\rf{resc} with $c\ra0$ when $x_\pm$ are given by
\be
x_\pm = b - \frac{\a+2}{a} \pm \frac 2a \sqrt{\a+1}~.
\label{penxy}
\ee
\eq{penchi} coincides with the susceptibility for the Penner
model~\cite{Tan91} and recovers the singularities~\rf{sing1} and \rf{sing2}.

Having the explicit formula~\rf{susc} for $\chi_0$, we can find out which
$\gamma_{str}$ is associated with each type of
the critical behavior~\rf{critical},
\rf{alpha=0} and \rf{alpha=-1}. Along the line~\rf{critical} where $\chi_0$
is not singular and equals to some value $\chi_0^c$ , one gets
\be
\chi_0-\chi_0^c \sim (x_--x_-^c) \,.
\label{4.14}
\ee
Since $(x_--x_-^c)\sim (\a_c-\a)^{1/2}$ near the critical line~\rf{critical},
one obtains $\gamma_{str}=-1/2$.

Near the critical lines~\rf{alpha=0} and \rf{alpha=-1}, where $\chi_0$ is
given by Eqs.~\rf{sing1} and \rf{sing2}, one gets $\gamma_{str}=0$.
While~\rf{sing1} is positive, \rf{sing2} is negative. For this reason we shall
not consider the continuum limit associated with the critical line~\rf{alpha=0}
as well as the tri-critical point~\rf{tri-critical0} since it corresponds to a
negative susceptibility.  The susceptibility in the vicinity of the
tri-critical point~\rf{tri-critical} is considered in Subsect.~3.3.

\newsection{The continuum limits}

The continuum theories are obtained by an expansion
near the critical points.
The continuum limit associated with $\gamma_{str}=0$
describes in a standard way $2D$ gravity plus $1D$ critical matter.
The continuum limit associated with $\gamma_{str}=-1/2$
looks like pure $2D$ gravity for local observables which
are defined at the same
site of the lattice. For another type of observables ---
extended Wilson loops ---  the continuum limit sets up
at distances $L\sim 1/\sqrt{\eps}$ with $\eps$ being a deviation from
the critical point. This is due to a singular behavior of the Itzykson--Zuber
correlator of gauge fields for this phase.
A Kosterlitz--Thouless phase transition separates the two phases.

\subsection{$\gamma_{srt}=-1/2$}

In order to find out what kind of continuum theory is associated with the
critical behavior~\rf{critical},
let us expand all quantities near the edge singularity of
the spectral density substituting
\be
z=z_c + \eps \sqrt \L~,~~~~~ z_c= \left( \frac{\b}{4} \right)^{\frac13}
\label{6.1}
\ee
where $\eps\ra0$ and $\L$ is to be identified with the cosmological constant of
$2D$ gravity. From Eqs.~\rf{stringequation}, \rf{roots} one gets
\be
\a_c-\a=3 \eps^2 \L~,~~~~~~\tx=-z_c+2 \eps \sqrt{\L}~,
{}~~~~~\ty=3z_c-\frac 1\b\,.
\label{6.2}
\ee
Introducing the continuum momentum variable, $\xi$, by
\be
\tl = -z_c + \eps \xi
\ee
one gets from \eq{zbone-cut}
\be
E_\l = -\frac{1}{2z_c} +\eps \xi - \eps^{3/2}
\left( \xi + \sqrt{\L} \right)
\sqrt{-\xi +2\sqrt{\L}}\frac{2}{\sqrt{z_c}\sqrt{16z_c^4-1}}  +
{\cal O}(\eps^2)
\,.
\ee
The last term on the r.h.s.\ determines the continuum spectral density
\be
\rho_c (\xi) = \frac 1\pi \left( \xi + \sqrt{\L} \right)
\sqrt{\xi - 2 \sqrt{\L}}
\label{rhocont}
\ee
and,
therefore, all continuum correlators of the trace
of powers of the (renormalized) field
$\Phi(x)$ at some point $x$, which is the standard set of observables of $2D$
gravity.

While the gravitational part of the system is continuous, this does not
necessarily mean that matter becomes critical. An example is the Ising
model on a random lattice where it is easy to construct the
$\gamma_{str}=-1/2 $ behavior which is associated with continuum $2D$
gravity and non-critical matter while matter becomes critical at a
tri-critical point changing the value of the string susceptibility to
$\gamma_{str}=-1/3 $.

A direct way to verify whether matter becomes critical at a given fixed
point is to investigate
observables which are associated with
extended objects --- the open-loop averages
\be
G_{\nu\l}(C_{xy})= \left\langle
\ntr{\Big(\frac{1}{\nu- \phi_x} U (C_{xy})
\frac{1}{\l- \phi_y}
U^\dagger(C_{xy}) \Big)} \right\rangle\,,
\label{KMG}
\ee
where $C_{xy}$ goes from $x$ to $y$ along some path on a
$D$-dimensional lattice
and the average is w.r.t.\ the same measure as in~\rf{spartition}.
$G_{\nu\l}(C_{xy})$ is symmetric in $\nu$ and $\l$
due to invariance of the Haar measure, $dU$, under
the transformation $U\ra U^\dagger$.

The averages~\rf{KMG} depend~\cite{Mak92} at large $N$
only on the algebraic length $L(C_{xy})$ of
the contour $C_{xy}$ (\ie the one after contracting backtrackings) and
can be calculated~\cite{DMS93} providing
$C(\nu,\l)$ --- the one-link Itzykson--Zuber correlator
of the gauge fields ---  is known.
The latter is expressed via the double discontinuity of
\be
G_{\nu\l} \equiv G_{\nu\l}(1)
\label{L=1}
\ee
across the cut:
\be
C(\nu,\l) \equiv \frac{1}{\pi^2 \rho(\nu) \rho(\l)}\,\hbox{Disc}_\nu\,
\hbox{Disc}_\l\, G_{\nu\l} \,,
\label{defC}
\ee
where the continuous and
discontinuous in $\nu$ parts of $G_{\nu\l}$ are defined by
\bea
& &\Di G_{\nu\l} \equiv \frac{G_{(\nu+i0)\l}-G_{(\nu-i0)\l}}{2i}~, \non
& &\Co G_{\nu\l} \equiv \frac{G_{(\nu+i0)\l}+G_{(\nu-i0)\l}}{2}
\label{5.22}
\eea
so that for a real $\l$ outside of the cut (cuts)
$\Di G_{\nu\l}$ coincides with
the imaginary part and $\Co G_{\nu\l}$ coincides with the real part.
In particular, $\Di E_\nu = -i\pi \rho(\nu)$.

For the potential~\rf{V} $G_{\nu\l}$ reads~\cite{Mak93}
\be
G_{\nu\l} =\frac{(b-\nu)E_\nu-(a+\l)E_\l+1}{(\l+a+E_\nu)(b-\nu)-\a}\,.
\label{3.8}
\ee
The few lower terms of the expansion of $G_{\nu\l}$ in $\eps$ are
\be
G_{\nu\l} =
\frac{2}{4z_c^2+1} - i \pi \sqrt{\eps}\; \frac{\rho_c(\xi_\nu) -
\rho_c(\xi_\l)}{\nu-\l}\; \frac{(4z_c^2-1)}{(4z_c^2+1)} + {\cal O}(\eps)
\label{ordereps}
\ee
which is similar to the one in a two-matrix model~\cite{Sta93}. The
double discontinuity of the first two terms on the r.h.s.\ vanishes so that one
needs the term at least ${\cal O}(\eps)$.

For this reason we start
directly from the exact $C(\nu,\l)$ which for the potential~\rf{V}
reads~\cite{Mak93}
\be
C(\nu,\l)= \frac{a+\l}{(b-\nu)[\l-r_+(\nu)][\l-r_-(\nu)]}
\label{diskC}
\ee
where
\be
r_\pm(\l)= \frac 12 \left\{ \frac{\a}{b-\l} -\frac{\a+1}{a+\l}+b-a
\right\} \pm i\pi \rho(\l)\,.
\label{rlog}
\ee
The r.h.s.\ of \eq{diskC} is symmetric in
$\nu$ and $\l$ due to the relation
\be
r_\pm(r_\mp(\l))=\l
\label{m}
\ee
which is satisfied by the solution~\rf{one-cut}~\cite{Mak93a}.  Analogously to
Ref.~\cite{Bou93}, it is easy to see that \eq{m} is satisfied by our explicit
continuum formulas.

To calculate the continuum limit of~\rf{diskC}, we expand
\be
r_\pm (\l) = -z_c + \eps \xi- \frac{3\eps^2 \xi^2}{z_c(16z_c^4-1)} +
{\cal O}(\eps^3) \pm i\pi \rho(\l)\,.
\ee
We put here $\L=0$ and keep terms of order $\eps^2$ which we shall need below.
One gets
\be
C(\nu,\l) = \frac{1}{\eps^2}
\frac{(4z_c^2-1)}{(4z_c^2+1)}\frac{(1+\eps
\frac{\xi_\nu}{z_c(4z_c^2-1)})(1+\eps \frac{\xi_\l}{z_c(4z_c^2-1)})}
{(\xi_\nu-\xi_\l)^2 + \frac{2\eps\xi_\nu\xi_\l(\xi_\nu+\xi_\l)}
{z_c(16z_c^4-1)}} \,.
\label{Ceps}
\ee
This expression comes from the ${\cal O}(\eps)$ term in the
expansion~\rf{ordereps} since the denominator in~\rf{defC} is ${\cal
O}(\eps^3)$.

In the continuum limit $\eps=0$ we get from \eq{Ceps}
\be
C_c(x,y) \propto \frac {1}{(x-y)^2}
\label{Cone-link}
\ee
which coincides with $C(x,y)$ for the quadratic potential~\cite{DMS93}%
\footnote{An analogous formula for the two-matrix model has been earlier
obtained in Ref.~\cite{Alf93}. }
\be
C(x,y) = \frac{\sqrt{\mu^2+4}+\mu}{2\,(x^2-\sqrt{\mu^2+4}\;xy+y^2+\mu)}
{}~~~~~~~\hbox{(quadratic potential)}
\label{quaC}
\ee
at $\mu=0$. While for the quadratic potential $\mu=0$ is possible only at $D=1$
where it is associated with the naive continuum limit~\cite{Gro92},
\eq{Cone-link} holds in our case at any $D$ along the line~\rf{critical}.

The expression~\rf{quaC} for the Gaussian $C(x,y)$ possesses a remarkable
convolution property~\cite{DMSW93} --- keeps its functional structure when
one combines the path of the length $L_1+L_2$ from the paths of the lengths
$L_1$ and $L_2$. Then
\be
C(x,y;L_1+L_2) = \int dt\, \rho(t)\, C(x,t;L_1)  C(t,y;L_2)
\label{comb}
\ee
where
\be
C(\nu,\l;L) \equiv \frac{1}{\pi^2 \rho(\nu) \rho(\l)}\,\hbox{Disc}_\nu\,
\hbox{Disc}_\l\, G_{\nu\l}(C_{xy}) \,.
\label{defCL}
\ee
Observables which are associated with matter reveal singularities at $\mu=0$
and become finite after an appropriate renormalization.

Since our expression~\rf{Cone-link} looks like the Gaussian one for $\mu =0$,
one might expect a similar property of matter correlators. However,
when~\rf{Cone-link} is substituted into \eq{comb}, the integral is divergent at
$t=x$ or $t=y$. To regularize, we keep ${\cal O}(\eps)$ term in the
denominator of~\rf{Ceps}. Now the integral is ${\cal O}(\eps^{-1/2})$ which
exactly cancels an extra ${\cal O}(\eps^{1/2})$ factor which emerges since each
of two $C$ in the integral~\rf{comb} is proportional to $1/\eps^2$ (according
to \eq{Ceps}) and the measure is proportional to $\eps^{5/2}$.  Therefore, one
gets
\be
C(x,y;2) \propto \frac{1}{\eps^2(x-y)^2}
\ee
and the functional
structure of $C_c$ is preserved.  The coefficient of proportionality is
calculated below.

An interesting question is whether \eq{comb} with $\rho$ given by the continuum
formula~\rf{rhocont} can have a solution with the same functional structure for
macroscopic loops as well.
A solution to this equation for $\L=0$ is given by
\be
C_c(x,y;\sqrt{u}) = \frac{2\sqrt{u}}{(x-y)^2+2u(x+y)xy+u^2x^2y^2}
\label{macrosolution}
\ee
which obeys the following convolution property
\be
\frac 1\pi\int_0^\infty dt\; {t}^{\frac 32} C_c(x,t;\sqrt{u}) C_c(t,y;\sqrt{v})
= C_c(x,y;\sqrt{u}+\sqrt{v}) \,.
\label{convolution}
\ee
This formula is proven in Appendix~B.

It is easy to see that the solution~\rf{macrosolution}
satisfies the initial condition~\rf{Cone-link} for $u\ra0$.
To find the relation between $u$ and the length $L$, let us first rescale \be
\xi \ra  \xi \frac{\sqrt{z_c(16z_c^4-1)}}{2}~,~~~~~~
u \ra  u \frac{\sqrt{z_c(16z_c^4-1)}}{2}\,.
\label{rescale}
\ee
Now the denominator in
\rf{macrosolution} coincides to order $\eps$ with the one in~\rf{Ceps}
providing
\be
u =  \frac{\eps}{4} \,.
\label{microsolution}
\ee
It follows then from \eq{convolution} that
\be
u = L^2 \frac{\eps}{4}
\label{uvsL}
\ee
where the length $L$ is measured in the lattice units.

One sees from Eqs.~\rf{macrosolution} and \rf{uvsL} that the continuum
limit of extended correlators is reached at distances
\be
L\sim \frac{1}{\sqrt{\eps}}
\label{distances}
\ee
rather than
$\sim 1/\eps$ as it might be naively expected. Therefore, a nontrivial
scale dimension of the matter field is developed at macroscopic distances.

The knowledge of the $L$-dependence of $C_c$ allows us to calculate correlators
of extended objects. The simplest one is that of the adjoint Wilson loop
\be
W_A(C)\equiv\LA \left( \left| \ntr{U(C)}\right|^2-1\right)  \RA =
\frac{1}{N^2} \int dt\; \rho(t)\, C(t,t;L) \,.
\label{adjloop}
\ee
Substituting~\rf{macrosolution}, one gets
\be
W_A(C) = \frac{2\sqrt{u}}{\eps \pi} \int_0^\infty
\frac{dt \;t^{3/2}}{4 u t^3 + u^2 t^4} = -\frac{1}{4\eps} \,.
\label{WA}
\ee
Since the integral is divergent as $t\ra0$, the result is obtained by an
analytic continuation. This divergence might be related to the fact that we put
$\L=0$.

Note, that the correlator \rf{WA} is divergent when $\eps\ra0$ similarly to the
one for the Gaussian case~\cite{DMSW93} as $\mu\ra0$. One concludes, therefore,
that matter could become critical at distances~\rf{distances}.  This mechanism
does {\it not}\/ work for the two-matrix model which is associated with $D=1/2$
in the above formulas. In order to have $L\ra \infty$ as $\eps \ra 0$, one
needs $D\geq1$. For $D=1$ one should choose a closed matrix chain
since otherwise the gauge field can be absorbed by a gauge transformation
of $\phi_x$.

Analogously to~\rf{WA} one can calculate the more general correlator
\be
G_\l(C)\equiv\LA \ntr{\left( \frac{1}{\l-\phi_x}U(C_{xx})\right)}
\ntr{U^\dagger(C)}  \RA =
\frac{1}{N^2} \int dt\; \rho(t)\, C(t,t;L) \frac{1}{\l-t}\,.
\ee
In the continuum limit one gets
\be
G_\xi(C) = \frac{2\sqrt{u}}{\eps^2 \pi} \int_0^\infty
\frac{dt\; t^{3/2}}{(4 u t^3 + u^2 t^4)} \frac{1}{(\xi-t)}
= \frac{1}{2\eps^2\sqrt{u}} \left\{
\frac{ (-\xi)^{-3/2} -\left( \frac{u}{4}\right)^{3/2}}{1+\frac{\xi u}{4}}
\right\}\,.
\label{Gl}
\ee
This expression is also divergent when $\eps\ra0$.

Finally, the continuum part of $G_{\nu\l}(C)$ is given by
\bea
G_{\zeta\xi}(C) = 2 \sqrt{\eps u} \frac{1}{\pi} \int_0^\infty dx dy
\frac{x^{3/2} y^{3/2}}{(x-y)^2+2u(x+y)xy+u^2x^2y^2}
\frac{1}{(\zeta-x)(\xi-y)} \non =
 2 \sqrt{\eps u}
\frac{(-\zeta)^{3/2} (-\xi)^{3/2}}{(\zeta-\xi)^2+2u(\zeta+\xi)\zeta\xi+
u^2\zeta^2\xi^2} + \ldots \,.
\eea
This expression is indeed ${\cal O}({\eps})$ for $L\sim1$ in agreement
with~\rf{ordereps} while it is ${\cal O}(\sqrt{\eps})$ for
$L\sim1/\sqrt{\eps}$.

\subsection{$\gamma_{str}=0$}

Another continuum limit is associated with the critical line $\a=-1$.
One substitutes the expansion near the edge singularity
which is quite similar to the one for the Penner model~\cite{CDL91}.
Introducing the cosmological constant, $\L$, by
\be
\a=-1+\eps^2  {\L}\, 4\b^2
\label{6.28}
\ee
where an extra factor is inserted for a latter convenience,
we get from \eq{1of3}
\be
z=\b - \frac{8 \eps^2 \b^3}{4\b^2-1}\L
\label{6.29}
\ee
and from \eq{roots}
\be
\hat{x}_\pm = \b - \frac{1}{2\b} \pm 2\eps\sqrt{\L}\,.
\label{6.30}
\ee
Defining the continuum momentum $\xi$ by
\be
\tl = \b-\frac{1}{2\b} +\eps \xi \,,
\ee
one gets from \eq{zbone-cut}
\be
E_\l = -2\b -2\b^2 \eps \xi -\eps \sqrt{\xi^2-4\L}\; 2\b^2 +{\cal O}(\eps^2)
\ee
which determines the continuum spectral density to be
\be
\rho_c(\xi) = \frac{1}{\pi} \sqrt{\xi^2-4\L} \,.
\label{6.33}
\ee

The one-link Itzykson--Zuber correlator~\rf{diskC} reads
\be
C(\nu,\l) = \frac{4 \b^2}{4\b^2-1} +{\cal O}(\eps) \,.
\label{6.34}
\ee
This expression does not depend on the continuum momenta $\xi_\nu$ and
$\xi_\l$ similar to the Gaussian expression~\rf{quaC} for $x,y \ll 1$.
Thus, nothing special happens with the Itzykson--Zuber correlators
in the $\gamma_{str}=0$
continuum limit far away from the tri-critical point
in contrast to the $\gamma_{str}=-1/2$ case.

\subsection{Kosterlitz--Thouless phase transition}

The above formulas are not applicable in the vicinity of the
tri-critical point $\b_c=1/2,$
\mbox{$\a_c=-1$} where the continuum system undergoes
a Kosterlitz--Thouless phase transition,
which was previously studied for a closed matrix chain~\cite{GK90,Yan90},
between the phases with $\gamma_{str}=-1/2$ and $\gamma_{str}=0$.
This domain is most interesting and should be treated separately.

Let us expand
\be
\b = \frac 12 + \delta \b~,~~~~
\a = -1 + \delta \a~,~~~~
z = \frac 12 + \delta z~,~~~~
\l = - \frac 12 + \delta \l~.
\ee
Then for the solution \rf{1of3} one gets
\be
\delta z = \frac 13 \delta \b +
\sqrt{\frac 49 (\delta \b)^2 - \frac 13 \delta \a }
\label{deltaz}
\ee
while the inequality \rf{discriminant} reduces to
\be
\delta \a \leq \frac 43 (\delta \b)^2 \,.
\label{inequa}
\ee

It is convenient to parametrize the vicinity of
the tri-critical point by the lines
\be
\delta \a  = (3\ka+1)(1-\ka) (\delta \b)^2
\label{lines}
\ee
which obey the inequality~\rf{inequa} for any $\ka$. We
choose $1/3\leq \ka \leq 1$ where the factor in \eq{lines}
is a monotone function of $\ka$. The cases
$\ka=1/3$ and $\ka=1$ are associated with the critical
lines~\rf{critical} and \rf{alpha=-1}, respectively.
Then, \rf{deltaz}
is rewritten as
\be
\delta z = \ka \delta \b
\label{linesz}
\ee
while \eq{roots} gives
\be
\hat{x}_\pm = -\frac 12 + (2+\ka) \delta \b\pm 2\sqrt{2(1-\ka^2)} \delta \b\,.
\label{linesroots}
\ee

Eqs.~\rf{lines}, \rf{linesz}, \rf{linesroots} recover Eqs.~\rf{6.1}, \rf{6.2}
or Eqs.~\rf{6.28}, \rf{6.29} \rf{6.30} if
\be
\ka = \frac 13 +\frac{\eps \sqrt{\L}}{\delta \b}
\label{ka1/3}
\ee
or
\be
\ka = 1 -\frac{\eps^2 \L}{4(\delta \b)^2}
\label{ka1}
\ee
with $\eps \ll \delta \b$, respectively.

The susceptibility \rf{susc} can easily be
expressed via $\delta \b$ and $\ka$ near the tri-critical point:
\be
\chi_0 = - \log{\left(\frac{1-\ka}{1+3\ka}\right)} -
2D \log{(\sqrt{2}\delta \b)} \,.
\label{singb}
\ee
The first term on the r.h.s.\ recovers \eq{sing1} in the limit~\rf{ka1}
while the second one which becomes singular when $\delta \b \ra 0$
is a new type of singularity which appears only at the tri-critical point.
Notice that this second term is $D$ dependent.
Using \eq{lines} one can rewrite~\rf{singb} alternatively via $\delta \a$.

The continuum spectral density can be obtained from \rf{zbone-cut} by
substituting the expansions~\rf{linesz}
and \rf{linesroots} which gives
\be
\rho_c(\delta \l) = \frac{1}{2\pi}
\sqrt{(\delta \l -(2+\ka) \delta \b)^2 - 8 (1-\ka^2)(\delta \b)^2}\,
\frac{\ka \delta \b + \delta \l}{\delta \b + \delta \l}\,.
\ee
For $\ka >1/3$ the zeros both of the numerator and of the denominator
lie outside of the eigenvalue support. In the limits~\rf{ka1/3} and
\rf{ka1} this expression recover Eqs.~\rf{rhocont} and \rf{6.33}, respectively.

The one-link Itzykzon--Zuber correlator~\rf{diskC}
near the tri-critical point reads
\be
C(\delta \nu, \delta \l))= \frac{(\delta\l+\delta\b)(\delta\nu+\delta\b)}
{{\cal D}(\delta \nu, \delta \l)}
\ee
where
\bea
{{\cal D}(\delta \nu, \delta \l)}=
\delta\l\delta\nu(\delta\l+\delta\nu)+((\delta\l)^2+(\delta\nu)^2)\delta\b
\non
+\ka(2-3\ka)(\delta\l+\delta\nu)(\delta\b)^2+2\ka^2(1-2\ka)(\delta\b)^3 \,.
\eea
This expression is $\sim 1/\delta \b$ in agreement with~\rf{Ceps}
and \rf{6.34} which are recovered when one substitutes
\be
\delta \l = -\frac 13 \delta \b + \eps \xi
\ee
and
\be
\delta \l = 3 \delta \b + \eps \xi\,,
\ee
respectively.

\newsection{Critical scaling in the large-$D$ limit \label{large-D} }

The Itzykson--Zuber integral
\be
I[\phi_x,\phi_y] \equiv \int dU \e^{Nc\tr{\phi_x U \phi_y U^\dagger }},
\ee
which enters the partition functions~\rf{spartition},
can easily be calculated at small $c$ (\ie at strong coupling):
\be
\log{I[\phi_x,\phi_y]} = c \tr{\phi_x}\tr{\phi_y} +
\frac{c^2 N^2}{2} \left[\ntr{\phi_x^2}-\left(\ntr{\phi_x}\right)^2\right]
\left[\ntr{\phi_y^2}-\left(\ntr{\phi_y}\right)^2\right] +{\cal O}(c^3) \,.
\label{IZ}
\ee

Let us consider the large-$D$ limit of the partition function.
Assuming that $V(\phi_x) \sim 1$ as $D\ra\infty$,%
\footnote{We differ at this point from Ref.~\cite{Bou93} where the
large-$D$ limit was considered for $c\sim1$ and $V(\phi_x)\sim D$
so that the solution with the minus sign in front of the
square root in \eq{mum0}
was chosen in order for $\mu$ to be $\sim 1$.
Contrary to the statement of Ref.~\cite{Bou93} that there is no
real scaling solutions at large $D$, we do have them in the limit~\rf{coD}
perturbing the Gaussian solution~\rf{mum0} with the plus sign
which agrees with the large-mass expansion.}
we see that the kinetic
term is of order $D$ (\ie of the same order as the potential) if
\be
c \sim \frac{1}{D} \,.
\label{coD}
\ee
The Itzykson--Zuber integral coincides in this limit with the first term
of the expansion~\rf{IZ} since the higher terms are suppressed.
Therefore, we can write down the partition function~\rf{spartition} as
\be
Z = \int \prod_x d\phi_x \e^{-N\sum_{x}\tr{V(\phi_x)}+ c \sum_{\{x,y\}}
\tr{\phi_x} \tr{\phi_y}} \,.
\label{exponent}
\ee

Further simplification occurs in the large-$N$ limit when we
can replace one trace
in the product of two traces in the exponent in~\rf{exponent} by the
average value due to factorization. One arrives, hence, to the one-matrix model
whose potential $\tV(\phi)$ is determined self-consistently from the equation
\be
\tV(\phi) = V(\phi) - 2c D \LA \ntr{\phi}\RA_{\tilde{V}} \phi \,.
\label{self-consistent}
\ee
We have assumed that the potential $V(\phi)$ is not symmetric so that
$\LA \tr{\phi} \RA \neq 0$. If $V(\phi)$ is symmetric ($V(\phi)=V(-\phi)$) and
$\LA \tr{\phi} \RA= 0$, then one should keep the second term on the r.h.s.\ of
\eq{IZ} which yields
\be
\tV(\phi) = V(\phi) -
c^2 D \LA \ntr{\phi^2}\RA_{\tilde{V}} \phi^2 ~~~~~
\hbox{(symmetric potential)}
\label{sself-consistent}
\ee
and
\be
c \sim \frac{1}{\sqrt{D}}~~~~~~
\hbox{(symmetric potential)}  \,.
\ee

\eq{sself-consistent} is analogous to that for the one-matrix model with a
symmetric potential which involves $(\tr{\phi^2})^2 $~\cite{DDSW90}.
An equation of the type~\rf{self-consistent} appears~\cite{Kor92,ABC93} for a
non-symmetric potential. Quite similar to these papers we shall see in a moment
that the KM model at large $D$ admits  scaling solutions with
$\gamma_{str}\geq0$.

Let us analyze to this aim \eq{self-consistent}.
For the one-cut solution of the Hermitean one-matrix model with the potential
$\tV$ one gets
\bea
\LA \ntr{\phi} \RA_{\tilde{V}} =
\int_{C_1} \frac{d\om}{4\pi i}
\frac{\tV'(\om)\, \om^2}{\sqrt{(\om-x_-)(\om-x_+)}}
- \frac{x_-+x_+}{2} \non=
\int_{C_1} \frac{d\om}{4\pi i} \tV'(\om)\,{\sqrt{(\om-x_-)(\om-x_+)}}
+ \frac{x_-+x_+}{2}  \,.
\eea
\eq{self-consistent} can then be written as the following equation for $\tV$:
\be
\tV(\l) = V(\l) -2 c D  \left[
\int_{C_1} \frac{d\om}{4\pi i} \tV'(\om)\,{\sqrt{(\om-x_-)(\om-x_+)}}
+ \frac{x_-+x_+}{2}  \right] \l \,.
\label{eqfortV}
\ee
Differentiating w.r.t.\ a parameter of the potential which is
associated with the
cosmological constant and w.r.t.\ $\l$, we get finally
\be
\dot{\tV}{}'(\l) = \dot{V}'(\l) -2 c D
\int_{C_1} \frac{d\om}{4\pi i} \dot{\tV}{}'(\om)\,{\sqrt{(\om-x_-)(\om-x_+)}}
\label{doteqfortVp} \,.
\ee

Let us identify the cosmological constant with $g_1$ --- the coupling in front
of the linear term of the potential. For the susceptibility in genus zero one
gets
\be
\chi_0 = \ci \dot{V}(\om)\dot{E}_\om = \dot{\tilde{g}}_1 \frac{(x_--x_+)^2}{16}
\label{ch}
\ee
while \eq{doteqfortVp} yields
\be
\dot{\tilde{g}}_1 = \frac{1}{1+c D\frac{(x_--x_+)^2}{8}} \,.
\label{dottg}
\ee

To obtain the critical behavior, we expand near $x_-=x_{-}^{c}$ which
gives for a $k$-th multicritical point of the one-matrix model~\cite{Kaz89}:
\be
x_--x_-^c \sim (\tilde{g}_1^c-\tilde{g}_1)^{1/k} \,.
\label{kth}
\ee
Under normal circumstances when \eq{4.14} holds,
one gets from~\rf{kth}
\be
\chi_0 -\chi_0^c \sim (\tilde{g}_1^c-\tilde{g}_1)^{1/k}
\ee
so that $\gamma_{str}=-1/k$
since
\be
(g_1^c-g_1)\sim (\tilde{g}_1^c-\tilde{g}_1)\,.
\ee

This is {\it not\/} the case, however, for
\be
c= - \frac{8}{D(x_--x_+)^2}
\label{ccric}
\ee
when the denominator in \eq{dottg} vanishes. At this point one has
\be
\dot{\tilde{g}}_1 \sim (x_--x_-^c)^{-1}
\ee
so that
\be
(g_1^c-g_1)\sim (\tilde{g}_1^c-\tilde{g}_1)^{(k+1)/k}
\sim (x_--x_-^c)^{k+1}\,.
\ee
For the susceptibility~\rf{chi} one gets at the point~\rf{critical}
\be
\chi_0 \sim (x_--x_-^c)^{-1} \sim  (g_1^c-g_1)^{-1/(k+1)}
\ee
which is associated with $\gamma_{str}=1/(k+1)$.

The formula~\rf{self-consistent}, which describes the reduction of the KM model
to an one-matrix model at large $N$ in the large-$D$ limit, can be
explicitly verified for the potential~\rf{Vc} when the exact solution is
known at any $D$. As $D\ra\infty$ with $c\ra0$ according to \eq{coD},
the potential $\tV$ of the one-matrix model is given by
the Penner potential~\rf{Pennerpot}. An explicit calculation which is based
on \eq{one-cut} yields
\be
\LA \ntr{\phi}\RA_{\tilde{V}} =b -\frac{\a+1}{a}
\ee
while the large-$D$ limit of the potential~\rf{Vc} is
\be
V(\phi)\ra -\a
\log{(b-\phi)}-a +2cD\left[b- \frac{(\a+1)}{a}\right] \phi
\ee
where~\rf{coD} is used.
It is easy to see now that \eq{self-consistent} is satisfied.

While the reduction to a one-matrix model with the Penner potential holds
for the KM model with the potential~\rf{Vc} in the large-$D$ limit,
the only possible scaling behavior is with $\gamma_{str}=0$ in a
perfect agreement with the results of Sect.~2. This seems to be
a limiting case of $\gamma_{str}=1/(k+1)$ which appear from the
critical behavior with $\gamma_{str}=-1/k$ of the one-matrix model with
the potential $\tV$.

Let us note finally that
nonvanishing results for continuum correlators  can be obtained in the
large-$D$ limit only for these of operators living at the same lattice site
while the Itzykson--Zuber correlator for a contour of the length $L$
is suppressed as
\be
C(\nu,\l;L) \sim c^L \sim D^{-L} \,.
\ee
Therefore extended correlators vanish in the large-$D$ limit.

\newsection{The gauged Potts versus KM models}

As is known, the KM model is equivalent
to the matrix model on a Bethe
tree~\cite{Bou92}. We propose in this section yet another matrix model
---  the gauged Potts model --- which is equivalent to the KM model at large
$N$ providing the coordination numbers coincide.
The gauged Potts model is convenient for studying a relation with discretized
random surfaces and for interpreting the results of the previous sections.
The proof of equivalence is given via loop equations which reduce at
large $N$ to a one-link equation whose different forms are considered.
This reduction holds in the strong coupling
phase where the vacuum expectation values of the closed Wilson loops
of the gauge field vanish except for those of vanishing minimal area.

\subsection{The partition function}

\begin{figure}[tbp]
\unitlength=1.00mm
\linethickness{0.6pt}
\begin{picture}(60.0,70.90)(-10,60)
\thicklines
\put(40.00,80.00){\line(1,2){20.00}}
\put(60.00,120.00){\line(2,-5){20.00}}
\put(80.00,70.00){\line(-4,1){40.00}}
\put(90.00,90.00){\line(-1,1){30.00}}
\put(80.00,70.00){\line(1,2){10.00}}
\put(60.00,120.00){\line(1,0){20.00}}
\put(90.00,90.00){\line(-1,3){10.00}} 
\thinlines
\put(40.00,80.00){\line(5,1){50.00}}
\put(80.00,70.00){\line(0,1){50.00}}
\put(40.00,80.00){\line(1,1){40.00}}

\put(36.00,79.00){\makebox(0,0)[cc]{1}}
\put(82.00,66.00){\makebox(0,0)[cc]{2}}
\put(58.00,123.00){\makebox(0,0)[cc]{3}}
\put(94.00,90.00){\makebox(0,0)[cc]{4}}
\put(83.00,123.00){\makebox(0,0)[cc]{5}}
\end{picture}
\caption[x]   {\hspace{0.2cm}\parbox[t]{13cm}
{\small
   The lattice in the form of a $q$-simplex (depicted for $q=5$).  }}
\label{symplex}
\end{figure}
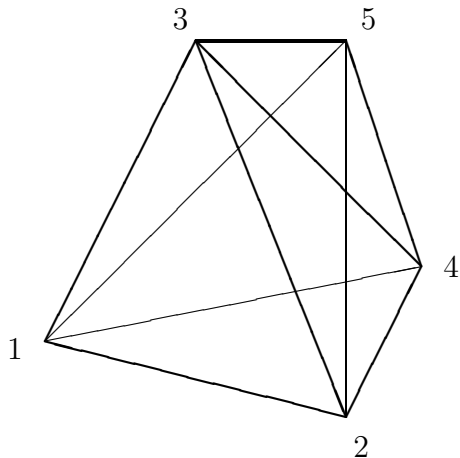

As is well-known~\cite{Kaz88}, the $q$-state Potts model on a random lattice is
equivalent to the matrix model
\be
Z_{Potts}=\int \prod_{x=1}^q
 d\phi_x \e^{  N \tr{}\left(- \sum_{x=1}^q V(\phi_x)+ c
\sum_{x>y}^q \phi_x \phi_{y} \right)}
\label{ppartition}
\ee
where the $N\times N$ Hermitean matrix $\phi_x$ lives on the lattice which form
a $q$-simplex (depicted in Fig.~\ref{symplex} for $q=5$).
The second term in the
action involves the sum over all the links (the link $\{x,y\}$ connects
the sites $x$ and $y$ which  are nothing but the vertices of the $q$-simplex).

We propose the following gauge-invariant extension of the
model~\rf{ppartition}
\be
Z_{GP}=\int \prod_{x>y} dU_{xy} \prod_{x=1}^q
 d\phi_x \e^{  N \tr{}\left(- \sum_{x=1}^q V(\phi_x)+ c
\sum_{x>y}^q \phi_x U_{xy} \phi_{y} U_{xy}^\dagger \right)}
\label{gppartition}
\ee
where the gauge variable $U_{xy}$ lives on the link $\{x,y\}$ and $d U_{xy}$
is the Haar measure on $U(N)$. This construction is quite similar to the
KM model~\cite{KM92} which is described by the partition
function~\rf{spartition} except the lattice is hypercubic in the latter case.

The relation of the partition function~\rf{ppartition} with discretized random
surfaces was studied in Ref.~\cite{Kaz88}. An analogous interpretation of
the gauged Potts model~\rf{gppartition} is based on the expansion~\rf{IZ}
of the Itzykson--Zuber integral. Now the terms of the type $(\tr \phi)^2$
generate vertices with touching surfaces~\cite{DDSW90}.

\subsection{Loop equations}

To investigate the model~\rf{gppartition}, let us introduce the extended
open-loop averages
\be
G_{\nu\l}(\C_{xy})= \left\langle
\ntr{\Big(\frac{1}{\nu- \phi_x} U (\C_{xy})
\frac{1}{\l- \phi_y}
U^\dagger(\C_{xy}) \Big)} \right\rangle
\label{gpG}
\ee
where $\C_{xy}$ goes from $x$ to $y$ along some path on the $q$-simplex
and the average is w.r.t.\ the same measure as in~\rf{gppartition}.
$G_{\nu\l}(\C_{xy})$ is symmetric in $\nu$ and $\l$
due to invariance of the Haar measure $dU$ under
the transformation $U\ra U^\dagger$.
These quantities are quite similar to those~\rf{KMG} for the
KM model.

The averages~\rf{gpG} obey quantum equations of motion which are known as
the Schwinger--Dyson \sloppy
or loop equations. Their derivation is analogous to the
one for the KM model which is presented in Appendix~C and the
resulting equation%
\footnote{We put $c=1$ in this section and in Appendix~C.}
\bea
 \left\langle
\ntr{}\Big(\frac{V^\p(\phi_x)}{\nu-\phi_x}
U(\C_{xy}) \frac{1}{\l-\phi_y}
U^\dagger(\C_{xy}) \Big) \right\rangle
\nonumber \\* -\sum_{\mu=1}^{q-1}
\left\langle \ntr{\Big( \phi_{x+\mu}
U(\C_{(x+\mu)x}) \frac{1}{\nu-\phi_x}U(\C_{xy})
 \frac{1}{\l-\phi_y}U^\dagger(\C_{(x+\mu)x}\C_{xy})}\Big)
 \right\rangle  \nonumber \\* =
  \left\langle \ntr{\Big(\frac{1}{\nu- \phi_x}\Big)}
\ntr{\Big(\frac{1}{\nu- \phi_x} U (\C_{xy})
\frac{1}{\l- \phi_y}
U^\dagger(\C_{xy}) \Big)} \right\rangle
\nonumber \\* + \delta_{xy}
\left\langle \ntr{}{\Big(
 \frac{1}{\nu-\phi_x} U(\C_{xy}) \frac{1}{\l-\phi_y}\Big)}
\ntr{}{ \Big(\frac{1}{\l-\phi_y}
U^\dagger(\C_{xy})\Big)}\right\rangle\,,
\label{gpAA}
\eea
where the sum over $\mu$ goes over
\be
\Delta = q-1
\label{Delta}
\ee
directions, coincides with \eq{AA} providing
\be
q-1=2D \,.
\label{equal}
\ee
Since $\Delta$ is nothing but the coordination number for the
$q$-simplex which equals $2D$ for the hypercubic lattice, the
equality~\rf{equal} simply means that the coordination numbers coincide in both
cases.

While the loop
equations which result from the variation of $\phi_x$ look similar,
the correlators in the second term on the r.h.s.\ of Eqs.~\rf{gpAA} and \rf{AA}
are, generally speaking, different. This term is proportional to $\delta_{xy}$
which does not vanish only for $x=y$, \ie only for a closed contour $\C_{xx}$.
However, the two models are {\it equivalent}\/ in the strong
coupling (or small-$c$) phase where
\bea
\left\langle
\ntr{}{\Big(\frac{1}{\nu- \phi_x}
U(\C_{xx})\frac{1}{\l-\phi_x}\Big)}
\ntr{}{\Big(U^\dagger(\C_{xx})
\frac{1}{\l-\phi_x}\Big)}\right\rangle  \nonumber \\*  =
\delta_{0A_{min}(\C)} \frac{1}{\nu-\l}(E_\l-E_\nu)E_\l
+{\cal O}\left({1\over N^2} \right)
\label{1overNp}
\eea
in both cases as $N\ra\infty$.
The r.h.s.\ of \eq{1overNp} does not vanish only for
the case of contractable contours $\C_{xx}=0$
with vanishing minimal area $A_{min}(\C)$.
Hence, the second term on the r.h.s.\ of the loop equation~\rf{gpAA}
vanishes for $\C_{xy}\neq0$ at $N=\infty$
independently of whether $C_{xy}$ is closed or open.

Finally, the second term on the l.h.s.\ of the loop equation~\rf{gpAA}
can be simplified at $N=\infty$ using the formula
\bea
\left\langle \ntr{\Big( \phi_{x+\mu}
U(\C_{(x+\mu)x}) \frac{1}{\nu-\phi_x}U(\C_{xy})
 \frac{1}{\l-\phi_y}U^\dagger(\C_{(x+\mu)x}\C_{xy})}\Big)
 \right\rangle  \nonumber \\* =
\left\langle \ntr{\Big(
\frac{F(\phi_x)}{\nu-\phi_x}U(\C_{xy})
 \frac{1}{\l-\phi_y}U^\dagger(\C_{xy})}\Big)
 \right\rangle
 \label{simplify}
\eea
which holds providing $\mu$ does not coincide with the direction
of the first link of the contour $\C_{xy}$ emanating from the point $x$.

The function
\be
F(\om)=\sum_{n=0}^\infty F_n \om^n\,,~~~~~~
F_0=\ntr{} \Big( \phi - \sum_{n=1}^\infty F_n \phi^n \Big)
\label{defF_0}
\ee
which enters \eq{simplify}
is determined by the pair correlator of the gauge fields
\be
\frac{\int dU\,\e^{\ntr{}( \phi U
\psi U^\dagger)} \ntr{} \Big(t^aU
\psi U^\dagger\Big)} {\int dU\,\e^{N
\tr{}(\phi U \psi U^\dagger)}}
=\sum_{n=1}^\infty  F_{n}
\ntr{}\left(t^a\phi^{n}\right)
\label{Lambda}
\ee
where $\phi$ and $\psi$ play
the role of external fields
and $t^a$ ($a=1,\ldots,N^2$--$1$) stand for the generators of the $SU(N)$.
\eq{Lambda} is based solely on the properties of the integral over
unitary matrices and
holds~\cite{Mig92a,Mig92d} at $N=\infty$.
The choice of $F_0$ which is not determined by \eq{Lambda}
is a matter of convenience~\cite{Mak93}.
How to calculate the function $F$ is explained in
the next subsection.

Using~\rf{simplify} and introducing
\be
\VVp(\om)\equiv V^\prime(\om)-(\Delta-1) F(\om)\,,
\label{defL}
\ee
we rewrite \eq{gpAA} and \eq{AA}
at large $N$, when
the factorization holds, in the same form
\be
 \int_{C_1} \frac{d \om}{2\pi i}
\frac{\VVp(\om)}{(\nu - \om)}\,G_{\om \l}(\C)=
E_{\nu}\, G_{\nu \l}(\C) + \l G_{\nu \l}(\C) - E_\nu +\delta_{0L(\C)}
\frac{1}{\nu-\l}(E_\l-E_\nu)E_\l
\label{sd}
\ee
where the contour $C_1$ --- the same as above --- encircles counterclockwise
singularities of the function $G_{\om\l}(\C)$.

The last term on the r.h.s.\ of \eq{sd} vanishes for $\C\neq0$ while
the explicit equation for $\C=0$ reads
\be
\int_{C_1}\frac {d\omega}{2\pi i}
\frac{\VVp(\omega)}{(\l-\omega)} E_{\omega}
- G_1(\l)=E_\l^2
\label{sd0}
\ee
where I have denoted
\be
E_\l \equiv \left\langle
\ntr{}\Big( \frac{1}{\l-\phi_x} \Big) \right\rangle
\label{defE}
\ee
analogously to \eq{2.19} for the KM model and defined the one-link average by
\be
G_1({\l}) \equiv \left\langle
\ntr{}\Big( \phi_{x+\mu} U(\C_{(x+\mu)x})
\frac{1}{\l-\phi_x}U^\dagger(\C_{(x+\mu)x}) \Big) \right\rangle
\ee
since the r.h.s.\ does not depend on $x$ and $\mu$.
Using~\eq{simplify} and noticing that
\be
\tilde{V}^\p(\om) =\VVp(\om)-F(\om)\,,
\ee
\eq{sd0} can be rewritten in the form of the loop equation of the Hermitean
one-matrix model
\be
\int_{C_1}\frac {d\omega}{2\pi i}
\frac{\tilde{V}^\p(\omega)}{(\l-\omega)} E_{\omega} = E_\l^2\,.
\ee

The explicit equation for the case when $\C$ coincides in~\eq{sd} with one
link reads
\be
 \int_{C_1} \frac{d \om}{2\pi i}
\frac{\VVp(\om)}{(\nu - \om)}\,G_{\om \l}=
E_{\nu}\, G_{\nu \l} + \l G_{\nu \l} - E_\nu \, ,
\label{main}
\ee
where $G_{\nu\l}$ is defined by \eq{L=1}.
\eq{sd0} is nothing but the $1/\l$ term of the expansion of \eq{main}
in $1/\l$.

\eq{main} coincides with that~\cite{DMS93} for the KM model
providing the relation~\rf{equal} holds.
It can be shown that \eq{sd} is satisfied for any contour $\C$ providing
\eq{main} is satisfied.
The open-loop averages~\rf{gpG} for the gauged Potts model coincide with
those~\rf{KMG} for the KM model providing the lengths of the contours
$\C_{xy}$ on a $q$-simplex and $C_{xy}$ on a $D$-dimensional lattice
coincide.

\subsection{The general solution}

\eq{main} looks the same as the loop equation
for the Hermitean two-matrix model~\cite{2mamo,Alf93,Sta93}.
This is because at $q=2$,
which is associated with the Hermitean two-matrix model,
$\Delta=1$ and the last
term on the r.h.s.\  of \eq{defL} disappears so that
one gets just ${\cal V}(\om)=V(\om)$.
To analyze it,
let us consider the Hermitean two-matrix model with the potential
\be
{\cal  V}(\Phi)=\sum_{m=1}^\infty \frac{g_{m}}{m} \phi^{m} \,.
\label{def L}
\ee

Expanding $G_{\nu\l}$ in $1/\nu$ \be
G_{\nu\l}=\frac{E_\l}{\nu} + \sum_{n=1}^\infty
\frac{G_n(\l)}{\nu^{n+1}} ~,~~~~~
G_n(\l)=\LA \ntr{}\Big( \phi^n_x U_{x(x+\mu)}\frac{1}{\l-\phi_{x+\mu}}
U^\dagger_{x(x+\mu)} \Big)\RA
\label{bcG}
\ee
and substituting into \eq{main}, one gets
\be
\sum_{m\geq1} g_{m}G_{m-1}(\l) = \l E_\l -1
\label{smart}
\ee
which determines $E_\l$ versus ${\cal  V}(\l)$.

The functions $G_n(\l)$ are expressed via $E_\l$ using the recurrence
relation
\be
G_{n+1}(\l)=\ci \frac{\VVp(\om)}{(\l-\om)}\, G_n(\om) - G_0(\l)\, G_n(\l)~,
{}~~~~~~G_0(\l)=E_\l
\label{recurrent}
\ee
which is obtained expanding \eq{main} in $1/\l$.
If ${\cal V}(\l)$ is a polynomial of degree $J$,
\eq{smart} contains $E_\l$ up to degree $J$ and the solution is
algebraic.
As is proven in Ref.~\cite{DMS93}:
\begin{itemize}
\item[i)] \vspace{-8pt}
 Equations which appear from
the next terms of the $1/\nu$-expansion of
\eq{main} are automatically
satisfied as a consequence of Eqs.~\rf{smart} and \rf{recurrent}.
\item[ii)] \vspace{-8pt}
$G_{\nu\l}$ is symmetric in $\nu$ and $\l$
for any solution of \eq{smart}.
The symmetry requirement can be used directly to determine $E_\l$
alternatively to \eq{smart}.
\end{itemize}
\vspace{-7pt}

The approach based on \eq{main} is equivalent~\cite{DMS93} to that
of Ref.~\cite{Mig92a}
which is based on the Riemann--Hilbert method.
To show this one takes the continuous and
discontinuous in $\nu$ parts of $G_{\nu\l}$ across the cut (cuts)
which are defined by Eqs.~\rf{5.22}.
The discontinuous part of \eq{main} then reads
\be
\Co G_{\nu\l} = 1 - \frac{1}{\Di E_\nu}
\Big( \l +\Co E_\nu -\VVp(\nu) \Big)\Di G_{\nu\l}~,~~~~~
\hbox{for \ } \nu\in \hbox{ cut}
\label{ME}
\ee
which coincides with the equation of Ref.~\cite{Mig92a}.

To obtain a formal solution to \eq{ME} for $ G_{\nu\l}$ versus $E_\nu$,
one notices that for any real $\nu$
\be
\frac{1-G_{(\nu+i0)\l}}{1-G_{(\nu-i0)\l}}
=\frac{1-\Co G_{\nu\l}- i\Di G_{\nu\l} }
{1-\Co G_{\nu\l}+i\Di G_{\nu\l}}
= \frac{\l+ \re E_\nu - i \im E_\nu - \VVp(\nu)}
{\l+ \re E_\nu + i \im E_\nu - \VVp(\nu)}
\label{RGproblem}
\ee
since $\Di G_{\nu\l}$ cancels at the cut (cuts) due to \eq{ME}.

The solution to the Riemann--Hilbert problem~\rf{RGproblem} for
$G_{\nu\l}$ can be expressed via $E_\l$ as follows~\cite{Mig92a,Gro92,Bou93}
\be
G_{\nu\l} = 1- \exp{\left\lbrace
- \ci \frac{1}{(\nu-\om)}\log{(\l-r_+(\om)})\right\rbrace}
\label{RGsolution}
\ee
where
\be
r_\pm(\l)= \frac{{\cal V}^\prime (\l)+F(\l)}{2} \pm i\pi \rho(\l) =
\left\{
\begin{array}{l}
{\cal V}^\prime (\l)-E_\l \\ F(\l)+E_\l
\end{array}
\right.\,.
\label{r}
\ee
\eq{RGsolution} solves \eq{RGproblem} and
holds for $\l$ outside of the cut where the asymptotic
expansion in $1/\l$ exists. It solves, hence, the recurrence
relation~\rf{recurrent}.

\subsection{Alternative one-link equations}

We have obtained in the previous subsection \eq{ME} from \eq{main}.
In order to show that \eq{main} can be inferred, in turn, from \eq{ME}, let us
consider the following auxiliary identity
\bea
\ci \frac{\Cow (E_\om G_{\om\l})}{\nu-\om} = E_\nu G_{\nu\l} -
\ci \frac{E_\om \Cow G_{\om\l}}{\nu-\om} \non = E_\nu G_{\nu\l} -E_\nu+
\ci \frac{(\l + \Cow E_\om -\VVp(\om))}{\nu-\om} G_{\om\l} = \non
E_\nu G_{\nu\l} - E_\nu + \l G_{\nu\l} -
\ci \frac{\VVp(\om)}{(\nu-\om)} G_{\om\l} +
\ci \frac{\Cow (E_\om G_{\om\l})}{\nu-\om}
\label{axulary}
\eea
which is based only on the analytic properties and \eq{ME}.
Canceling the l.h.s.\ with the last term on the r.h.s., one arrives at
\eq{main}.

Starting from \eq{ME}, one can derive also slightly different form of
\eq{main}. Let us consider the inverse quantity
\be
\tG_{\nu\l} = \frac{G_{\nu\l}}{1-G_{\nu\l}} \,.
\label{deftG}
\ee
It is convenient to introduce
\be
{\cal T}_{\nu\l} = G_{\nu\l} -1
\label{T}
\ee
and
\be
\tilde{{\cal T}}_{\nu\l} = \tilde{G}_{\nu\l}+1=\frac{1}{1- G_{\nu\l}}
\label{calT}
\ee
so that
\be
\tilde{{\cal T}}_{\nu\l} = - {\cal T}_{\nu\l}^{-1} \,.
\label{TT}
\ee
Eq.~\rf{main} can be written in terms of ${\cal T}_{\nu\l}$ as
\be
 \int_{C_1} \frac{d \om}{2\pi i}
\frac{\VVp(\om)}{(\nu - \om)}\,{\cal T}_{\om\l}=
E_{\nu}\, {\cal T}_{\nu\l} + \l {\cal T}_{\nu\l} +\l\, ,
\label{mainT}
\ee

It follows from \eq{TT} that
\bea
\Di \tilde{{\cal T}}_{\nu\l}
&=& \frac{ \Di {\cal T}_{\nu\l} }{|{\cal T}_{\nu\l}|^2}\,, \non
\Co \tilde{{\cal T}}_{\nu\l}
&=& -\frac{ \Co {\cal T}_{\nu\l} }{|{\cal T}_{\nu\l}|^2}
\label{inverse}
\eea
so that \eq{ME} can be rewritten as
\be
\Co \tilde{{\cal T}}_{\nu\l} =  \frac{1}{\Di E_\nu}
\Big( \l -\Co E_\nu -F(\nu) \Big)\Di \tilde{{\cal T}}_{\nu\l}~,~~~~~
\hbox{for \ } \nu\in \hbox{ cut}\,.
\label{FME}
\ee

The solution to the Riemann--Hilbert problem~\rf{FME} which is analogous
to~\rf{RGsolution} reads
\be
\tilde{{\cal T}}_{\nu\l} = \exp{\left\lbrace
- \ci \frac{1}{(\nu-\om)}\log{(\l-r_-(\om)})\right\rbrace} \,.
\label{FRGsolution}
\ee
Moreover, these two solution coincide due to the relation~\rf{TT}.

The condition for the r.h.s.\ of \eq{RGsolution} (or~\rf{FRGsolution}) to be
symmetric in $\nu$ and $\l$ is given~\cite{Bou93} by \eq{m}.
Since $G_{\nu\l}$ is symmetric in $\nu$ and $\l$, the master
field equation~\cite{Mig92a}
\be
E_\l = \pm \ci \log{(\l-r_\pm(\om)})\,,
\label{MFE}
\ee
obtained as the $1/\nu$ term of \eq{RGsolution},
will be satisfied as a consequence of \eq{m}
which guarantees the symmetry.

To derive an analogue of \eq{main} starting from \eq{MFE}, one proceeds
similarly to~\rf{axulary}:
\bea
\ci \frac{\Cow (E_\om \tilde{{\cal T}}_{\om\l})}{\nu-\om} =
E_\nu \tilde{{\cal T}}_{\nu\l} -
\ci \frac{E_\om \Cow \tilde{{\cal T}}_{\om\l}}{\nu-\om} \non =
E_\nu \tilde{{\cal T}}_{\nu\l} +
 \ci \frac{(-\l + \Cow E_\om + F(\om))}{\nu-\om} \tilde{{\cal T}}_{\om\l}
= \non E_\nu \tilde{{\cal T}}_{\nu\l} - \l \tilde{{\cal T}}_{\nu\l} +\l
+ \ci \frac{F(\om)}{(\nu-\om)} \tilde{{\cal T}}_{\om\l} + \ci \frac{\Cow E_\om
\tilde{{\cal T}}_{\om\l}}{\nu-\om} \,.
\label{Faxulary}
\eea
Hence, we get
\be
 \int_{C_1} \frac{d \om}{2\pi i}
\frac{F(\om)}{(\nu - \om)}\,\tilde{{\cal T}}_{\om\l}=
- E_{\nu}\, \tilde{{\cal T}}_{\nu\l} + \l \tilde{{\cal T}}_{\nu\l} -\l
\label{FmainT}
\ee
which yields
\be
 \int_{C_1} \frac{d \om}{2\pi i}
\frac{F(\om)}{(\nu - \om)}\,\tilde{G}_{\om \l}=
- E_{\nu}\, \tilde{G}_{\nu \l} + \l \tilde{G}_{\nu \l} - E_\nu
\label{Fmain}
\ee
which is an analog of \eq{main} with $\VVp$ replaced by $F$. Note that
with the relation~\rf{deftG} \eq{Fmain} differs from \eq{main}
only by the sign
of the first term on the r.h.s..

Eqs.~\rf{smart} and \rf{Fsmart} are
equivalent. It is a matter of practical
convenience which equation to solve.
\eq{main} is more convenient when $\VVp(\om)$ is a polynomial when it reduces
to an algebraic equation for $E_\nu$.
\eq{Fmain} in turn is more convenient if $F(\om)$ is a polynomial when it
reduces to an algebraic equation for $E_\nu$
\be
\sum_{m\geq1} F_{m}\tilde{G}_{m-1}(\l) = \l E_\l -1
\label{Fsmart}
\ee
and $\tilde{G}_{m}(\l)$, which are defined via the asymptotic expansion
\be
\tilde{G}_{\nu\l}=\frac{E_\l}{\nu} + \sum_{n=1}^\infty
\frac{\tilde{G}_n(\l)}{\nu^{n+1}}\,,
\label{FbcG}
\ee
can be expressed in terms of $E_\l$ using the recurrence relation
\be
\tilde{G}_{n+1}(\l)=\ci \frac{F(\om)}{(\l-\om)}\, \tilde{G}_n(\om)
+ \tilde{G}_0(\l)\, \tilde{G}_n(\l)~,
{}~~~~~~\tilde{G}_0(\l)=E_\l \,.
\label{Frecurrent}
\ee

The explicit solutions exist at  any $\Delta$
for the quadratic potential~\cite{Gro92}
and the logarithmic potential~\rf{V} when one gets~\cite{Mak93}
\be
\VVp(\l)=\frac{\a}{b-\l}-a~,~~~~ F(\l)=b-\frac{\a+1}{a+\l} \,.
\ee
It is easy to verify that~\eq{Fmain} is satisfied for $\tilde{G}_{\nu\l}$
given by~\rf{deftG} and \rf{3.8} analogously to \eq{main}.

\newsection{Discussion}

The KM model with the logarithmic potential~\rf{V} is a very nice, explicitly
solvable example of how a critical behavior can exist for $D>1$.
The analytic properties of the solution and the position of the eigenvalue
support are for $\a<0$ of the type advocated in Ref.~\cite{Mig92a}.
{}From this point of view it illustrates how such unusual analytic properties
can
emerge in matrix models.

The continuum theories which are obtained in the vicinity of the critical lines
are associated with the $\gamma_{str}=0$ and $\gamma_{str}=-1/2$ phases of
$2D$ gravity with matter independently of the fact that one starts from the
matrix model on a $D$-dimensional lattice. Similar results can be obtained for
$D=1$ if one considers a closed matrix chain
(for an open one the gauge field can be absorbed by a gauge transformation
of $\phi_x$). Therefore, an exact solution to this $D=1$
problem is found for the potential~\rf{V}.

While the phase with $\gamma_{str}=0$ seems to
coincide with the standard $d=1$ string,
a Kosterlitz--Thouless phase transition which occurs at the tri-critical point
separates it from the phase with $\gamma_{str}=-1/2$ which is, however, a novel
one since the continuum limit of matter at large distances sets up due to
a special behavior of the Itzykson--Zuber correlator of the gauge fields.
This phase never shows up, say, in the case of an open matrix chain
where the gauge field can be gauged away.

The existence of these two continuum limits does not mean that a $d>1$ phase
is impossible for the KM model with the logarithmic potential
on a $D$-dimensional lattice. An example is
a two-matrix model where a nontrivial phase with $\gamma_{str}=-1/3$
realizes at a tri-critical point while generically it has $\gamma_{str}=-1/2$.
It is most interesting to investigate the KM model with the potential~\rf{V} in
the vicinity of the tri-critical point which is not done in
the present paper. The point is that a singular behavior of the
Itzykson--Zuber correlator, which occurs at the tri-critical point, may
induce~\cite{DKSW93} a phase transition to a $d>1$ stringy phase.

It could be, however, that the logarithmic potential~\rf{V}
is too simple to exhibit a
nontrivial scaling behavior except for $\gamma_{str}=0$ and
$\gamma_{str}=-1/2$. From this point of view the results of Sect.~4 about
$\gamma_{str}>0$ in the large-$D$ limit for other potentials, including
the quartic one, may be interesting. I do not think, however, that an exact
solution can be obtained for these potentials. One should look rather for a
scaling solution just near the critical point. In particular, $1/D$-corrections
is the simplest problem to study. I hope that the fact that the mechanism of
Ref.~\cite{DDSW90} of getting $\gamma_{str}>0$ realizes for the KM model in the
large-$D$ limit can answer the question whether it corresponds to
branch polymer or stringy phases.

\subsection*{Acknowledgements}

I am indebted to  J.~Ambj{\o}rn, D.~Boulatov, C.~Kristjansen, A.~Migdal,
G.~Semenoff, N.~Weiss and K.~Zarembo for very useful discussions.


\setcounter{section}{0}
\setcounter{subsection}{0}
\appendix{The eigenvalue support for $\a<0$ \label{appB} }

To find the position of the contour $C$ in the complex $\l$ plane along which
the spectral density, $\rho(\l)$, has the support for $\a<0$, we apply the
criterion~\cite{Dav91} which is based
for the one-cut solution~\rf{one-cut} on the following integral
\be
G(\l) = \int_{\tx}^{\tl} d \tau \left[ \tV'(\tau) - 2 E_\tau \right] \,.
\label{G(l)}
\ee
The eigenvalue support is along the
curve from $\tx$ to $\ty$  determined by
\be
\re G(\l) =0~~~~~~~\hbox{for }~~\l\in C
\label{criterion}
\ee
which is embedded into the domain where $\re G(\l) <0$.
The spectral density is given by
\be
 \rho(\l) =\frac{1}{2 \pi i}
 \frac{d G(\l)}{d \l} ~~~~~~~\hbox{for }~~\l\in C   \,.
\ee

Let us consider in some detail the case $\a=-1$ when
\be
z=\b~,~~~~~~~\tx=\ty = \b -\frac{1}{2\b}
\label{C4}
\ee
according to Eqs.~\rf{alongalpha=-1} and \rf{roots}.
Since the points $\tx$ and $\ty$
coincide,  $C$ forms a closed loop which passes through the point
$\b-1/2\b$ and encircles the point $\b$.

The one-cut solution~\rf{zbone-cut} has for $\a=-1$ the following explicit
form
\be
E_\l = \frac{1}{2(\tl-\b)} - \b \pm \left[ \frac{1}{2(\tl-\b)} + \b \right] =
\left\{
\begin{array}{l}
\frac{1}{\tl-\b}  \\ -2\b
\end{array}
\right.
\label{array}
\ee
where the plus sign should be substituted when $\l$ is outside of the loop $C$
and the minus sign is associated with $\l$ which is inside $C$.
The expression~\rf{array} has the same expansion in $1/\l$ as an analytic
continuation of the solution for positive $\a$ to $\a=-1$ which were give
for $E_\l$ simply $1/(\tl-\b)$ with the pole at $\tl=\b$. The true
solution~\rf{array} is analytic at $\tl=\b$.

Substituting \rf{array} in to \eq{G(l)} one gets
\be
G(\l) = \pm \left[-2\b\tl -\log{(\tl-\b)} +2\b^2-1-\log{2\b} \right]\,.
\label{GG}
\ee
Introducing the new variable $\ka$ by
\be
\tl = \b + \frac{1}{2\b} \left( \ka -\frac 12  \right)
\label{defka}
\ee
which coincides with $\tl$ at $\b=1/2$,
one rewrites \eq{GG} in the form
\be
G(\ka) = \mp \left[ \ka+\frac 12 +\log{\left(\ka-\frac 12 \right)} \right]
\ee
which does not depend explicitly on $\b$.
Denoting
\be
\re{\ka} = u~,~~~~~~\im{\ka}=v\,,
\label{defuv}
\ee
we find the solution to the equation $\re G(\ka)=0$  to be
\be
v^2 = \e^{-2u-1} -\left( u - \fr 12 \right)^2 \,.
\label{curve}
\ee

The curve~\rf{curve} is depicted in
Fig.~\ref{support}
\begin{figure}[tbp]
\centering
\setlength{\unitlength}{0.240900pt}
\ifx\plotpoint\undefined\newsavebox{\plotpoint}\fi
\sbox{\plotpoint}{\rule[-0.500pt]{1.000pt}{1.000pt}}%
\begin{picture}(1500,900)(0,0)
\font\gnuplot=cmr10 at 10pt
\gnuplot
\sbox{\plotpoint}{\rule[-0.500pt]{1.000pt}{1.000pt}}%
\put(220.0,495.0){\rule[-0.500pt]{292.934pt}{1.000pt}}
\put(828.0,113.0){\rule[-0.500pt]{1.000pt}{184.048pt}}
\put(220.0,113.0){\rule[-0.500pt]{4.818pt}{1.000pt}}
\put(198,113){\makebox(0,0)[r]{$-1.5$}}
\put(1416.0,113.0){\rule[-0.500pt]{4.818pt}{1.000pt}}
\put(220.0,240.0){\rule[-0.500pt]{4.818pt}{1.000pt}}
\put(198,240){\makebox(0,0)[r]{$-1$}}
\put(1416.0,240.0){\rule[-0.500pt]{4.818pt}{1.000pt}}
\put(220.0,368.0){\rule[-0.500pt]{4.818pt}{1.000pt}}
\put(198,368){\makebox(0,0)[r]{$-0.5$}}
\put(1416.0,368.0){\rule[-0.500pt]{4.818pt}{1.000pt}}
\put(220.0,495.0){\rule[-0.500pt]{4.818pt}{1.000pt}}
\put(198,495){\makebox(0,0)[r]{$0$}}
\put(1416.0,495.0){\rule[-0.500pt]{4.818pt}{1.000pt}}
\put(220.0,622.0){\rule[-0.500pt]{4.818pt}{1.000pt}}
\put(198,622){\makebox(0,0)[r]{$0.5$}}
\put(1416.0,622.0){\rule[-0.500pt]{4.818pt}{1.000pt}}
\put(220.0,750.0){\rule[-0.500pt]{4.818pt}{1.000pt}}
\put(198,750){\makebox(0,0)[r]{$1$}}
\put(1416.0,750.0){\rule[-0.500pt]{4.818pt}{1.000pt}}
\put(220.0,877.0){\rule[-0.500pt]{4.818pt}{1.000pt}}
\put(198,877){\makebox(0,0)[r]{$1.5$}}
\put(1416.0,877.0){\rule[-0.500pt]{4.818pt}{1.000pt}}
\put(292.0,113.0){\rule[-0.500pt]{1.000pt}{4.818pt}}
\put(292,68){\makebox(0,0){$-1.5$}}
\put(292.0,857.0){\rule[-0.500pt]{1.000pt}{4.818pt}}
\put(470.0,113.0){\rule[-0.500pt]{1.000pt}{4.818pt}}
\put(470,68){\makebox(0,0){$-1$}}
\put(470.0,857.0){\rule[-0.500pt]{1.000pt}{4.818pt}}
\put(649.0,113.0){\rule[-0.500pt]{1.000pt}{4.818pt}}
\put(649,68){\makebox(0,0){$-0.5$}}
\put(649.0,857.0){\rule[-0.500pt]{1.000pt}{4.818pt}}
\put(828.0,113.0){\rule[-0.500pt]{1.000pt}{4.818pt}}
\put(828,68){\makebox(0,0){$0$}}
\put(828.0,857.0){\rule[-0.500pt]{1.000pt}{4.818pt}}
\put(1007.0,113.0){\rule[-0.500pt]{1.000pt}{4.818pt}}
\put(1007,68){\makebox(0,0){$0.5$}}
\put(1007.0,857.0){\rule[-0.500pt]{1.000pt}{4.818pt}}
\put(1186.0,113.0){\rule[-0.500pt]{1.000pt}{4.818pt}}
\put(1186,68){\makebox(0,0){$1$}}
\put(1186.0,857.0){\rule[-0.500pt]{1.000pt}{4.818pt}}
\put(1364.0,113.0){\rule[-0.500pt]{1.000pt}{4.818pt}}
\put(1364,68){\makebox(0,0){$1.5$}}
\put(1364.0,857.0){\rule[-0.500pt]{1.000pt}{4.818pt}}
\put(220.0,113.0){\rule[-0.500pt]{292.934pt}{1.000pt}}
\put(1436.0,113.0){\rule[-0.500pt]{1.000pt}{184.048pt}}
\put(220.0,877.0){\rule[-0.500pt]{292.934pt}{1.000pt}}
\put(45,495){\makebox(0,0){Im$\,\tl$}}
\put(828,23){\makebox(0,0){\shortstack{\mbox{} \\ Re$\,\tl$}}}
\put(470,571){\makebox(0,0)[r]{$G(\lambda)>0$}}
\put(1043,444){\makebox(0,0)[r]{$G(\lambda)<0$}}
\put(506,775){\makebox(0,0)[l]{$G(\lambda)<0$}}
\put(1007,291){\makebox(0,0)[l]{\shortstack{eigenvalue \\ support}}}
\put(1112.5,495){\makebox(0,0)[r]{{\footnotesize )}}}
\put(220.0,113.0){\rule[-0.500pt]{1.000pt}{184.048pt}}
\multiput(986.68,329.00)(-0.498,0.540){100}{\rule{0.120pt}{1.343pt}}
\multiput(986.92,329.00)(-54.000,56.213){2}{\rule{1.000pt}{0.671pt}}
\put(935,388){\vector(-1,1){0}}
\sbox{\plotpoint}{\rule[-0.300pt]{0.600pt}{0.600pt}}%
\multiput(336.99,871.75)(0.501,-0.969){9}{\rule{0.121pt}{1.264pt}}
\multiput(334.75,874.38)(7.000,-10.376){2}{\rule{0.600pt}{0.632pt}}
\multiput(344.00,858.81)(0.500,-0.934){19}{\rule{0.121pt}{1.250pt}}
\multiput(341.75,861.41)(12.000,-19.406){2}{\rule{0.600pt}{0.625pt}}
\multiput(356.00,837.02)(0.500,-0.890){19}{\rule{0.121pt}{1.200pt}}
\multiput(353.75,839.51)(12.000,-18.509){2}{\rule{0.600pt}{0.600pt}}
\multiput(368.00,816.35)(0.500,-0.818){21}{\rule{0.121pt}{1.119pt}}
\multiput(365.75,818.68)(13.000,-18.677){2}{\rule{0.600pt}{0.560pt}}
\multiput(381.00,795.43)(0.500,-0.802){19}{\rule{0.121pt}{1.100pt}}
\multiput(378.75,797.72)(12.000,-16.717){2}{\rule{0.600pt}{0.550pt}}
\multiput(393.00,776.43)(0.500,-0.802){19}{\rule{0.121pt}{1.100pt}}
\multiput(390.75,778.72)(12.000,-16.717){2}{\rule{0.600pt}{0.550pt}}
\multiput(405.00,757.74)(0.500,-0.737){21}{\rule{0.121pt}{1.027pt}}
\multiput(402.75,759.87)(13.000,-16.869){2}{\rule{0.600pt}{0.513pt}}
\multiput(418.00,738.64)(0.500,-0.758){19}{\rule{0.121pt}{1.050pt}}
\multiput(415.75,740.82)(12.000,-15.821){2}{\rule{0.600pt}{0.525pt}}
\multiput(430.00,720.85)(0.500,-0.714){19}{\rule{0.121pt}{1.000pt}}
\multiput(427.75,722.92)(12.000,-14.924){2}{\rule{0.600pt}{0.500pt}}
\multiput(442.00,703.85)(0.500,-0.714){19}{\rule{0.121pt}{1.000pt}}
\multiput(439.75,705.92)(12.000,-14.924){2}{\rule{0.600pt}{0.500pt}}
\multiput(454.00,687.31)(0.500,-0.616){21}{\rule{0.121pt}{0.888pt}}
\multiput(451.75,689.16)(13.000,-14.156){2}{\rule{0.600pt}{0.444pt}}
\multiput(467.00,671.26)(0.500,-0.626){19}{\rule{0.121pt}{0.900pt}}
\multiput(464.75,673.13)(12.000,-13.132){2}{\rule{0.600pt}{0.450pt}}
\multiput(479.00,656.26)(0.500,-0.626){19}{\rule{0.121pt}{0.900pt}}
\multiput(476.75,658.13)(12.000,-13.132){2}{\rule{0.600pt}{0.450pt}}
\multiput(491.00,641.50)(0.500,-0.575){21}{\rule{0.121pt}{0.842pt}}
\multiput(488.75,643.25)(13.000,-13.252){2}{\rule{0.600pt}{0.421pt}}
\multiput(504.00,626.47)(0.500,-0.582){19}{\rule{0.121pt}{0.850pt}}
\multiput(501.75,628.24)(12.000,-12.236){2}{\rule{0.600pt}{0.425pt}}
\multiput(516.00,612.68)(0.500,-0.538){19}{\rule{0.121pt}{0.800pt}}
\multiput(513.75,614.34)(12.000,-11.340){2}{\rule{0.600pt}{0.400pt}}
\multiput(528.00,599.68)(0.500,-0.538){19}{\rule{0.121pt}{0.800pt}}
\multiput(525.75,601.34)(12.000,-11.340){2}{\rule{0.600pt}{0.400pt}}
\multiput(539.00,588.50)(0.538,-0.500){19}{\rule{0.800pt}{0.121pt}}
\multiput(539.00,588.75)(11.340,-12.000){2}{\rule{0.400pt}{0.600pt}}
\multiput(552.00,576.50)(0.494,-0.500){19}{\rule{0.750pt}{0.121pt}}
\multiput(552.00,576.75)(10.443,-12.000){2}{\rule{0.375pt}{0.600pt}}
\multiput(564.00,564.50)(0.494,-0.500){19}{\rule{0.750pt}{0.121pt}}
\multiput(564.00,564.75)(10.443,-12.000){2}{\rule{0.375pt}{0.600pt}}
\multiput(576.00,552.50)(0.541,-0.501){17}{\rule{0.805pt}{0.121pt}}
\multiput(576.00,552.75)(10.330,-11.000){2}{\rule{0.402pt}{0.600pt}}
\multiput(588.00,541.50)(0.653,-0.501){15}{\rule{0.930pt}{0.121pt}}
\multiput(588.00,541.75)(11.070,-10.000){2}{\rule{0.465pt}{0.600pt}}
\multiput(601.00,531.50)(0.541,-0.501){17}{\rule{0.805pt}{0.121pt}}
\multiput(601.00,531.75)(10.330,-11.000){2}{\rule{0.402pt}{0.600pt}}
\multiput(613.00,520.50)(0.671,-0.501){13}{\rule{0.950pt}{0.121pt}}
\multiput(613.00,520.75)(10.028,-9.000){2}{\rule{0.475pt}{0.600pt}}
\multiput(625.00,511.50)(0.653,-0.501){15}{\rule{0.930pt}{0.121pt}}
\multiput(625.00,511.75)(11.070,-10.000){2}{\rule{0.465pt}{0.600pt}}
\multiput(638.00,501.50)(0.888,-0.501){9}{\rule{1.179pt}{0.121pt}}
\multiput(638.00,501.75)(9.554,-7.000){2}{\rule{0.589pt}{0.600pt}}
\multiput(650.00,496.99)(0.764,0.501){11}{\rule{1.050pt}{0.121pt}}
\multiput(650.00,494.75)(9.821,8.000){2}{\rule{0.525pt}{0.600pt}}
\multiput(662.00,504.99)(0.764,0.501){11}{\rule{1.050pt}{0.121pt}}
\multiput(662.00,502.75)(9.821,8.000){2}{\rule{0.525pt}{0.600pt}}
\multiput(674.00,512.99)(0.833,0.501){11}{\rule{1.125pt}{0.121pt}}
\multiput(674.00,510.75)(10.665,8.000){2}{\rule{0.563pt}{0.600pt}}
\multiput(687.00,520.99)(0.888,0.501){9}{\rule{1.179pt}{0.121pt}}
\multiput(687.00,518.75)(9.554,7.000){2}{\rule{0.589pt}{0.600pt}}
\multiput(699.00,527.99)(0.888,0.501){9}{\rule{1.179pt}{0.121pt}}
\multiput(699.00,525.75)(9.554,7.000){2}{\rule{0.589pt}{0.600pt}}
\multiput(711.00,534.99)(0.969,0.501){9}{\rule{1.264pt}{0.121pt}}
\multiput(711.00,532.75)(10.376,7.000){2}{\rule{0.632pt}{0.600pt}}
\multiput(724.00,541.99)(1.066,0.501){7}{\rule{1.350pt}{0.121pt}}
\multiput(724.00,539.75)(9.198,6.000){2}{\rule{0.675pt}{0.600pt}}
\multiput(736.00,547.99)(1.066,0.501){7}{\rule{1.350pt}{0.121pt}}
\multiput(736.00,545.75)(9.198,6.000){2}{\rule{0.675pt}{0.600pt}}
\multiput(748.00,553.99)(1.066,0.501){7}{\rule{1.350pt}{0.121pt}}
\multiput(748.00,551.75)(9.198,6.000){2}{\rule{0.675pt}{0.600pt}}
\multiput(760.00,559.99)(1.475,0.502){5}{\rule{1.710pt}{0.121pt}}
\multiput(760.00,557.75)(9.451,5.000){2}{\rule{0.855pt}{0.600pt}}
\multiput(773.00,564.99)(1.953,0.503){3}{\rule{1.950pt}{0.121pt}}
\multiput(773.00,562.75)(7.953,4.000){2}{\rule{0.975pt}{0.600pt}}
\multiput(785.00,568.99)(1.350,0.502){5}{\rule{1.590pt}{0.121pt}}
\multiput(785.00,566.75)(8.700,5.000){2}{\rule{0.795pt}{0.600pt}}
\multiput(797.00,573.99)(2.141,0.503){3}{\rule{2.100pt}{0.121pt}}
\multiput(797.00,571.75)(8.641,4.000){2}{\rule{1.050pt}{0.600pt}}
\multiput(810.00,577.99)(1.953,0.503){3}{\rule{1.950pt}{0.121pt}}
\multiput(810.00,575.75)(7.953,4.000){2}{\rule{0.975pt}{0.600pt}}
\put(822,581.25){\rule{2.550pt}{0.600pt}}
\multiput(822.00,579.75)(6.707,3.000){2}{\rule{1.275pt}{0.600pt}}
\put(834,584.25){\rule{2.550pt}{0.600pt}}
\multiput(834.00,582.75)(6.707,3.000){2}{\rule{1.275pt}{0.600pt}}
\put(846,587.25){\rule{2.750pt}{0.600pt}}
\multiput(846.00,585.75)(7.292,3.000){2}{\rule{1.375pt}{0.600pt}}
\put(859,589.75){\rule{2.891pt}{0.600pt}}
\multiput(859.00,588.75)(6.000,2.000){2}{\rule{1.445pt}{0.600pt}}
\put(871,591.75){\rule{2.891pt}{0.600pt}}
\multiput(871.00,590.75)(6.000,2.000){2}{\rule{1.445pt}{0.600pt}}
\put(883,593.25){\rule{3.132pt}{0.600pt}}
\multiput(883.00,592.75)(6.500,1.000){2}{\rule{1.566pt}{0.600pt}}
\put(896,594.25){\rule{2.891pt}{0.600pt}}
\multiput(896.00,593.75)(6.000,1.000){2}{\rule{1.445pt}{0.600pt}}
\put(908,595.25){\rule{2.891pt}{0.600pt}}
\multiput(908.00,594.75)(6.000,1.000){2}{\rule{1.445pt}{0.600pt}}
\put(957,595.25){\rule{2.891pt}{0.600pt}}
\multiput(957.00,595.75)(6.000,-1.000){2}{\rule{1.445pt}{0.600pt}}
\put(969,593.75){\rule{3.132pt}{0.600pt}}
\multiput(969.00,594.75)(6.500,-2.000){2}{\rule{1.566pt}{0.600pt}}
\put(982,591.75){\rule{2.891pt}{0.600pt}}
\multiput(982.00,592.75)(6.000,-2.000){2}{\rule{1.445pt}{0.600pt}}
\put(994,589.25){\rule{2.550pt}{0.600pt}}
\multiput(994.00,590.75)(6.707,-3.000){2}{\rule{1.275pt}{0.600pt}}
\multiput(1006.00,587.50)(1.953,-0.503){3}{\rule{1.950pt}{0.121pt}}
\multiput(1006.00,587.75)(7.953,-4.000){2}{\rule{0.975pt}{0.600pt}}
\multiput(1018.00,583.50)(2.141,-0.503){3}{\rule{2.100pt}{0.121pt}}
\multiput(1018.00,583.75)(8.641,-4.000){2}{\rule{1.050pt}{0.600pt}}
\multiput(1031.00,579.50)(1.350,-0.502){5}{\rule{1.590pt}{0.121pt}}
\multiput(1031.00,579.75)(8.700,-5.000){2}{\rule{0.795pt}{0.600pt}}
\multiput(1043.00,574.50)(0.888,-0.501){9}{\rule{1.179pt}{0.121pt}}
\multiput(1043.00,574.75)(9.554,-7.000){2}{\rule{0.589pt}{0.600pt}}
\multiput(1055.00,567.50)(0.833,-0.501){11}{\rule{1.125pt}{0.121pt}}
\multiput(1055.00,567.75)(10.665,-8.000){2}{\rule{0.563pt}{0.600pt}}
\multiput(1068.00,559.50)(0.599,-0.501){15}{\rule{0.870pt}{0.121pt}}
\multiput(1068.00,559.75)(10.194,-10.000){2}{\rule{0.435pt}{0.600pt}}
\multiput(1081.00,547.47)(0.500,-0.582){19}{\rule{0.121pt}{0.850pt}}
\multiput(1078.75,549.24)(12.000,-12.236){2}{\rule{0.600pt}{0.425pt}}
\multiput(1093.00,530.98)(0.500,-1.110){19}{\rule{0.121pt}{1.450pt}}
\multiput(1090.75,533.99)(12.000,-22.990){2}{\rule{0.600pt}{0.725pt}}
\put(920.0,597.0){\rule[-0.300pt]{8.913pt}{0.600pt}}
\multiput(336.99,113.00)(0.501,0.969){9}{\rule{0.121pt}{1.264pt}}
\multiput(334.75,113.00)(7.000,10.376){2}{\rule{0.600pt}{0.632pt}}
\multiput(344.00,126.00)(0.500,0.934){19}{\rule{0.121pt}{1.250pt}}
\multiput(341.75,126.00)(12.000,19.406){2}{\rule{0.600pt}{0.625pt}}
\multiput(356.00,148.00)(0.500,0.890){19}{\rule{0.121pt}{1.200pt}}
\multiput(353.75,148.00)(12.000,18.509){2}{\rule{0.600pt}{0.600pt}}
\multiput(368.00,169.00)(0.500,0.818){21}{\rule{0.121pt}{1.119pt}}
\multiput(365.75,169.00)(13.000,18.677){2}{\rule{0.600pt}{0.560pt}}
\multiput(381.00,190.00)(0.500,0.802){19}{\rule{0.121pt}{1.100pt}}
\multiput(378.75,190.00)(12.000,16.717){2}{\rule{0.600pt}{0.550pt}}
\multiput(393.00,209.00)(0.500,0.802){19}{\rule{0.121pt}{1.100pt}}
\multiput(390.75,209.00)(12.000,16.717){2}{\rule{0.600pt}{0.550pt}}
\multiput(405.00,228.00)(0.500,0.737){21}{\rule{0.121pt}{1.027pt}}
\multiput(402.75,228.00)(13.000,16.869){2}{\rule{0.600pt}{0.513pt}}
\multiput(418.00,247.00)(0.500,0.758){19}{\rule{0.121pt}{1.050pt}}
\multiput(415.75,247.00)(12.000,15.821){2}{\rule{0.600pt}{0.525pt}}
\multiput(430.00,265.00)(0.500,0.714){19}{\rule{0.121pt}{1.000pt}}
\multiput(427.75,265.00)(12.000,14.924){2}{\rule{0.600pt}{0.500pt}}
\multiput(442.00,282.00)(0.500,0.714){19}{\rule{0.121pt}{1.000pt}}
\multiput(439.75,282.00)(12.000,14.924){2}{\rule{0.600pt}{0.500pt}}
\multiput(454.00,299.00)(0.500,0.616){21}{\rule{0.121pt}{0.888pt}}
\multiput(451.75,299.00)(13.000,14.156){2}{\rule{0.600pt}{0.444pt}}
\multiput(467.00,315.00)(0.500,0.626){19}{\rule{0.121pt}{0.900pt}}
\multiput(464.75,315.00)(12.000,13.132){2}{\rule{0.600pt}{0.450pt}}
\multiput(479.00,330.00)(0.500,0.626){19}{\rule{0.121pt}{0.900pt}}
\multiput(476.75,330.00)(12.000,13.132){2}{\rule{0.600pt}{0.450pt}}
\multiput(491.00,345.00)(0.500,0.575){21}{\rule{0.121pt}{0.842pt}}
\multiput(488.75,345.00)(13.000,13.252){2}{\rule{0.600pt}{0.421pt}}
\multiput(504.00,360.00)(0.500,0.582){19}{\rule{0.121pt}{0.850pt}}
\multiput(501.75,360.00)(12.000,12.236){2}{\rule{0.600pt}{0.425pt}}
\multiput(516.00,374.00)(0.500,0.538){19}{\rule{0.121pt}{0.800pt}}
\multiput(513.75,374.00)(12.000,11.340){2}{\rule{0.600pt}{0.400pt}}
\multiput(528.00,387.00)(0.500,0.538){19}{\rule{0.121pt}{0.800pt}}
\multiput(525.75,387.00)(12.000,11.340){2}{\rule{0.600pt}{0.400pt}}
\multiput(539.00,401.00)(0.538,0.500){19}{\rule{0.800pt}{0.121pt}}
\multiput(539.00,398.75)(11.340,12.000){2}{\rule{0.400pt}{0.600pt}}
\multiput(552.00,413.00)(0.494,0.500){19}{\rule{0.750pt}{0.121pt}}
\multiput(552.00,410.75)(10.443,12.000){2}{\rule{0.375pt}{0.600pt}}
\multiput(564.00,425.00)(0.494,0.500){19}{\rule{0.750pt}{0.121pt}}
\multiput(564.00,422.75)(10.443,12.000){2}{\rule{0.375pt}{0.600pt}}
\multiput(576.00,437.00)(0.541,0.501){17}{\rule{0.805pt}{0.121pt}}
\multiput(576.00,434.75)(10.330,11.000){2}{\rule{0.402pt}{0.600pt}}
\multiput(588.00,448.00)(0.653,0.501){15}{\rule{0.930pt}{0.121pt}}
\multiput(588.00,445.75)(11.070,10.000){2}{\rule{0.465pt}{0.600pt}}
\multiput(601.00,458.00)(0.541,0.501){17}{\rule{0.805pt}{0.121pt}}
\multiput(601.00,455.75)(10.330,11.000){2}{\rule{0.402pt}{0.600pt}}
\multiput(613.00,468.99)(0.671,0.501){13}{\rule{0.950pt}{0.121pt}}
\multiput(613.00,466.75)(10.028,9.000){2}{\rule{0.475pt}{0.600pt}}
\multiput(625.00,478.00)(0.653,0.501){15}{\rule{0.930pt}{0.121pt}}
\multiput(625.00,475.75)(11.070,10.000){2}{\rule{0.465pt}{0.600pt}}
\multiput(638.00,487.99)(0.888,0.501){9}{\rule{1.179pt}{0.121pt}}
\multiput(638.00,485.75)(9.554,7.000){2}{\rule{0.589pt}{0.600pt}}
\multiput(650.00,492.50)(0.764,-0.501){11}{\rule{1.050pt}{0.121pt}}
\multiput(650.00,492.75)(9.821,-8.000){2}{\rule{0.525pt}{0.600pt}}
\multiput(662.00,484.50)(0.764,-0.501){11}{\rule{1.050pt}{0.121pt}}
\multiput(662.00,484.75)(9.821,-8.000){2}{\rule{0.525pt}{0.600pt}}
\multiput(674.00,476.50)(0.833,-0.501){11}{\rule{1.125pt}{0.121pt}}
\multiput(674.00,476.75)(10.665,-8.000){2}{\rule{0.563pt}{0.600pt}}
\multiput(687.00,468.50)(0.888,-0.501){9}{\rule{1.179pt}{0.121pt}}
\multiput(687.00,468.75)(9.554,-7.000){2}{\rule{0.589pt}{0.600pt}}
\multiput(699.00,461.50)(0.888,-0.501){9}{\rule{1.179pt}{0.121pt}}
\multiput(699.00,461.75)(9.554,-7.000){2}{\rule{0.589pt}{0.600pt}}
\multiput(711.00,454.50)(0.969,-0.501){9}{\rule{1.264pt}{0.121pt}}
\multiput(711.00,454.75)(10.376,-7.000){2}{\rule{0.632pt}{0.600pt}}
\multiput(724.00,447.50)(1.066,-0.501){7}{\rule{1.350pt}{0.121pt}}
\multiput(724.00,447.75)(9.198,-6.000){2}{\rule{0.675pt}{0.600pt}}
\multiput(736.00,441.50)(1.066,-0.501){7}{\rule{1.350pt}{0.121pt}}
\multiput(736.00,441.75)(9.198,-6.000){2}{\rule{0.675pt}{0.600pt}}
\multiput(748.00,435.50)(1.066,-0.501){7}{\rule{1.350pt}{0.121pt}}
\multiput(748.00,435.75)(9.198,-6.000){2}{\rule{0.675pt}{0.600pt}}
\multiput(760.00,429.50)(1.475,-0.502){5}{\rule{1.710pt}{0.121pt}}
\multiput(760.00,429.75)(9.451,-5.000){2}{\rule{0.855pt}{0.600pt}}
\multiput(773.00,424.50)(1.953,-0.503){3}{\rule{1.950pt}{0.121pt}}
\multiput(773.00,424.75)(7.953,-4.000){2}{\rule{0.975pt}{0.600pt}}
\multiput(785.00,420.50)(1.350,-0.502){5}{\rule{1.590pt}{0.121pt}}
\multiput(785.00,420.75)(8.700,-5.000){2}{\rule{0.795pt}{0.600pt}}
\multiput(797.00,415.50)(2.141,-0.503){3}{\rule{2.100pt}{0.121pt}}
\multiput(797.00,415.75)(8.641,-4.000){2}{\rule{1.050pt}{0.600pt}}
\multiput(810.00,411.50)(1.953,-0.503){3}{\rule{1.950pt}{0.121pt}}
\multiput(810.00,411.75)(7.953,-4.000){2}{\rule{0.975pt}{0.600pt}}
\put(822,406.25){\rule{2.550pt}{0.600pt}}
\multiput(822.00,407.75)(6.707,-3.000){2}{\rule{1.275pt}{0.600pt}}
\put(834,403.25){\rule{2.550pt}{0.600pt}}
\multiput(834.00,404.75)(6.707,-3.000){2}{\rule{1.275pt}{0.600pt}}
\put(846,400.25){\rule{2.750pt}{0.600pt}}
\multiput(846.00,401.75)(7.292,-3.000){2}{\rule{1.375pt}{0.600pt}}
\put(859,397.75){\rule{2.891pt}{0.600pt}}
\multiput(859.00,398.75)(6.000,-2.000){2}{\rule{1.445pt}{0.600pt}}
\put(871,395.75){\rule{2.891pt}{0.600pt}}
\multiput(871.00,396.75)(6.000,-2.000){2}{\rule{1.445pt}{0.600pt}}
\put(883,394.25){\rule{3.132pt}{0.600pt}}
\multiput(883.00,394.75)(6.500,-1.000){2}{\rule{1.566pt}{0.600pt}}
\put(896,393.25){\rule{2.891pt}{0.600pt}}
\multiput(896.00,393.75)(6.000,-1.000){2}{\rule{1.445pt}{0.600pt}}
\put(908,392.25){\rule{2.891pt}{0.600pt}}
\multiput(908.00,392.75)(6.000,-1.000){2}{\rule{1.445pt}{0.600pt}}
\put(957,392.25){\rule{2.891pt}{0.600pt}}
\multiput(957.00,391.75)(6.000,1.000){2}{\rule{1.445pt}{0.600pt}}
\put(969,393.75){\rule{3.132pt}{0.600pt}}
\multiput(969.00,392.75)(6.500,2.000){2}{\rule{1.566pt}{0.600pt}}
\put(982,395.75){\rule{2.891pt}{0.600pt}}
\multiput(982.00,394.75)(6.000,2.000){2}{\rule{1.445pt}{0.600pt}}
\put(994,398.25){\rule{2.550pt}{0.600pt}}
\multiput(994.00,396.75)(6.707,3.000){2}{\rule{1.275pt}{0.600pt}}
\multiput(1006.00,401.99)(1.953,0.503){3}{\rule{1.950pt}{0.121pt}}
\multiput(1006.00,399.75)(7.953,4.000){2}{\rule{0.975pt}{0.600pt}}
\multiput(1018.00,405.99)(2.141,0.503){3}{\rule{2.100pt}{0.121pt}}
\multiput(1018.00,403.75)(8.641,4.000){2}{\rule{1.050pt}{0.600pt}}
\multiput(1031.00,409.99)(1.350,0.502){5}{\rule{1.590pt}{0.121pt}}
\multiput(1031.00,407.75)(8.700,5.000){2}{\rule{0.795pt}{0.600pt}}
\multiput(1043.00,414.99)(0.888,0.501){9}{\rule{1.179pt}{0.121pt}}
\multiput(1043.00,412.75)(9.554,7.000){2}{\rule{0.589pt}{0.600pt}}
\multiput(1055.00,421.99)(0.833,0.501){11}{\rule{1.125pt}{0.121pt}}
\multiput(1055.00,419.75)(10.665,8.000){2}{\rule{0.563pt}{0.600pt}}
\multiput(1068.00,430.00)(0.599,0.501){15}{\rule{0.870pt}{0.121pt}}
\multiput(1068.00,427.75)(10.194,10.000){2}{\rule{0.435pt}{0.600pt}}
\multiput(1081.00,439.00)(0.500,0.582){19}{\rule{0.121pt}{0.850pt}}
\multiput(1078.75,439.00)(12.000,12.236){2}{\rule{0.600pt}{0.425pt}}
\multiput(1093.00,453.00)(0.500,1.110){19}{\rule{0.121pt}{1.450pt}}
\multiput(1090.75,453.00)(12.000,22.990){2}{\rule{0.600pt}{0.725pt}}
\put(920.0,393.0){\rule[-0.300pt]{8.913pt}{0.600pt}}
\end{picture}
\caption[x]   {\hspace{0.2cm}\parbox[t]{13cm}
{\small
   The eigenvalue support for $\a=-1$ and $\b=1/2$:
   the loop to the right of $\re \tl=-1/2$. The whole curve represents
   $G(\l)=0$.
   }}
\label{support}
\end{figure}
and is quite similar to the one for the Penner
model~\cite{CDL91}. The contour $C$ which represents the eigenvalue support
is the closed loop to the right or $u=-1/2$. The lines to the left of
$u=-1/2$ separate regions of positive and negative $G(\l)$ which is
positive to the left and negative to the right of these lines.
Since $\ka=\tl$ at $\b=1/2$, Fig.~\ref{support} represents
the eigenvalue support for the $\b=1/2$ while that for
an arbitrary $\b>1/2$ can be restored using \eq{defka}.

The variables~\rf{defuv} are convenient to perform calculations with the
spectral density which is determined by Eqs.~\rf{array} and \rf{curve} to be
\be
\rho(\tl) = \frac{i\b}{\pi}\left\{ \e^{2u+1}(u-\fr 12 - iv)+1 \right\}\,.
\ee
While this expression is complex, the element of probability
\be
d\tl\rho(\tl)=du \frac{\e^{-u-1/2}+2u}{2\pi \sqrt{\e^{-2u-1}
-\left( u - \fr 12 \right)^2}}
\label{prob}
\ee
with $v$ given by \eq{curve} is real and positive. It is easy to verify the
normalization condition
\be
\int_{-1/2}^{u_*} du
\frac{\e^{-u-1/2}+2u}{\pi \sqrt{\e^{-2u-1} -\left( u - \fr 12 \right)^2}} =1
\ee
where $u_*=0.77846454$ is the solution of the equation
\be
\e^{-u_*-1/2} -  u_* + \fr 12 =0\,.
\ee

\appendix{The convolution formula \label{appD} }

We prove in this appendix that~\rf{macrosolution} satisfies \eq{convolution}.

Let us first change the variables $x\ra1/x$, $y\ra1/y$ and rewrite
\eq{macrosolution} as
\be
C_c(\frac 1x, \frac 1y; \sqrt{u}) = x^2 y^2
\hC(x, y; \sqrt{u})
\ee
with
\be
\hC(x, y; \sqrt{u})
 = \frac{2\sqrt{u}}{(x-y)^2+2u(x+y)+u^2} \,.
\label{hatC}
\ee
Substituting $t\ra 1/t$, we rewrite \eq{convolution} as
\be
\frac 1\pi \int_0^\infty dt \sqrt{t} \hC(x,t;\sqrt{u})  \hC(t,y;\sqrt{v}) =
\hC(x,y;\sqrt{u}+\sqrt{v}) \,.
\label{hatconvolution}
\ee

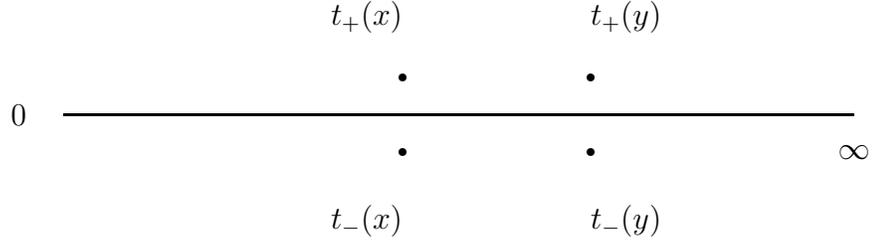
\begin{figure}[tbp]
\unitlength=1.00mm
\linethickness{0.6pt}
\begin{picture}(100.0,70.00)(-10,70)
\put(85.00,105.00){\circle*{1.00}}
\put(85.00,95.00){\circle*{1.00}}
\put(110.00,105.00){\circle*{1.00}}
\put(110.00,95.00){\circle*{1.00}}
\put(35.00,100.00){\makebox(0,0)[rc]{$0$}}
\put(145.00,95.00){\makebox(0,0)[cc]{$\infty$}}
\put(110.00,112.00){\makebox(0,0)[lb]{$t_+(y)$}}
\put(110.00,88.00){\makebox(0,0)[lt]{$t_-(y)$}}
\put(85.00,112.00){\makebox(0,0)[rb]{$t_+(x)$}}
\put(85.00,88.00){\makebox(0,0)[rt]{$t_-(x)$}}
\put(10.00,130.00){\circle{8.00}}
\put(10.00,130.00){\makebox(0,0)[cc]{$t$}}
\thicklines
\put(40.00,100.00){\line(1,0){105.00}}
\end{picture}
\caption[x]   {\hspace{0.2cm}\parbox[t]{13cm}
{\small
   The position of the cut of $\rho_c(t)$ in the $t$-plane (the bold line)
   and the positions of the poles  of the integrand in
   Eq.~(\ref{convintegral}), $t_\pm(x)$ and $t_\pm(y)$, which are given
   by Eqs.~\rf{tpmx} and \rf{tpmy}, respectively. }} \label{Fig.cutt}
\end{figure}

To prove \eq{hatconvolution}, let us calculate
the integral on the l.h.s.\ taking
residues at $t=t_\pm(x)$ and $t=t_\pm(y)$ with
\be
t_\pm(x) = x - u \pm 2i\sqrt{xu}
\label{tpmx}
\ee
and
\be
t_\pm(y) = y - v \pm 2i\sqrt{yv}
\label{tpmy}
\ee
being the poles of $\hC(x,t;\sqrt{u})$ or $\hC(t,y;\sqrt{v})$, respectively.
Since $t_+(x)$ and $t_-(x)$ (or $t_+(y)$ and $t_-(y)$) lie on the
opposite sides of the branch cut of $\sqrt{t}$ which goes along positive real
axis as in depicted in Fig.~\ref{Fig.cutt},
we get
\be
\sqrt{t_+(x)} = \sqrt{x} +i\sqrt{u}~,~~~~~~
\sqrt{t_-(x)} = -\sqrt{x} +i\sqrt{u}
\ee
and
\be
\sqrt{t_+(y)} = \sqrt{y} +i\sqrt{v}~,~~~~~~
\sqrt{t_-(y)} = -\sqrt{y} +i\sqrt{v}~.
\ee

Collecting the contribution from all four poles,
factoring the common denominator as
\bea
[(x-y)^2 + 2(x+y)(u+v) + (u-v)^2]^2 - 64 xy\, uv    \non =
[(x-y)^2 + 2(\sqrt{u}+\sqrt{v})^2(x+y)+(\sqrt{u}+\sqrt{v})^4] \non
\cdot\;[(x-y)^2+2(\sqrt{u}-\sqrt{v})^2 (x+y)+(\sqrt{u}-\sqrt{v})^4 ]
\eea
and cancelling the last factor with the numerator, we get finally \bea
\frac 1\pi \int_0^\infty \frac{dt
\sqrt{t}}{[(x-t)^2+2u(x+t)+u^2][(y-t)^2+2v(y+t)+v^2]}  \non=
\frac{\sqrt{u}+\sqrt{v}}{2\sqrt{uv}}
\frac{1}{[(x-y)^2+2(\sqrt{u}+\sqrt{v})^2(x+y)+(\sqrt{u}+\sqrt{v})^4]} \,.
\label{convintegral}
\eea
This completes the proof of \eq{hatconvolution}.

\appendix{Derivation of the loop equations}

Loop equations of the KM model or the gauged Potts model result,
as usual, from the invariance of the
integration measure under an infinitesimal shift of fields. One
distinguishes equations resulting from the shift
of the matter fields $\phi_x$  (that gives, in particular, the
lattice Klein--Gordon equation)
and that of the gauge
field $U_{xy}$ (that gives, in particular, the lattice Maxwell equation
in the case of the standard lattice gauge theory).

\subsection{Matter loop equation}

Let us consider an equation which results from the invariance of the measure
over $\phi$ in the open-loop average~\rf{KMG}
under an infinitesimal shift
\be
\phi_x\ra\phi_x+\xi_x
\label{xi}
\ee
of $\phi_x$ at the given site $x$ with $\xi_x$ being an infinitesimal
Hermitean matrix. For $\phi_x$ being a matrix from the adjoint representation
of $SU(N)$,
one should impose $\tr{\xi_x}=0$ in order for the shifted matrix to belong to
the adjoint representation as well. For the general Hermitean matrices
$\phi_x$,  $\xi_x$ is arbitrary Hermitean.

These two cases can be considered simultaneously
introducing $N^2$ generators
\be
[t^A]_{ij}=\Big( \delta_{ij},\, [t^a]_{ij} \Big)
\label{t^A}
\ee
with $t^a$ ($a=1,\ldots,N^2$--$1$) being the standard generators of
$SU(N)$.  The generators~\rf{t^A} obey the following normalization
\be
 \ntr{t^A t^B}=\delta^{AB}
\label{normalization}
\ee
and completeness condition
\be
[t^A]_{ij}[t^A]_{kl}=N \delta_{il} \delta_{kj} \,.
\label{completeness}
\ee
An arbitrary $N\times N$ Hermitean matrix $\phi$ can be represented as
\be
\phi=t^A \phi^A \hbox{ \ \ \ \ where \ \ \ }
\phi^A=\Big(  \ntr{\phi},\,  \ntr{t^a\phi} \Big)
\ee
with $\phi^0=\ntr{\phi}$ vanishing if $\phi$ is taken in the
adjoint representation of $SU(N)$.

To derive the loop equation I apply a trick similar to that used in deriving
loop equations of QCD~\cite{Mig83}. Let us consider the loop average
\be
\LA \ntr{\Big(t^A \frac{1}{\nu-\phi_x} U(C_{xy}) \frac{1}{\l-\phi_y}
U^\dagger(C_{xy})\Big)}  \RA =0\,,
\label{vanishing}
\ee
where the averaging is taken with the same measure as in \eq{spartition},
which vanishes due to the gauge invariance. Performing the
shift~\rf{xi} of $\phi_x$, using the invariance of the measure and
calculating $\d/\d \phi^B_x$, one gets
\bea
\LA \ntr{\Big(t^B V^\p(\phi_x)\Big)}
\ntr{\Big(t^A \frac{1}{\nu-\phi_x} U(C_{xy}) \frac{1}{\l-\phi_y}
U^\dagger(C_{xy})\Big)} \RA  \non
-\sum_{\mu=-D\atop \mu\neq0}^D \LA \ntr{\Big(
t^B U_{x(x+\mu)}\phi_{x+\mu} U_{x(x+\mu)}^\dagger
\Big)} \ntr{\Big( t^A \frac{1}{\nu-\phi_x} U(C_{xy})
\frac{1}{\l-\phi_y} U^\dagger(C_{xy}) \Big)} \RA \nonumber \\
=
 \LA \frac{\tr{}}{\hbox{N}^3} {\Big(t^A
 \frac{1}{\nu-\phi_x}t^B \frac{1}{\nu-\phi_x}
U(C_{xy})\frac{1}{\l-\phi_y} U^\dagger(C_{xy})\Big)} \RA \non +
\delta_{xy} \LA \frac{\tr{}}{\hbox{N}^3} {\Big(t^A \frac{1}{\nu-\phi_x}
U(C_{xy})\frac{1}{\l-\phi_y}
t^B \frac{1}{\l-\phi_y} U^\dagger(C_{xy})\Big)} \RA \;.
\label{AB}
\eea
The l.h.s.\ of this equation results from the variation of the action while
the r.h.s.\ represents the commutator term resulting from the variation of
the integrand.

The averaging over the gauge group picks up two nonvanising invariant
equations for the Hermitean matrices. The first one can be obtained
contracting \eq{AB} by $\delta^{AB}$ ($A,B=0,\ldots,N^2$--$1$)
while the second one is given by the $A,B=0$ component.

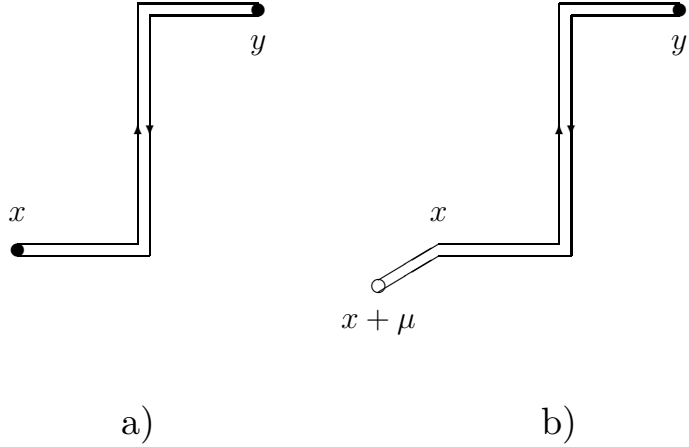
\begin{figure}[tbp]
\begin{picture}(120.00,190.00)(-34,20)
\unitlength=0.80mm
\linethickness{0.5pt}
\put(30.00,50.00){\line(3,0){20.00}}
\put(50.00,90.00){\line(3,0){20.00}}
\put(100.00,50.00){\line(3,0){20.00}}
\put(120.00,90.00){\line(3,0){20.00}}
\put(30.00,55.00){\makebox(0,0)[cc]{$x$}}
\put(70.00,83.00){\makebox(0,0)[cc]{$y$}}
\put(100.00,55.00){\makebox(0,0)[cc]{$x$}}
\put(140.00,83.00){\makebox(0,0)[cc]{$y$}}
\put(90.00,44.00){\line(5,3){10.00}}
\put(90.00,37.00){\makebox(0,0)[cc]{$x+\mu$}}
\put(50.00,20.00){\makebox(0,0)[cc]{{\large a)}}}
\put(120.00,20.00){\makebox(0,0)[cc]{{\large b)}}}
\put(50.00,50.00){\vector(0,1){20.00}}
\put(50.00,90.00){\line(0,-3){20.00}}
\put(120.00,50.00){\vector(0,1){20.00}}
\put(120.00,70.00){\line(0,3){20.00}}
\put(30.00,49.00){\circle*{2.00}}
\put(70.00,89.00){\circle*{2.00}}
\put(52.00,88.00){\vector(0,-1){20.00}}
\put(52.00,48.00){\line(0,3){20.00}}
\put(30.00,48.00){\line(3,0){22.00}}
\put(52.00,88.00){\line(3,0){18.00}}
\put(90.00,43.00){\circle{2.00}}
\put(140.00,89.00){\circle*{2.00}}
\put(122.00,88.00){\line(3,0){18.00}}
\put(122.00,88.00){\vector(0,-1){20.00}}
\put(122.00,48.00){\line(0,3){20.00}}
\put(100.00,48.00){\line(3,0){22.00}}
\put(90.00,42.00){\line(5,3){10.00}}
\end{picture}
\caption[x]   {\hspace{0.2cm}\parbox[t]{13cm}
{\small
   The graphic representation for $G_{\nu\l}(C_{xy})$ (a)
   and $G_{\nu\l}(C_{(x+\mu)x}C_{xy})$ (b) entering \eq{sd}. The
   empty circle represents  $\phi_{x+\mu}$ while the filled
   ones represent $\frac{1}{\nu-\phi_x}$ or
   $\frac{1}{\l-\phi_y}$.
   The oriented solid lines represent the
   path-ordered products $U(C_{xy})$ and $U(C_{(x+\mu)x}C_{xy})$.  The color
   indices are contracted according to the arrows.}}
\label{fig.a1}
\end{figure}

The first equation for the Hermitean model reads
\bea
 \left\langle
\ntr{}\Big(\frac{V^\p(\phi_x)}{\nu-\phi_x}
U(C_{xy}) \frac{1}{\l-\phi_y}
U^\dagger(C_{xy}) \Big) \right\rangle
\nonumber \\* -\sum_{\mu=-D\atop \mu\neq0}^D
\left\langle \ntr{\Big( \phi_{x+\mu}
U(C_{(x+\mu)x}) \frac{1}{\nu-\phi_x}U(C_{xy})
 \frac{1}{\l-\phi_y}U^\dagger(C_{(x+\mu)x}C_{xy})}\Big)
 \right\rangle  \nonumber \\* =
  \left\langle \ntr{\Big(\frac{1}{\nu- \phi_x}\Big)}
\ntr{\Big(\frac{1}{\nu- \phi_x} U (C_{xy})
\frac{1}{\l- \phi_y}
U^\dagger(C_{xy}) \Big)} \right\rangle
\nonumber \\* + \delta_{xy}
\left\langle \ntr{}{\Big(
 \frac{1}{\nu-\phi_x} U(C_{xy}) \frac{1}{\l-\phi_y}\Big)}
\ntr{}{ \Big(\frac{1}{\l-\phi_y}
U^\dagger(C_{xy})\Big)}\right\rangle
\label{AA}
\eea
where the path $C_{(x+\mu)x}C_{xy}$
on the l.h.s.\ is obtained by attaching the link $(x,x+\mu)$ to the path
$C_{xy}$ at the end point $x$ as is depicted in Fig.~\ref{fig.a1}.
Using the definition~\rf{KMG}, this equation can be
written finally at large $N$ in the form~\rf{sd}.

The second equation which is given by the $A,B=0$ component of \eq{AB} reads
\bea
 \LA \Big(\ntr{}
 V^\p(\phi_x) -\sum_{\mu=-D\atop \mu\neq0}^D
 \ntr{}{\phi_{x+\mu}}\Big)
 \ntr{}\frac{1}{\nu-\phi_x} U(C_{xy})
\frac{1}{\l-\phi_y} U^\dagger(C_{xy}) \RA \non
 =
 \LA\frac{\tr{}}{\hbox{N}^3} \frac{1}{\Big(\nu-\phi_x\Big)^2} U(C_{xy})
\frac{1}{( \l-\phi_y )} U^\dagger(C_{xy})\RA
\non
+ \delta_{xy}\LA\frac{\tr{}}{\hbox{N}^3} \frac{1}{(\nu-\phi_x)} U(C_{xy})
\frac{1}{\Big( \l-\phi_y\Big)^2} U^\dagger(C_{xy})\RA\,.
\non
\label{00}
\eea
In the large-$N$ limit when the factorization holds, \eq{00} is
automatically satisfied as a consequence of the ${\cal O}((\nu\l)^{-1})$ term
in \eq{AA}.

When the matrices $\phi_x$ belong to
the {\it adjoint\/} representation of $SU(N)$,
there exists only one invariant
which results from contracting \eq{AB} by
$\delta^{ab}$ ($a,b=1,\ldots,N^2$--$1$).
The point is that the $B=0$ component does not appear since solely the
 variation $\d/\d \phi^b_x$ is permitted for the
adjoint-representation matrices.  Therefore,
in order to obtain an analog of \eq{AA} for
the adjoint scalars, one should subtract the l.h.s.\ of
\eq{00} from its l.h.s.\ and the r.h.s.\ of \eq{00} from the r.h.s.\
to kill the $A$$=$$B$$=$$0$ component. The result
differs from \eq{AA} by contact terms which enter \eq{00} and
should not survive as $N\ra\infty$.
The vanishing of these contact terms in the
large-$N$ limit, when the factorization holds,
is obvious for an even potential $V(\phi)=V(-\phi)$.

\subsection{Maxwell open-loop equation}

The open-loop averages \rf{KMG} obey one more loop equation which results
from the invariance of the Haar measure over $U_{x(x+\mu)}$ under the shift
\be
U_{x(x+\mu)}\ra (1+i\epsilon_{x(x+\mu)})U_{x(x+\mu)}
\label{epsilon}
\ee
of $U_{x(x+\mu)}$ at the link $(x,\mu)$ with $\epsilon_{x(x+\mu)}$
being an infinitesimal traceless Hermitean matrix.

The loop equation can again be obtained by the trick~\cite{Mig83}.
Let us consider the loop average
\be \LA \ntr{\Big(
\frac{1}{\nu-\phi_x} U(C_{xz}) t^A U^\dagger(C_{xz})\Big)}
 \ntr{\Big( t^B U(C_{zy})
\frac{1}{\l-\phi_y} U^\dagger(C_{zy})\Big)}  \RA =0\,,
\label{mvanishing} \ee
where the averaging is taken with the same measure as
in \eq{spartition}, which vanishes due to the gauge invariance. Performing
the shift~\rf{epsilon} of $U_{z(z+\mu)}$ at some link  $(z,\mu)\in C_{xy}$ and
using the invariance of the Haar measure, one gets
\bea
\LA  \ntr{\Big(
\frac{1}{\nu-\phi_x}U(C_{xz}) t^A U^\dagger(C_{xz})\Big)}
\ntr{\Big( [t^B,t^c] U(C_{zy})
\frac{1}{\l-\phi_y} U^\dagger(C_{zy})\Big)} \right. \non
+  \ntr{\Big(
\frac{1}{\nu-\phi_x} U(C_{xz}) t^A U^\dagger(C_{xz})\Big)}
 \ntr{\Big( t^B U(C_{zy})
\frac{1}{\l-\phi_y} U^\dagger(C_{zy})\Big)} \non
\left. \cdot N \ntr{\Big([ \phi_z,t^c] U_\mu(z)
\phi_{z+\mu} U_\mu^\dagger(z)\Big)} \RA =0
\label{mABc}
\eea
where the contours are depicted in Fig.~\ref{fig.a2}.
\begin{figure}[tbp]
\begin{picture}(120.00,190.00)(-34,20)
\unitlength=0.80mm
\linethickness{0.5pt}
\put(30.00,50.00){\line(3,0){20.00}}
\put(50.00,90.00){\line(3,0){20.00}}
\put(30.00,55.00){\makebox(0,0)[cc]{$x$}}
\put(70.00,83.00){\makebox(0,0)[cc]{$y$}}
\put(50.00,25.00){\makebox(0,0)[cc]{{\large a)}}}
\put(50.00,50.00){\vector(0,1){9.00}}
\put(50.00,90.00){\line(0,-3){9.00}}
\put(50.00,71.00){\vector(0,1){10.00}}
\put(50.00,58.00){\line(0,3){10.00}}
\put(30.00,49.00){\circle*{2.00}}
\put(70.00,89.00){\circle*{2.00}}
\put(52.00,68.00){\vector(0,-1){11.00}}
\put(52.00,48.00){\line(0,3){9.00}}
\put(52.00,71.00){\line(0,3){8.00}}
\put(52.00,88.00){\vector(0,-1){9.00}}
\put(30.00,48.00){\line(3,0){22.00}}
\put(52.00,88.00){\line(3,0){18.00}}
\put(55.00,69.00){\makebox(0,0)[lc]{$z$}}
\put(48.00,67.00){\makebox(0,0)[rc]{$t^B$}}
\put(48.00,72.00){\makebox(0,0)[rc]{$t^A$}}
\put(51.00,68.00){\circle{2.00}}
\put(51.00,71.00){\circle{2.00}}
\put(121.00,71.00){\circle{2.00}}
\put(121.00,79.00){\circle{2.00}}
\put(120.00,71.00){\vector(0,1){5.00}}
\put(120.00,79.00){\line(0,-3){3.00}}
\put(122.00,79.00){\vector(0,-1){5.00}}
\put(122.00,71.00){\line(0,3){3.00}}
\put(125.00,66.00){\makebox(0,0)[cc]{$z$}}
\put(121.00,84.00){\makebox(0,0)[cc]{$z+\mu$}}
\put(117.00,67.00){\makebox(0,0)[cc]{$t^c$}}
\put(121.00,25.00){\makebox(0,0)[cc]{{\large b)}}}
\put(55.00,81.00){\makebox(0,0)[lc]{$C_{yz}$}}
\put(48.00,81.00){\makebox(0,0)[rc]{$C_{zy}$}}
\put(55.00,58.00){\makebox(0,0)[lc]{$C_{zx}$}}
\put(48.00,58.00){\makebox(0,0)[rc]{$C_{xz}$}}
\put(125.00,75.00){\makebox(0,0)[lc]{$C_{(z+\mu)z}$}}
\put(117.00,75.00){\makebox(0,0)[rc]{$C_{z(z+\mu)}$}}
\end{picture}
\caption[x]   {\hspace{0.2cm}\parbox[t]{13cm}
{\small
   The contours $C_{xz}$ ($C_{zx}$)
   or $C_{zy}$ ($C_{yz}$) (a) and $C_{(x+\mu)x}$ ($C_{x(x+\mu)}$) (b)
   for \eq{mABc}.
   }}
\label{fig.a2}
\end{figure}

The desired loop equation can be obtained contracting \eq{mABc} by the
structure constant $f^{abc}$ which is defined by the commutator
\be
[t^a,t^b]=i f^{abc} t^c\,.
\label{structureconstant}
\ee
The normalizations are fixed by \eq{normalization} which gives
\be
f^{abc} f^{dbc}= 2N \delta^{ab}\,.
\label{ff}
\ee
I shall utilize as well the following formulas \be i f^{abc}[t^a]_{ij}
[t^b]_{kl} =  [t^c]_{il} \delta_{kj} - [t^c]_{kj} \delta_{il} \label{ftt} \ee
and
\be
i f^{abc}[t^a]_{ij} [t^b]_{kl} [t^c]_{mn} = N (\delta_{il} \delta_{kn}
\delta_{mj} -\delta_{in} \delta_{kj} \delta_{ml})\,.
\label{fttt}
\ee
\eq{fttt} can be obtained from \eq{fttt} multiplying by $[t^c]_{mn}$ and
substituting the completeness condition \eq{completeness}. \eq{fttt} was
used, in particular, in studies~\cite{MM81} of loop equations of QCD.

The result of contracting \eq{mABc} by $i f^{abc}$ reads
\bea
\LA \ntr{\Big(
\frac{1}{\nu-\phi_x} U(C_{xy})
\frac{1}{\l-\phi_y} U^\dagger(C_{zy})\Big)}
-\ntr{\frac{1}{\nu-\phi_x}}
\ntr{\frac{1}{\l-\phi_y}}
\RA \non
=\frac 12  \LA
 \ntr{\Big( \frac{1}{\nu-\phi_x}
 \Big[ U(C_{xz)})\phi_z
U(C_{z(z+\mu)})\phi_{z+\mu} U(C_{(z+\mu)y})
\frac{1}{\l-\phi_y}U^\dagger(C_{xy}) \right. \non
- U(C_{x(z+\mu)})\phi_{z+\mu}
U^\dagger(C_{z(z+\mu)})\phi_z U(C_{zy})
\frac{1}{\l-\phi_y} U^\dagger(C_{xy})  \non
+U(C_{xy})\frac{1}{\l-\phi_y}
U^\dagger(C_{(z+\mu)y})\phi_{z+\mu} U^\dagger(C_{z(z+\mu)})
\phi_z U^\dagger(C_{x(z+\mu)}) \non
\left.  - U(C_{xy})\frac{1}{\l-\phi_y} U^\dagger(C_{zy})
\phi_z U(C_{z(z+\mu)})
\phi_{z+\mu} U^\dagger(C_{x(z+\mu)}) \Big] \Big)}\RA \;.
\label{mfabc}
\eea
where the contours entering the r.h.s.\ are depicted in Fig.~\ref{fig.a3}.
\begin{figure}[tbp]
\begin{picture}(120.00,290.00)(-34,-80)
\unitlength=0.80mm
\linethickness{0.5pt}
\put(30.00,49.00){\circle*{2.00}}
\put(30.00,50.00){\line(3,0){20.00}}
\put(50.00,50.00){\vector(0,1){13.00}}
\put(50.00,63.00){\line(0,1){4.00}}
\put(50.00,68.00){\circle{2.00}}
\put(50.00,69.00){\line(0,1){6.00}}
\put(50.00,76.00){\circle{2.00}}
\put(50.00,90.00){\line(0,-3){13.00}}
\put(50.00,90.00){\line(3,0){20.00}}
\put(70.00,89.00){\circle*{2.00}}
\put(52.00,88.00){\line(3,0){18.00}}
\put(52.00,88.00){\vector(0,-1){27.00}}
\put(52.00,48.00){\line(0,3){13.00}}
\put(30.00,48.00){\line(3,0){22.00}}
\put(102.00,49.00){\circle*{2.00}}
\put(102.00,50.00){\line(3,0){18.00}}
\put(120.00,50.00){\vector(0,1){13.00}}
\put(120.00,63.00){\line(0,3){13.00}}
\put(121.00,76.00){\circle{2.00}}
\put(122.00,76.00){\line(0,-3){8.00}}
\put(123.00,68.00){\circle{2.00}}
\put(124.00,68.00){\line(0,3){22.00}}
\put(124.00,90.00){\line(3,0){18.00}}
\put(142.00,89.00){\circle*{2.00}}
\put(126.00,88.00){\line(3,0){16.00}}
\put(126.00,88.00){\vector(0,-1){27.00}}
\put(126.00,48.00){\line(0,3){13.00}}
\put(102.00,48.00){\line(3,0){24.00}}
\put(30.00,55.00){\makebox(0,0)[cc]{$x$}}
\put(30.00,-15.00){\makebox(0,0)[cc]{$x$}}
\put(70.00,83.00){\makebox(0,0)[cc]{$y$}}
\put(70.00,13.00){\makebox(0,0)[cc]{$y$}}
\put(102.00,55.00){\makebox(0,0)[cc]{$x$}}
\put(102.00,-15.00){\makebox(0,0)[cc]{$x$}}
\put(142.00,83.00){\makebox(0,0)[cc]{$y$}}
\put(142.00,13.00){\makebox(0,0)[cc]{$y$}}
\put(46.00,68.00){\makebox(0,0)[rc]{$z$}}
\put(46.00,76.00){\makebox(0,0)[rc]{$z+\mu$}}
\put(56.00,-2.00){\makebox(0,0)[lc]{$z$}}
\put(56.00,6.00){\makebox(0,0)[lc]{$z+\mu$}}
\put(123.00,64.00){\makebox(0,0)[ct]{$z$}}
\put(123.00,-6.00){\makebox(0,0)[ct]{$z$}}
\put(116.00,76.00){\makebox(0,0)[rc]{$z+\mu$}}
\put(130.00,6.00){\makebox(0,0)[lc]{$z+\mu$}}
\put(86.00,71.00){\makebox(0,0)[cc]{{\large ---}}}
\put(86.00,1.00){\makebox(0,0)[cc]{{\large ---}}}
\put(16.00,1.00){\makebox(0,0)[cc]{{\large +}}}
\put(30.00,-21.00){\circle*{2.00}}
\put(30.00,-20.00){\line(3,0){20.00}}
\put(50.00,-20.00){\vector(0,1){13.00}}
\put(50.00,-7.00){\line(0,1){27.00}}
\put(50.00,20.00){\line(3,0){20.00}}
\put(70.00,19.00){\circle*{2.00}}
\put(52.00,18.00){\line(3,0){18.00}}
\put(52.00,18.00){\line(0,-1){11.00}}
\put(52.00,6.00){\circle{2.00}}
\put(52.00,-1.00){\line(0,1){6.00}}
\put(52.00,-2.00){\circle{2.00}}
\put(52.00,-3.00){\vector(0,-1){6.00}}
\put(52.00,-22.00){\line(0,3){13.00}}
\put(30.00,-22.00){\line(3,0){22.00}}
\put(102.00,-21.00){\circle*{2.00}}
\put(102.00,-20.00){\line(3,0){18.00}}
\put(120.00,-20.00){\vector(0,1){13.00}}
\put(120.00,-7.00){\line(0,3){27.00}}
\put(120.00,20.00){\line(3,0){22.00}}
\put(142.00,19.00){\circle*{2.00}}
\put(122.00,18.00){\line(3,0){20.00}}
\put(122.00,18.00){\line(0,-3){20.00}}
\put(125.00,6.00){\circle{2.00}}
\put(123.00,-2.00){\circle{2.00}}
\put(124.00,-2.00){\line(0,3){8.00}}
\put(126.00,6.00){\vector(0,-1){15.00}}
\put(126.00,-22.00){\line(0,3){13.00}}
\put(102.00,-22.00){\line(3,0){24.00}}
\end{picture}
\caption[x]   {\hspace{0.2cm}\parbox[t]{13cm}
{\small
   The graphic representation of the r.h.s.\ of \eq{mfabc}.
   }}
\label{fig.a3}
\end{figure}

Note that quantities of a new type appears on the r.h.s.\ of \eq{mfabc}
so that the set of equations is not closed. The meaning of this equation is
that it expresses the one-link correlator of $UUU^\dagger U^\dagger$ via
the correlator of $UU^\dagger$, \ie via $C(x,y)$.

\eop

\end{document}